\documentclass[12pt] {article}
\usepackage{latexsym}
\usepackage{amsmath}
\usepackage{mathrsfs}
\usepackage{amsfonts}
\usepackage{graphicx}
\usepackage{epsfig}

\usepackage[totalheight = 23cm, totalwidth = 17cm]{geometry}

\newcommand{\bea}{\begin{eqnarray}}
\newcommand{\eea}{\end{eqnarray}}
\newcommand{\be}{\begin{equation}}
\newcommand{\ee}{\end{equation}}

\def\non{\nonumber}

\newcommand{\ba}{\begin{align}}
\newcommand{\ea}{\end{align}}

\newcommand{\A}[1]{A^{(#1)}}

\newcommand{\G}[1]{G^{(#1)}}
\newcommand{\CHI}[1]{\chi^{(#1)}}

\newcommand{\PSI}[1]{\Psi^{(#1)}}

\begin{document}
\begin{titlepage}  
\rightline{\tt gr-qc/0508105}
\rightline{CERN-PH-TH/2005-154, ACT-06-05, MIFP-05-19}
\vspace{0.01cm}
\begin{center}
{\bf {\large Liouville Cosmology at Zero and Finite Temperatures}}
\end{center}
\begin{center}
\vskip 0.03in
{\bf John~Ellis}$^1$,~ {\bf Nikolaos E. Mavromatos}$^{2}$,~ 
{\bf Dimitri V. Nanopoulos}$^{3}$ \\ and \\ {\bf Michael Westmuckett}$^{2}$
\vskip 0.5cm
{\it
$^1${TH Division, Physics Department, CERN, CH-1211 Geneva 23, Switzerland}\\
$^2${Theoretical Physics, Physics Department, 
King's College London, Strand WC2R 2LS, UK}\\
$^3${George P. and Cynthia W. Mitchell Institute for Fundamental 
Physics, \\ Texas A\&M
University, College Station, TX 77843, USA; \\
Astroparticle Physics Group, Houston
Advanced Research Center (HARC),
Mitchell Campus,
Woodlands, TX~77381, USA; \\
Academy of Athens,
Division of Natural Sciences, 28~Panepistimiou Avenue, Athens 10679,
Greece}\\
} 

\vspace{0.5cm}
{\bf Abstract}
\end{center}

We discuss cosmology in the context of Liouville strings, characterized by
a central-charge deficit $Q^2$, in which target time is identified with
(the world-sheet zero mode of the) Liouville field: {\it Q-Cosmology}.  
We use a specific example of colliding brane worlds to illustrate the
phase diagram of this cosmological framework. The collision provides the
necessary initial cosmological instability, expressed as a departure from
conformal invariance in the underlying string model. The brane motion
provides a way of breaking target-space supersymmetry, and leads to
various phases of the brane and bulk Universes.  Specifically, we find a
hot metastable phase for the bulk string Universe soon after the brane
collision in which supersymmetry is broken, which we describe by means of
a subcritical world-sheet $\sigma$ model dressed by a space-like
Liouville field, representing finite temperature (Euclidean time). This
phase is followed by an inflationary phase for the brane Universe, in
which the bulk string excitations are cold. This is described by a
super-critical Liouville string with a time-like Liouville mode, whose
zero mode is identified with the Minkowski target time. Finally, we
speculate on possible ways of exiting the inflationary phase, either by
means of subsequent collisions or by deceleration of the brane Universe
due to closed-string radiation from the brane to the bulk. While phase
transitions from hot to cold configurations occur in the bulk string
universe, stringy excitations attached to the brane world remain
thermalized throughout, at a temperature which can be relatively high. 
The late-time behaviour
of the model results in dilaton-dominated dark energy and 
present-day acceleration of the expansion of the Universe, asymptoting 
eventually to zero.

\vspace{0.5cm}
\leftline{CERN-PH-TH/2005-154}
\leftline{August 2005}
\end{titlepage}

\section{Introduction} 

Formal developments in string theory~\cite{gsw} over the 
past decade, in particular the discovery of a consistent
way of studying quantum  domain wall structures (D-branes)~\cite{polchinski},
have opened up
novel ways of looking at both the microcosmos and the 
macrocosmos.
In the {\it microcosmos}, there are novel ways of compactification,
either via the observation~\cite{add} that extra dimensions
that are large compared to the string scale may be consistent with the 
foundations of string theory, or by 
viewing our four-dimensional world as a brane embedded in 
a bulk space-time, allowing for large extra dimensions
that might even be infinite in size~\cite{randal},
in a manner consistent with a large hierarchy between the Planck scale 
and the electroweak or supersymmetry 
breaking scale. In this modern approach, 
fields in the gravitational (super)multiplet of the (super)string,
or more generally those neutral under the Standard Model (SM) group,
are allowed to 
propagate in the bulk. This is not the case for non-Abelian gauge fields,
nor fields charged under the SM group, 
which are attached
to the brane world. 
In this approach, the weakness of  
gravity compared to the rest of the interactions is a result of 
the large compact dimensions, the compactification 
not necessarily being achieved
through conventional means, i.e., closing up the 
extra dimensions in spatial compact manifolds, 
but perhaps also through the involvement of shadow brane worlds 
with special reflecting properties, such as orientifolds,
which restrict the bulk dimension~\cite{ibanez}.
In such approaches the string scale $M_s$ is not necessarily 
identical to the four-dimensional
Planck mass scale $M_P$. Instead, the
two scales are related through the large compactification volume $V_6$:
\begin{equation}
M_P^2 = \frac{8M_s^2 V_6}{g_s^2}.
\label{planckstring}
\end{equation}
As for the {\it macrocosmos}, 
this modern approach has offered new insights 
into the cosmic evolution of our Universe. 
Novel ways of discussing cosmology
in brane worlds have been discovered over the past few years, 
which may revolutionize our way of approaching
issues such as inflation~\cite{langlois,ekpyrotic} and the present 
acceleration of the expansion of the Universe~\cite{emn04}. 

In parallel, mounting experimental evidence from diverse
astrophysical sources presents some important puzzles that string theory
must address if it is to provide a realistic description of Nature.
Observations of distant Type-1a supernovae~\cite{snIa}, as well as 
detailed
studies of the cosmic microwave background fluctuations by the WMAP
satellite~\cite{wmap}, indicate that our Universe is currently in an
accelerating epoch, and that 73\% of its energy density consists of dark
energy that does not cluster, but is present in `empty space'.
These issues are highly significant for string theory, motivating a novel
perspective on the treatment and understanding of string dynamics.  If the
dark energy turns out to be a true cosmological constant, leading to an
asymptotic de Sitter horizon, then the entire concept of the scattering
S-matrix, upon which perturbative string theory is built, breaks
down~\footnote{This is also true in more general scenarios for the vacuum
energy with a de Sitter horizon.}. This would cast doubt on the foundations of
string theory, at least as they are conventionally 
formulated. On the
other hand, even if models for relaxing the vacuum energy are invoked,
leading asymptotically to a vanishing vacuum energy density at large
cosmic times and consistency with an S-matrix, there is still the open
issue of embedding such models in (perturbative) string theory. In
particular, the formal question arises how to formulate consistently a
world-sheet $\sigma$-model description of strings propagating in such
time-dependent space-time backgrounds.

The standard world-sheet conformal invariance conditions of critical
string theory~\cite{gsw}, which are equivalent to target-space equations
of motion for the background fields on which the string propagates, are
very restrictive, allowing only vacuum solutions of critical strings. The
main problem may be illustrated as follows.  Consider the graviton
world-sheet $\beta$ function, which is simply the Ricci tensor of the
target space-time background to lowest order in $\alpha '$:
\begin{equation} 
\beta_{\mu\nu} = \alpha ' R_{\mu\nu}~,
\label{beta} \end{equation}
in the absence of other fields. Conformal invariance would require that
$\beta_{\mu\nu}=0$, implying a Ricci-flat background,
which is a solution to the {\it vacuum} Einstein equations. {\it A
priori}, a cosmological-constant vacuum solution is inconsistent with this
conformal invariance in strings, since it has a Ricci tensor $R_{\mu\nu} =
\Lambda g_{\mu\nu}$, where $g_{\mu\nu}$ is the metric tensor. The question
then arises how to describe cosmological backgrounds for strings that are
not vacuum solutions, but require the presence of a matter fluid yielding
a non-flat Ricci tensor.

One proposal for obtaining a non-zero cosmological constant in string
theory was made in~\cite{fischler}, according to which dilaton tadpoles on
higher-genus world-sheet surfaces produce additional modular infinities,
whose regularization leads to extra world-sheet structures in the $\sigma$
model. Since they do not appear at the world-sheet tree level, they lead
to modifications of the $\beta$ function such that the Ricci tensor of the
space-time background is now that of an (anti) de Sitter Universe, with a
cosmological constant fixed by the dilaton tadpole graph: $J_D >0$ ($J_D <
0$). The problem with this approach is the above-mentioned existence of an
asymptotic horizon in the de Sitter case, which prevents the proper
definition of asymptotic states, and hence a scattering matrix.  Since the
perturbative world-sheet formalism is based on such an S-matrix, there is
\emph{a priori} an inconsistency in the approach.

It was proposed in~\cite{aben} that one way out of this dilemma would be
to assume a time-dependent dilaton background, with a linear
dependence on time in the so-called $\sigma$-model frame. Such
backgrounds, even when the $\sigma$-model metric is flat, lead to exact
solutions (to all orders in $\alpha '$) of the conformal invariance
conditions of the pertinent stringy $\sigma$-model, and so are acceptable
solutions from a perturbative viewpoint. It was argued in~\cite{aben} that
such backgrounds describe linearly-expanding Robertson-Walker Universes,
which were shown to be exact conformal-invariant solutions, corresponding
to Wess-Zumino models on appropriate group manifolds.

The pertinent $\sigma$-model action in a background with graviton $G$,
antisymmetric tensor $B$ and dilaton $\Phi$ reads~\cite{gsw}:
\begin{equation} S_\sigma = \frac{1}{4\pi\alpha '} \int_\Sigma d^2\xi
[\sqrt{-\gamma} G_{\mu\nu} \partial_\alpha X^\mu \partial^\alpha X^\nu +
i\epsilon^{\alpha\beta} B_{\mu\nu} \partial_\alpha X^\mu \partial_\beta
X^\nu + \alpha '\sqrt{-\gamma}R^{(2)}\Phi], \label{sigmamodel}
\end{equation} where $\Sigma$ denotes the world-sheet, with metric
$\gamma$ and the topology of a sphere, $\alpha$ are world-sheet indices,
and $\mu,\nu $ are target-space-time indices. The important point
of~\cite{aben} was the role of target time $t$ as a specific dilaton
background, linear in that coordinate, of the form
\begin{equation}
\label{lineardil}
\Phi = {\rm const} - \frac{1}{2}Q~t,
\end{equation}
where $Q$ is a constant and $Q^2 > 0$ is the $\sigma$-model central-charge
deficit, allowing this {\it supercritical} string theory to be formulated
in some number of dimensions different from the critical number.
Consistency of the underlying world-sheet conformal field
theory, as well as modular invariance of the string scattering amplitudes,
required {\it discrete} values of $Q^2$, when expressed in units of the
string length $M_s$~\cite{aben}.
This was the first example of a non-critical string cosmology, with the
spatial target-space coordinates $X^i$, $i=1, \dots D-1$, playing the
r\^oles of $\sigma$-model fields. This non-critical string was not
conformally invariant, and hence required Liouville dressing~\cite{ddk}.
The Liouville field had time-like signature in target space, since the
central charge deficit $Q^2 > 0$ in the model
of~\cite{aben}, and its zero mode played the r\^ole of target time.

As a result of the non-trivial dilaton field, the 
Einstein term in the effective $D$-dimensional low-energy
field theory action is conformally rescaled by $e^{-2\Phi}$.
This requires a redefinition 
of the $\sigma$-model-frame space-time metric $g_{\mu\nu}^\sigma$ to
the `physical' Einstein metric  $g_{\mu\nu}^E$:
\begin{equation}
g_{\mu\nu}^E = e^{-\frac{4\Phi}{D-2}}G_{\mu\nu}~.
\label{smodeinst}
\end{equation}
Target time must also be rescaled, so that the
metric acquires the standard Robertson-Walker (RW) form in the normalized
Einstein frame for the effective action:
\begin{equation}
ds^2_E = -dt_E^2 + a_E^2(t_E) \left(dr^2 + r^2 d\Omega^2 \right),
\end{equation}
where we show the example of a spatially-flat RW metric for definiteness,
and $a_E(t_E)$ is an appropriate scale factor, which is a function of $t_E$
alone in the homogeneous cosmological backgrounds we assume throughout.

The Einstein-frame time is related to the 
time in the $\sigma$-model frame~\cite{aben} by:
\begin{equation}\label{einsttime}
dt_E = e^{-2\Phi/(D-2)}dt \qquad \to \qquad t_E = \int ^t e^{-2\Phi(t')/(D-2)}
dt'~. 
\end{equation} 
The linear dilaton background (\ref{lineardil}) yields
the following relation between the Einstein and $\sigma$-model frame 
times:
\begin{equation} 
t_E = c_1 + \frac{D-2}{Q}e^{\frac{Q}{D-2}t},
\end{equation}
where $c_{1,0}$ are appropriate (positive) constants.
Thus, a dilaton background (\ref{lineardil}) that is
linear in the $\sigma$-model time scales logarithmically with 
the Einstein time (Robertson-Walker cosmic time) $t_E$:
\begin{equation}\label{dil2}
\Phi (t_E) =({\rm const.}') - \frac{D-2}{2}{\rm ln}(\frac{Q}{D-2}t_E).
\end{equation} 
In this regime, the string coupling~\cite{gsw}: 
\begin{equation}
g_s = {\rm exp}\left(\Phi(t)\right)
\label{defstringcoupl}
\end{equation}
varies with the cosmic time $t_E$ as 
$g_s^2 (t_E) \equiv e^{2\Phi} \propto \frac{1}{t_E^{D-2}}$,
thereby implying a vanishing effective string coupling 
asymptotically in cosmic time. 
In the linear dilaton background of~\cite{aben}, the asymptotic
space-time metric in the Einstein frame reads:
\begin{equation} \label{metricaben}
ds^2 = -dt_E^2 + a_0^2 t_E^2 \left(dr^2 + r^2 d\Omega^2 \right)
\end{equation}
where $a_0$ a constant.
Clearly, there is no acceleration in the expansion of the Universe 
(\ref{metricaben}).

The effective low-energy action on the four-dimensional 
brane world for the gravitational multiplet 
of the string in the Einstein frame reads~\cite{aben}:
\begin{equation}
S_{\rm eff}^{\rm brane} = \int d^4x\sqrt{-g}\{ R  - 2(\partial_\mu \Phi)^2 
- \frac{1}{2} e^{4\Phi}( \partial_\mu b)^2 - \frac{2}{3}e^{2\Phi}\delta c 
\},
\label{effaction}
\end{equation}
where, as we discuss below, $b$ is the four-dimensional axion field
associated with a four-dimensional representation of the antisymmetric
tensor, and $\delta c = C_{\rm int} - c^*$,
where $C_{\rm int}$ is the central charge of the conformal world-sheet
theory corresponding to the transverse (internal) string dimensions, and
$c^*=22 (6)$ is the critical value of this internal central charge of the
(super)string theory for flat four-dimensional space-times.  The linear
dilaton configuration (\ref{lineardil}) corresponds, in this language, to
a background charge $Q$ of the conformal theory, which contributes a term
$-3Q^2$ (in our normalization)  to the total central charge. The latter
includes the contributions from the four uncompactified dimensions of our
world.  In the case of a flat four-dimensional Minkowski space-time, one
has $C_{\rm total} = 4 -3Q^2 + C_{\rm int} = 4 - 3Q^2 + c^* + \delta c$,
which should equal 26 (10). This implies that $C_{\rm int} = 22 + 3Q^2~(6 
+
3Q^2)$ for bosonic (supersymmetric) strings.  

An important result in~\cite{aben} was the discovery of an exact conformal
field theory corresponding to the dilaton background (\ref{dil2}) and a
constant-curvature (Milne) static metric in the $\sigma$-model frame (or,
equivalently, a linearly-expanding Robertson-Walker Universe in the
Einstein frame).  The conformal field theory corresponds to a
Wess-Zumino-Witten two-dimensional world-sheet model on a group manifold
$O(3)$ with appropriate constant curvature, whose coordinates correspond
to the spatial components of the four-dimensional metric and antisymmetric
tensor fields, together with a free world-sheet field corresponding to the
target time coordinate. The total central charge in this more general case
reads $C_{\rm total} = 4 - 3Q^2 - \frac{6}{k+2}+ C_{\rm int}$, where $k$
is a positive integer corresponding to the level of the Kac-Moody algebra
associated with the WZW model on the group manifold. The value of $Q$ is
chosen in such a way that the overall central charge $c=26$ and the theory
is conformally invariant. Since such unitary conformal field theories have
{\it discrete} values of their central charges, which accumulate to
integers or half-integers from {\it below}, it follows that the values of
the central charge deficit $\delta c$ are {\it discrete} and {\it finite}
in number.  From a physical point of view, this implies that the
linear-dilaton Universe may either stay in such a state for ever, for a
given $\delta c$, or tunnel between the various discrete levels before
relaxing to a critical $\delta c =0$ theory. It was argued in~\cite{aben}
that, due to the above-mentioned finiteness of the set of allowed discrete
values of the central charge deficit $\delta c$, the Universe could reach
flat four-dimensional Minkowski space-time, and thus exit from the
expanding phase, after a finite number of phase transitions.

The analysis in~\cite{aben} also showed, as we discuss below, that there
are tachyonic mass shifts of order $-Q^2$ in the bosonic string
excitations, but not in the fermionic ones. This implies the appearance of
tachyonic instabilities and the breaking of target-space supersymmetry in
such backgrounds, as far as the excitation spectrum is concerned. The
instabilities could trigger the cosmological phase transitions, since they
correspond to relevant renormalization-group world-sheet operators, and
hence initiate the flow of the internal unitary conformal field theory
towards minimization of its central charge, in accordance with the
Zamolodchikov $c$-theorem~\cite{zam}. As we discuss later on, in
semi-realistic cosmological models~\cite{dgmpp} such tachyons decouple
from the spectrum relatively quickly. On the other hand, as a result of
the form of the dilaton in the Einstein frame (\ref{dil2}), we observe
that the dark-energy density for this (four-dimensional)  Universe,
$\Lambda \equiv e^{2\Phi}\delta c$, is relaxing to zero with a
$1/t_E^{(D-2)}$ dependence on the Einstein-frame time for each of the
equilibrium values of $\delta c$.  Therefore, the breaking of
supersymmetry induced by the linear dilaton is only an
obstruction~\cite{witten}, rather than a spontaneous breaking, in the
sense that it appears only temporarily in the boson-fermion mass
splittings between the excitations, whilst the vacuum energy of the
asymptotic equilibrium theory vanishes.

In~\cite{emn} we went one step beyond the analysis in~\cite{aben}, and
considered more complicated $\sigma$-model metric backgrounds that did not
satisfy the $\sigma$-model conformal-invariance conditions, and therefore
needed Liouville dressing~\cite{ddk} to restore conformal invariance. Such
backgrounds could even be time-dependent, living in $(d+1)$-dimensional
target space-times. Various mathematically-consistent forms of
non-criticality can be considered, for instance cosmic catastrophes such
as the collision of brane worlds~\cite{gravanis,brany}. Such models lead
to supercriticality of the associated $\sigma$ models describing stringy
excitations on the brane worlds.  The Liouville dressing of such
non-critical models results in $(d+2)$-dimensional target spaces with two
time directions.  An important point in~\cite{emn} was the identification
of the (world-sheet zero mode of the) Liouville field with the target
time, thereby restricting the Liouville-dressed $\sigma$ model to a
$(d+1)$-dimensional hypersurface of the $(d+2)$-dimensional target space 
and maintaining the initial target space-time dimensionality. We
stress that this identification is possible only in cases where
the initial $\sigma$ model is supercritical, so that the Liouville mode
has time-like signature~\cite{aben,ddk}. In certain
models~\cite{gravanis,brany}, such an identification was proven to be
energetically preferable from a target-space viewpoint, since it minimized
certain effective potentials in the low-energy field theory corresponding
to the string theory at hand.

\begin{figure}[tb]
\begin{center}
\includegraphics[width=7cm]{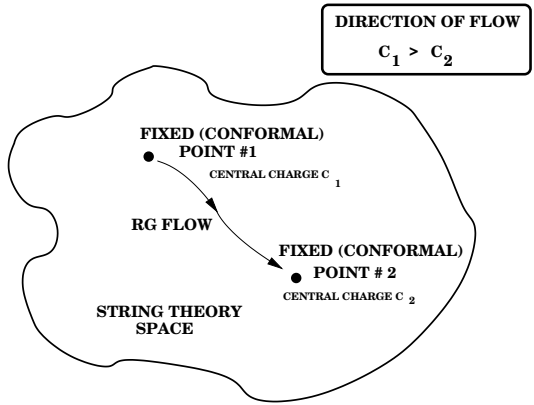}
\end{center}
\caption{\it 
A schematic view of string theory space, which is an infinite-dimensional
manifold endowed with a (Zamolodchikov) metric. The dots denote conformal
string backgrounds. A non-conformal string flows (in a two-dimensional
renormalization-group sense)  from one fixed point to another, either of
which could be a hypersurface in theory space.  The direction of the flow
is irreversible, and is directed towards the fixed point with a lesser
value of the central charge, for unitary theories, or, for general
theories, towards minimization of the degrees of freedom of the system.}
\label{fig:flow}
\end{figure}

All such cosmologies require some initial physical reason for the initial
departure from the conformal invariance of the underlying $\sigma$ model
that describes string excitations in such Universes. The reason could be
an initial quantum fluctuation, or, in brane models, a catastrophic cosmic
event such as the collision of two or more brane worlds.  Such
non-critical $\sigma$ models relax asymptotically to conformal $\sigma$
models, which may be viewed as equilibrium points in string theory space,
as illustrated in Fig.~\ref{fig:flow}. In some interesting cases of
relevance to cosmology~\cite{dgmpp}, which are particularly generic, the
asymptotic conformal field theory is that of~\cite{aben} with a linear
dilaton and a flat Minkowski target-space metric in the $\sigma$-model
frame. In others, the asymptotic theory is characterized by a constant
dilaton and a Minkowskian space-time~\cite{gravanis}.
Since, as we discuss below, the evolution of the central-charge deficit of
such a non-critical $\sigma$ model, $Q^2(t)$, plays a crucial r\^ole in
inducing the various phases of the Universe, including an inflationary
phase, graceful exit from it, thermalization and a contemporary phase of
accelerating expansion, we term such Liouville-string-based cosmologies 
{\it Q-Cosmologies}.

The use of Liouville strings to describe the evolution of our Universe has
a broad motivation, since non-critical strings are associated with
non-equilibrium situations, as are likely to have occurred in the early
Universe. The space of non-critical string theories is much larger than
that of critical strings. It is therefore remarkable that the departure
from criticality may enhance the predictability of string theory to the
extent that a purely stringy quantity such as the string coupling $g_s$
may become accessible to experiment via its relation to the present-era
cosmic acceleration parameter: $g_s^2 = -q^{0}$~\cite{emn04}.  Another
example arises in a non-critical string approach to inflation, if the Big
Bang is identified with the collision~\cite{ekpyrotic} of two
D-branes~\cite{brany}. In such a scenario, astrophysical observations may
place important bounds on the recoil velocity of the brane worlds after
the collision, and lead to an estimate of the separation of the branes at
the end of the inflationary period.

In such a framework, the identification of target time with a world-sheet
renormalization-group scale, the zero mode of the Liouville
field~\cite{emn}, provides a novel way of selecting the ground state of
the string theory. This is not necessarily associated with minimization of
energy, but could simply be a result of cosmic chance. It may be a random
global event that the initial state of our cosmos corresponds to a certain
Gaussian fixed point in the space of string theories, which is then
perturbed into a Big Bang by some relevant (in a world-sheet sense)
deformation, which makes the theory non-critical, and hence out of
equilibrium from a target space-time viewpoint. The theory then flows, as
indicated in Fig.~\ref{fig:flow}, along some specific
renormalization-group trajectory, heading asymptotically to some ground
state that is a local extremum
corresponding to an infrared fixed point of this perturbed world-sheet 
$\sigma$-model theory. This approach allows for many `parallel
universes' to be implemented, and our world might be just one of these.
Each Universe may flow between different fixed points, its trajectory
following a perturbation by a different operator.  It seems to us that
this scenario is more attractive and specific than the landscape
scenario~\cite{sussk}, which has recently been advocated as an framework
for parametrizing our ignorance of the true nature of string/M theory.

In this article we describe the main features of such non-critical string
cosmological models. The structure of the article is as follows:  in the
next Section we review briefly the basic properties of Liouville strings
at zero temperature, emphasizing the r\^ole of the Liouville mode as
target time~\cite{emn}.  We start in Section 2.1 with a comprehensive
description of the generic properties of Liouville dressing, and proceed
in Section 2.2 to present the basic features of (compactified)
cosmological models of non-critical strings upon which we rely later. We
give physical reasons for the departure from conformal invariance,in
Section 2.3 to discuss inflation in this framework, identifying the Hubble
parameter with the central-charge deficit $Q^2$ of the corresponding
supercritical $\sigma$ model describing string excitations in the
pertinent non-conformal cosmological backgrounds. In Section 2.4 we
discuss the late stages of such universes, in particular the r\^ole of the
(time-dependent) string coupling of the non-critical string in inducing
the present-day acceleration of the Universe. In Section 2.5 we discuss
the inclusion of matter in the late-stage evolution of the Universe, 
demonstrating the
differences of our non-equilibrium Liouville formalism from standard
Friedmann-Robertson-Walker (FRW) Cosmologies. 

In Section 3 we present a concrete example of non-critical strings, that
of colliding brane worlds~\cite{gravanis,emw}, where the departure from
criticality results from the cosmic collision of the branes. Specific
scenarios of this type are discussed in Sections 3.1 and 3.2, relevant
aspects of Type-IIA supergravity are discussed in Section 3.3 and
compactification issues in Section 3.4. Then, in Section 3.5 we present
within this framework some scenarios for supersymmetry breaking at zero
temperature, associated with either the presence of moving branes or the
existence of magnetic fields in internal manifolds of the compactified
space of brane worlds. The latter is compatible with the present value of
the dark energy, as inferred from observations.  In these models the dark
energy is viewed as a relaxation energy of the brane world, which was
excited after the collision.

In Section 4 we present a finite-temperature analysis of the early
Universe in the context of Liouville strings. We commence our analysis in
Section 4.1 with a review of the hot phase soon after the brane collision,
and give estimates of the bulk and brane excitation energies. In Section
4.2 we review the Liouville approach to finite-temperature heterotic
strings, whereby the temperature is associated with a space-like Liouville
mode in a subcritical string describing the thermal vacuum of the
heterotic string~\cite{kounnas}.  In Section 4.3 we discuss the
finite-temperature properties of the Type-IIA vacuum which characterises
the specific colliding-brane cosmological model mentioned above. In
Section 4.4 we discuss metastability properties of the Type-IIA thermal
supergravity vacuum, which, in contrast to the heterotic string case, is
an unstable vacuum, leading to the exit of our Universe from the hot phase
soon after the collision. In Section 4.5 we describe in some detail the
various phases of the colliding brane scenario and the associated phase
transitions, namely the transition from an initial hot phase to a cold
inflationary phase, and its subsequent exit from it.
We also pay attention to the fact that in these models the brane world
appears thermalized throughout, leading to decelerating brane motion in
the bulk, as a result of gravitational radiation leading towards thermal
equilibrium between the brane and bulk worlds. This deceleration is the
essential mechanism for the exit from the inflationary phase. We also
discuss in Section 4.5 several open questions associated with this phase,
in the context of our colliding-brane models, such as the possibility of a
second collision and some delicate issues concerning nucleosynthesis in
these models.

Finally, in Section 5 we present our conclusions and the outlook for
future work in Liouville Q-cosmologies.

\section{Non-Critical Liouville String Q-Cosmologies} 

\subsection{Zero-Temperature Liouville Formalism} 

We consider a $\sigma$-model action deformed by 
a family of 
vertex operators $V_i$, corresponding to 
`couplings' $g^i$, which represent \emph{non-conformal} background space-time
fields from the massless string multiplet, such as 
gravitons, $G_{\mu\nu}$, antisymmetric tensors, $B_{\mu\nu}$,
dilatons $\Phi$, their supersymmetric partners, \emph{etc.}:
\begin{equation}
S=S_{0}\left( X\right) +\sum_{i}g^{i}\int d^{2}z\,V_{i}\left( 
X \right)~,  
\label{action}
\end{equation}
where $S_0$ represents a conformal $\sigma$ model describing 
an equilibrium situation.
The non-conformality of the background means 
that the pertinent $\beta^i$ function $\beta^i \equiv dg^i/d{\rm ln}\mu 
\ne 0$, where $\mu$ is a world-sheet renormalization scale. 
Conformal invariance would imply restrictions on the 
background and couplings $g^i$, corresponding to 
the constraints $\beta^i = 0$, which are equivalent to equations
of motion derived from a target-space effective action for the corresponding
fields $g^i$.  The entire low-energy 
phenomenology and model building of critical string theory is based on such
restrictions~\cite{gsw}.

In the non-conformal case $\beta^i \ne 0$, the theory is in need of
dressing by the Liouville field $\phi$ in order to restore conformal
symmetry~\cite{ddk}. The field $\phi $ acquires dynamics through the
integration over world-sheet covariant metrics in the path integral, and
may be viewed as a local dynamical scale on the world sheet~\cite{emn}.  
If the central charge of the (supersymmetric)  matter theory is $c_{m}>25
(9)$ (i.e., supercritical), the signature of the kinetic term of the
Liouville coordinate in the dressed $\sigma$-model is opposite to that of
the $\sigma$-model fields corresponding to the other target-space
coordinates.  As mentioned previously, this opens the way to the important
step of interpreting the Liouville field physically by identifying its
world-sheet zero mode $\phi _{0}$ with the target time in supercritical
theories~\cite{emn}.  Such an identification emerges naturally from the
dynamics of the target-space low-energy effective theory by minimizing the
effective potential~\cite{gravanis}.

The action of the Liouville mode $\phi$ reads~\cite{ddk}:
\begin{equation}
S_L = S_{0}\left( X\right) 
+ \frac{1}{8\pi} \int_\Sigma d^2\xi \sqrt{\widehat \gamma} 
[ \pm (\partial \phi)^2 - QR^{(2)}\phi] + \int_\Sigma d^2\xi 
\sqrt{\widehat \gamma} g^i(\phi)V_i(X)~,
\label{liouvilleact} 
\end{equation} 
where ${\widehat \gamma}$ is a fiducial world-sheet metric, and 
the plus (minus) sign in front of the kinetic term of the 
Liouville mode pertains to subcritical (supercritical) strings. 
The dressed couplings $g^i(\phi)$  are obtained by the following procedure: 
\begin{equation}\label{dressing}
\int d^{2}z\,\,g_{i} \,V_{i}\left(X\right) \rightarrow \int
d^{2}z\,\,g_{i}(\phi)\,\,e^{\alpha _{i}\phi }\,V_{i}\left(X\right)~,
\end{equation}
where $\alpha_i$ is the 
``gravitational'' anomalous dimension. If the original non-conformal 
vertex operator 
has anomalous scaling dimension $\Delta_i -2$ (for closed strings, 
to which we restrict
ourselves for definiteness), where $\Delta_i$ is the conformal dimension,
and the central charge surplus
of the theory is $Q^2 = \frac{c_m - c^*}{3} > 0 $ (for bosonic strings
$c^*=25$, for superstrings $c^*=9$), then the condition
that the dressed operator is marginal on the world sheet implies the
relation:
\begin{equation}
\alpha_i (\alpha_i + Q) = 2 - \Delta_i~. 
\end{equation}
Imposing appropriate boundary conditions in the limit 
$\phi \to \infty$~\cite{ddk}, the acceptable solution is:
\begin{equation}
\alpha_i = -\frac{Q}{2} + \sqrt{\frac{Q^2}{4} + 2 - \Delta_i}~.
\end{equation}
The gravitational dressing is trivial for marginal couplings, 
$\Delta_i=2$, as it should be. 
This dressing applies also to higher orders in the 
perturbative $g^i$ expansion. For instance,
at the next order, where the deviation from marginality 
in the deformations of the undressed $\sigma$model 
is due to the operator product expansion coefficients 
$c^i_{jk}$ in the $\beta^i$ function, the Liouville-dressing
procedure implies the replacement~\cite{schmid}:
\begin{equation}
g^i \qquad \to \qquad g^ie^{\alpha_i\phi} + \frac{\pi~\phi}{Q \pm 
2\alpha_i}c^i_{jk}g^jg^ke^{\alpha_i \phi} ,
\label{dressing2}
\end{equation}
in order for the dressed operator to become marginal to this order
(the $\pm$ sign originates in (\ref{liouvilleact})).

In terms of the Liouville renormalization-group scale, one
has the following equation relating Liouville-dressed
couplings $g^i$ and $\beta$ functions in the non-critical string case:
\begin{equation}
{\ddot  g}^i + Q{\dot g}i = \mp\beta^i(g_j)~,
\label{liouvilleeq}
\end{equation}
where the - (+) sign in front of the $\beta$-functions 
on the right-hand-side applies to super(sub)critical strings, 
the overdot denotes differentiation with respect to the 
Liouville zero mode, $\beta^i$ is the world-sheet renormalization-group
$\beta$ function (but with the renormalized couplings replaced 
by the Liouville-dressed ones as defined by the procedure in
(\ref{dressing}), (\ref{dressing2})), 
and the minus sign on the right-hand side (r.h.s.) of (\ref{liouvilleeq}) 
is due to the time-like signature of the Liouville field.
Formally, the $\beta^i$ of the r.h.s.\ of (\ref{liouvilleeq})
may be viewed as 
power series in the (weak) couplings $g^i$. 
The covariant (in theory space)
${\cal G}_{ij}\beta^j$ function may be expanded as:
\begin{equation}
{\cal G}_{ij}\beta^j = 
\sum_{i_n} \langle V_i^L V_{i_1}^L \dots V_{i_n}^L \rangle_\phi g^{i_1} 
\dots g^{i_n}~,
\label{betaexp}
\end{equation}
where $V_i^L$ indicates Liouville dressing  \`a la (\ref{dressing}),
$ \langle \dots \rangle_\phi = \int d\phi d\vec{r}~{\rm exp}(-S(\phi, \vec{r}, g^j))$
denotes a functional average including Liouville integration, and 
$S(\phi, \vec{r}, g^i)$ is the Liouville-dressed $\sigma$-model
action, including the Liouville action~\cite{ddk}.

In the case of stringy $\sigma$ models, 
the diffeomorphism invariance of the target space results in the 
replacement of (\ref{liouvilleeq} by: 
\begin{equation} 
  {\ddot g}^i + Q(t){\dot g}^i = \mp{\tilde \beta}^i ,
\label{liouvilleeq2}
\end{equation} 
where the ${\tilde \beta}^i$ are the Weyl anomaly coefficients of the 
stringy $\sigma$ model in the background $\{ g^i \}$, which differ 
from the ordinary world-sheet renormalization-group $\beta^i$ functions
by terms of the form:
\begin{equation}
{\tilde \beta}^i = \beta^i + \delta g^i 
\end{equation}
where $\delta g^i$ denote transformations of the 
background field $g^i$ under infinitesimal general coordinate
transformations, e.g., for gravitons~\cite{gsw} 
${\tilde \beta}^G_{\mu\nu} = 
\beta^G_{\mu\nu} + \nabla_{(\mu} W_{\nu)}$, 
with $W_\mu = \nabla _\mu \Phi$,
and $\beta^G_{\mu\nu} = R_{\mu\nu}$ to order $\alpha '$ 
(one $\sigma$-model loop).

The set of equations (\ref{liouvilleeq}),(\ref{liouvilleeq2})  defines the
\emph{generalized conformal invariance conditions}, expressing the
restoration of conformal invariance by the Liouville mode.  The solution
of these equations, upon the identification of the Liouville zero mode
with the original target time, leads to constraints in the space-time
backgrounds~\cite{emn,gravanis}, in much the same way as the conformal
invariance conditions $\beta^i = 0$ define consistent space-time
backgrounds for critical strings~\cite{gsw}.  It is important to
remark~\cite{emninfl, emn} that the equations (\ref{liouvilleeq2}) can be
derived from an action.  This follows from general properties of the
Liouville renormalization group, which guarantee that the appropriate
Helmholtz conditions in the string-theory space $\{ g^i \}$ for the
Liouville-flow dynamics to be derivable from an action principle are
satisfied.

To be specific, consider the case of the $\sigma$ model (\ref{sigmamodel}),
and the ${\cal O}(\alpha ')$ Weyl anomaly coefficients~\cite{aben}, 
assuming that the $\sigma$-model target space is a $D$-dimensional 
space-time $(X^0,{\vec X})$. The pertinent $\beta$ functions are 
 \begin{eqnarray} 
&& {\tilde \beta}^G_{\mu\nu} = \alpha ' \left( R_{\mu\nu} + 2 \nabla_{\mu} 
\partial_{\nu} \Phi 
 - \frac{1}{4}H_{\mu\rho\sigma}H_{\nu}^{\rho\sigma}\right)~, \nonumber \\
&& {\tilde \beta}^B_{\mu\nu} = \alpha '\left(-\frac{1}{2}\nabla_{\rho} H^{\rho}_{\mu\nu} + 
H^{\rho}_{\mu\nu}\partial_{\rho} \Phi \right)~, \nonumber \\
&& {\tilde \beta}^\Phi = \beta^\Phi - \frac{1}{4}G^{\rho\sigma}\beta^G_{\rho\sigma} = 
\frac{1}{6}\left( C^{(D)} - 26 \right)~, \\
&& C^{(D)} = D -\frac{3}{2}\alpha ' [R - \frac{1}{12}H^2 - 4(\nabla \Phi)^2 + 4\Box \Phi ]~,
\label{bfunct}
\end{eqnarray} 
where $\alpha '$ is the Regge slope~\cite{gsw},
the Greek indices are D-dimensional,
and $H_{\mu\nu\rho}= \partial_{[\mu}B_{\nu\rho]}$ is the
field strength of the $B$ field.

Dressing this $\sigma$ model with a Liouville mode  
results in the 
appropriate equations (\ref{liouvilleeq2}), and it is straightforward
to show that these 
can be \emph{derived} from the following $(D+1)$-dimensional 
action in the $\sigma$-model (string) frame~\cite{schmid}: 
\begin{eqnarray}
&& I^{(D+1)} = \int d\phi d^DX \sqrt{G}\sqrt{|G_{\phi\phi}|} 
e^{-2\Phi}\{ C^{(D)} - 25 + \nonumber \\ 
&& 3G^{\phi\phi}[({\dot \Phi} - 
\frac{1}{2}G^{\mu\nu}{\dot G}_{\mu\nu})^2  - \frac{1}{4}
G^{\mu\nu}G^{\rho\lambda}\left({\dot G}_{\mu\rho}{\dot G}_{\nu\lambda}
+ {\dot B}_{\mu\rho}{\dot B}_{\nu\lambda}\right)]\}~,
\label{dplus1action} 
\end{eqnarray} 
where $G,B,\Phi$ are all $D$-dimensional fields, depending
in general on $X^0,\phi,{\vec X}$, 
and the overdot denotes 
differentiation with respect to the Liouville zero mode $\phi$.
In the Liouville dressing procedure~\cite{ddk} 
we employ in this work one has the 
normalization $G_{\phi\phi} = -1$.  This justifies the 
presence of only spatial components of the metric and antisymmetric tensor 
fields in the terms inside the $[ \dots ]$ in (\ref{dplus1action}). 
This action may be schematically represented in the form~\cite{emninfl}:
\begin{equation}\label{lambdaaction}  
I^{(D+1)} = \int d\phi d^DX e^{-\varphi} \{ C^{(D)}(X) - 25 -3[{\dot \varphi}^2 
-\frac{1}{4}\left({\dot \lambda}^I {\cal G}_{IJ} {\dot \lambda}^J\right)]\}~,
\end{equation} 
where 
$\lambda_I = \{ G, B\}$, ${\cal G}_{IJ} =
G^{\mu\rho}G^{\nu\lambda}$ is a Zamolodchikov metric
in $\lambda_I$ space~\footnote{Note that this is compatible with the
definition of this metric in string space as a two-point correlation
function of appropriate vertex operators, as explicit ${\cal O}(\alpha ')$ 
computations have demonstrated~\cite{mm}.}, 
and $\varphi \equiv 2\Phi -{\rm ln}\sqrt{G}$ is a rescaled
dilaton~\cite{schmid}, which guarantees the diagonalization of the appropriate 
Zamolodchikov metric in the string theory space $(\Phi, \lambda_I)$,
with ${\cal G}_{\varphi\varphi} = 1$.

Upon the identification of $\phi$ with the target time $X^0$, the 
$(D+1)$-dimensional action is constrained onto a $D$-dimensional
hypersurface $(X^0=\phi,~{\vec X})$. In that case the 
resulting $D$-dimensional target-space-time action 
reads:
\begin{eqnarray} 
&& I^{D} = -\frac{3}{2}\alpha '\int d^DX \sqrt{G} e^{-2\Phi}\left(R +  4(\nabla \Phi)^2 
-\frac{1}{12}H^2 
- \frac{2}{3\alpha '}(D - 25)\right)  + {\cal I}_\phi~, 
\nonumber \\
&& {\cal I}_\phi \equiv  
\int d^DX \sqrt{G} e^{-2\Phi}\left(
-3[({\dot \Phi} - \frac{1}{2}G^{\mu\nu}{\dot G}_{\mu\nu})^2 - \frac{1}{4}
G^{\mu\nu}G^{\rho\lambda}\left({\dot G}_{\mu\rho}{\dot G}_{\nu\lambda}
+ {\dot B}_{\mu\rho}{\dot B}_{\nu\lambda}\right)]\}\right).
\label{ddimaction}
\end{eqnarray}
The extra piece, ${\cal I}_\phi$, 
as compared with a standard string theory target-space effective action, 
describes the non-equilibrium effects associated with the 
Liouville flow. 
In general, one may have critical target-space dimensionality $D=25$, but 
deviations from conformal invariance, due for example
to the recoil of brane worlds during collisions or other
catastrophic cosmic events, will be the topic 
of interest to us here. 
The non-critical central-charge term proportional 
to $\int d^DX\sqrt{G}e^{-2\Phi}(D-25)$ in the action (\ref{ddimaction}) 
may then be written in the form
\begin{equation} 
\frac{3\alpha '}{2}\int d^DX \sqrt{G} e^{-2\Phi} \frac{2}{\alpha '}Q^2 
 \label{centralcharge}
\end{equation}
in the normalization of the Einstein term appearing in (\ref{ddimaction}), 
where $Q^2$ represents the central charge deficit (whatever its origin) 
of the appropriate
$\sigma$ model: $Q^2 = (C- c^*)/3$, 
describing closed-string excitations in an
appropriate non-conformal background.

\subsection{Cosmological Liouville Models: Generic Features} 

There are many cosmological models that fall in the above category
of Liouville strings, notably models  
with catastrophic cosmic events such as the collision of
two brane worlds~\cite{gravanis,emw}, which we
concentrate upon later in this work.
When one considers 
a brane moving in a bulk space in the presence of other brane
worlds~\cite{emw}, there are in general non-trivial 
potentials between the moving branes: only the static configuration
and certain other special configurations are supersymmetric in target 
space, with vanishing ground-state energy~\cite{emw}.
When one considers string theory excitations in such Universes,
the corresponding $\sigma$ model is \emph{non-critical}. 
In the model of \cite{emw}, as we discuss later, 
there are various phases for the bulk string Universe, 
which involve a passage from subcritical to supercritical strings,
due to a change in sign of the pertinent supersymmetry-breaking 
potential of the moving brane world.

In such situations, the resulting $\sigma$ model describing low-energy
string excitations in the bulk lives in a $(d+1)$-dimensional space-time,
$d$ denoting the number of spatial dimensions, and is not conformal. As
already discussed, to restore the conformal symmetry required for
consistency of the path-integral quantization, one needs Liouville
dressing~\cite{ddk}, resulting in equations of the form
(\ref{liouvilleeq2}) for the background fields under consideration.
Liouville dressing results in a critical string in $(d+2)$ dimensions,
with restored conformal symmetry expressed by the vanishing of the
$(d+2)$-target-dimensional $\sigma$-model $\beta$ functions. Eventually,
dynamics which we review in due course results in the
identification of the Liouville mode in supercritical situations with the
target time, so the final target-space dimensionality of the dressed
$\sigma$ model remains $(d+1)$.

One such model was considered in detail in \cite{dgmpp}. The
model is based on a specific string theory, namely ten-dimensional
Type-0~\cite{type0}, which leads to a non-supersymmetric target-space
spectrum as a result of a special projection of the supersymmetric
partners out of the spectrum. Nevertheless, the basic properties of its
cosmology are sufficiently generic to be extended to the bosonic sector of
any other effective low-energy string-inspired supersymmetric field
theory. The model also involves flux compactification to four dimensions,
which, as was pointed out in \cite{dgmpp}, plays an important r\^ole in
ensuring the existence of large stable bulk dimensions.

The ten-dimensional metric configuration considered in~\cite{dgmpp} 
was: 
\begin{equation}
G_{MN}=\left(\begin{array}{ccc}g^{(4)}_{\mu\nu} \qquad 0 \qquad 0 \\
0 \qquad e^{2\sigma_1} \qquad 0 \\ 0 \qquad 0 \qquad
e^{2\sigma_2} I_{5\times 5} \end{array}\right)
\label{metriccomp}
\end{equation}
where lower-case Greek indices are four-dimensional space-time
indices, and $I_{5\times 5}$ denotes the $5\times 5$ unit matrix.
We have chosen two different scales for internal space. The field
$\sigma_{1}$ sets the scale of the fifth dimension, while the
$\sigma_{2}$ parametrize a flat five-dimensional space. In the
context of the cosmological models we deal with here, the
fields $g_{\mu\nu}^{(4)}$, $\sigma_{i},~i=1,2$ are assumed to
depend on the time $t$ only.
Type-0 string theory, as well as its supersymmetric
versions appearing in brane models,
contains appropriate form fields 
with non-trivial gauge fluxes (flux-form fields), 
which live in the 
higher-dimensional bulk space. In the specific model of~\cite{type0}, just
one such field was allowed to be 
non-trivial.
As was demonstrated in~\cite{dgmpp}, a consistent background choice
for the flux-form field has the flux parallel 
to the fifth dimension $\sigma_1$. This implies  
that the internal space is stabilized 
in such a way that this dimension is much larger than the 
remaining four $\sigma_2$. This demonstrates the physical
importance of the flux fields for large radii of compactification.

Considering the fields to be time-dependent only, i.e., considering
spherically-symmetric homogeneous backgrounds, restricting ourselves to
the compactification (\ref{metriccomp}), and assuming a Robertson-Walker
form of the four-dimensional metric with scale factor $a(t)$, the
generalized conformal-invariance conditions and the Curci-Pafutti
$\sigma$-model renormalizability constraint~\cite{curci} imply the set of
differential equations (\ref{liouvilleeq2}), which were solved numerically
in~\cite{dgmpp}. The set of $\{ g^i \}$ contains the graviton, dilaton,
tachyon, flux and moduli fields $\sigma_{1,2}$ whose vacuum expectation
values control the sizes of the extra dimensions.

The detailed analysis of \cite{dgmpp} indicated that the moduli 
fields $\sigma_i$ freeze quickly to their equilibrium values, so, together 
with the
tachyon field which also decays to a constant value rapidly, they decouple
from the four-dimensional fields at very early stages in the evolution of
this string Universe~\footnote{The presence of the tachyonic instability
in the spectrum of the model of~\cite{dgmpp} is due to the fact that in
Type-0 strings there is no target-space supersymmetry by construction.  
In other models with supersymmetry breaking~\cite{gravanis,emw}, due to
either thermalization or other instabilities, e.g., brane motion, there
are also tachyonic modes reflecting the broken supersymmetric spectrum.  
From a cosmological viewpoint such tachyon fields are not necessarily bad
features, since they may provide the initial instability leading to
cosmic expansion.}. There is an inflationary phase in this scenario and
dynamical exit from it. The important point to guarantee the exit is the
fact that the central charge deficit $Q^2$ is a time-dependent entity in
this approach, which obeys specific relaxation laws determined by the
underlying conformal field theory~\cite{dgmpp,gravanis,brany}. In fact,
the central charge runs with the local world-sheet renormalization-group
scale, the zero mode of the Liouville field, which is
identified~\cite{emn} with the target time in the $\sigma$-model frame.  
The supercriticality~\cite{aben} $Q^2 > 0$ of the underlying $\sigma$
model is crucial, as already mentioned.  Physically, the non-critical
string provides a framework for {\it non-equilibrium} dynamics, which may
be the result of some catastrophic cosmic event, such as a collision of
two brane worlds~\cite{ekpyrotic,gravanis,brany}, or an initial quantum
fluctuation~\cite{emninfl,dgmpp}.  It also provides, as we discuss below,
a unified mathematical framework for analyzing various phases of string
cosmology, from the early inflationary phase, graceful exit from
it and reheating, until the current and future eras of accelerated
cosmologies. Interestingly, one can constrain string parameters such as
the separation of brany worlds at the end of inflation, as well as the
recoil velocity of the branes after the collision, by fits to current
astrophysical data~\cite{brany}.

\subsection{Liouville Inflation: the General Picture} 

As discussed in~\cite{emninfl,gravanis,brany}, 
a constant central-charge deficit
$Q^2$ in a stringy $\sigma$ model may be associated with an initial 
inflationary phase with
\begin{equation}
\label{centraldeficit} 
Q^2 = 9 H^2 > 0~,
\end{equation}
where the Hubble parameter $H$ can be fixed in terms of other parameters
of the model. One may
consider various scenarios for such a departure from criticality. For 
example,
in the model of~\cite{gravanis,brany} this was due to 
the
collision of two brane worlds. In such a scenario, as we now review
briefly, it is possible to obtain an initial {\it
supercritical } central charge deficit, and hence a time-like Liouville
mode in the theory. For instance, in the specific colliding-brane model
of~\cite{gravanis,brany}, $Q$ (and thus $H$) 
is proportional to the square of the
relative velocity of the colliding branes, $Q \propto u^2$ during the
inflationary era.  As is evident from (\ref{centraldeficit}) and discussed
in more detail below, in a phase of constant $Q$ one obtains an
inflationary de Sitter Universe.

However, catastrophic non-critical string scenarios for cosmology, such as
that in~\cite{gravanis}, allow in general for a time-dependent deficit
$Q^2(t)$ that relaxes to zero. This may occur in such a way that, although
during the inflationary era $Q^2$ is (for all practical purposes)
constant, as in (\ref{centraldeficit}), eventually $Q^2$ decreases with
time so that, at the present era, one obtains compatibility with the
current accelerating expansion of the Universe. As already mentioned, such
relaxing quintessential scenarios~\cite{gravanis,dgmpp,brany} have the
advantage of asymptotic states that can be defined properly as $t \to
\infty$, as well as a string scattering $S$-matrix~\footnote{As mentioned
in the Introduction, another string scenario for inducing a de Sitter
Universe envisages generating the inflationary space-time from string
loops (dilaton tadpoles)~\cite{fischler}, but in such models a string
$S$-matrix cannot be properly defined.}.

The specific normalization in (\ref{centraldeficit})  is 
due to the 
identification of the time $t$ with
the zero mode of the Liouville field $-\varphi$ of the 
{\it supercritical} $\sigma$ model. The minus 
sign may be understood both mathematically, as
due to properties of the Liouville mode, and physically by the requirement
of the relaxation of the deformation of the space-time following
the distortion induced by the recoil. With this identification, the 
general equation of motion for the couplings $\{ g_i \}$ of the 
$\sigma$-model 
background modes is given by (\ref{liouvilleeq})~\cite{emn}:
\begin{equation}
{\ddot g}^i + Q{\dot g}^i 
= -{\tilde \beta}^i (g) = -{\cal G}^{ij} \partial C[g]/\partial g^j~,
\label{liouveq}
\end{equation}
where the dot denotes a derivative with respect to the Liouville
world-sheet zero mode $\varphi$, i.e., target time, 
and ${\cal G}^{ij}$ is an inverse
Zamolodchikov metric in the space of string theory couplings $\{ g^i 
\}$~\cite{zam}. When
applied to scalar, inflaton-like, string modes, (\ref{liouveq})  would
yield standard field equations for scalar fields in de Sitter
(inflationary)  space-times, provided the normalization
(\ref{centraldeficit}) is valid, implying a `Hubble' expansion
parameter $H=-Q/3$~\footnote{The gradient-flow property of
the $\beta$ functions makes the analogy with the inflationary case even
more profound, with the running central charge $C[g]$~\cite{zam} playing
the r\^ole of the inflaton potential in conventional inflationary field
theory.}.  The minus sign in $Q=-3H$ is due to the fact that, as we
discuss below, one identifies the target time $t$ with the world-sheet
zero mode of $-\varphi$~\cite{emn}.

The relations (\ref{liouveq}) replace the conformal invariance conditions
$\beta^i = 0$ of the critical string theory, and express the conditions
necessary for the restoration of conformal invariance by the Liouville
mode~\cite{ddk}. Interpreting the latter as an extra target dimension, the
conditions (\ref{liouveq}) may also be viewed as conformal invariance
conditions of a {\it critical} $\sigma$ model in (D+1) target space-time
dimensions, where D is the target dimension of the non-critical $\sigma$
model before Liouville dressing.  In most Liouville approaches, one treats
the Liouville mode $\varphi$ and time $t$ as independent coordinates.  In
our approach~\cite{emn,dgmpp,gravanis}, however, we take a further step,
basing ourselves on dynamical arguments which restrict this extended
(D+1)-dimensional space-time to a hypersurface determined by the
identification $\varphi = -t$. This means that, as time flows, one is
restricted to this D-dimensional subspace of the full (D+1)-dimensional
Liouville space-time.

In the work of~\cite{gravanis,brany} which invoked a brane collision
as a source of departure from criticality, this restriction arose because
the potential between massive particles, in an effective field theory
context, was found to be proportional to ${\rm cosh}(t + \varphi)$, which
is minimized when $\varphi = -t$. However, the flow of the Liouville mode 
opposite
to that of target time may be given a deeper mathematical interpretation.  
It may be viewed as a consequence of a specific treatment of the area
constraint in non-critical (Liouville) $\sigma$ models~\cite{emn},
which involves the evaluation of the Liouville-mode path integral via an
appropriate steepest-descent contour.  In this way, one obtains a
`breathing' world-sheet evolution, in which the world-sheet area starts
from a very large value (serving as an infrared cutoff), shrinks to a very
small one (serving as an ultraviolet cutoff), and then inflates again
towards very large values (returning to an infrared cutoff). Such a
situation may then be interpreted~\cite{emn} 
as a world-sheet `bounce' back to the
infrared, implying that the
physical flow of target time is opposite to that of the world-sheet scale
(Liouville zero mode).

We now become more specific.  We consider a non-critical $\sigma$ model in
metric ($G_{\mu\nu}$), antisymmetric tensor ($B_{\mu\nu}$), and dilaton
($\Phi$) backgrounds. These have the following ${\cal O}(\alpha ')$ 
$\beta$ functions (\ref{bfunct}), 
where $\alpha '$ is the Regge slope~\cite{gsw}:
 \begin{eqnarray} 
&& \beta^G_{\mu\nu} = \alpha ' \left( R_{\mu\nu} + 2 \nabla_{\mu} 
\partial_{\nu} \Phi 
 - \frac{1}{4}H_{\mu\rho\sigma}H_{\nu}^{\rho\sigma}\right)~, \nonumber \\
&& \beta^B_{\mu\nu} = \alpha '\left(-\frac{1}{2}\nabla_{\rho} H^{\rho}_{\mu\nu} + 
H^{\rho}_{\mu\nu}\partial_{\rho} \Phi \right)~, \nonumber \\
&& {\tilde \beta}^\Phi = \beta^\Phi - \frac{1}{4}G^{\rho\sigma}\beta^G_{\rho\sigma} = 
\frac{1}{6}\left( C - 26 \right).
\label{bfunctions}
\end{eqnarray} 
The Greek indices are four-dimensional, including target-space-time
components $\mu, \nu, ...= 0,1,2,3$ on the D3-branes
of~\cite{gravanis}, and $H_{\mu\nu\rho}= \partial_{[\mu}B_{\nu\rho]}$ is the
field strength.
We consider the following representation of the four-dimensional 
field strength in terms of a pseudoscalar (axion-like) field $b$: 
\begin{equation}
H_{\mu\nu\rho} = \epsilon_{\mu\nu\rho\sigma}\partial^\sigma b ~,
\label{axion}
\end{equation}
where $\epsilon_{\mu\nu\rho\sigma}$ is the four-dimensional antisymmetric
symbol. Next, we choose an axion background that is linear in the
time $t$~\cite{aben}:
\begin{equation} 
b = b(t) = \beta t~, \quad  \beta={\rm constant} ,
\label{axion2}
\end{equation}
which yields a constant field strength with spatial indices only: $H_{ijk}
= \epsilon_{ijk}\beta$, $H_{0jk}= 0$.  This implies that such a background
is a conformal solution of the full ${\cal O}(\alpha')$ $\beta$ function
for the four-dimensional antisymmetric tensor. We also consider a dilaton
background that is linear in the time $t$~\cite{aben}:
\begin{equation}
\Phi (t,X) = {\rm const} + ({\rm const})' t .
\label{constdil}
\end{equation}
This background does not contribute to the $\beta$ functions 
for the antisymmetric tensor and metric.

Suppose now that only the metric is a non-conformal background, due to 
some initial quantum fluctuation or catastrophic event, such as the 
collision of two branes discussed above, 
which results in an initial central charge deficit $Q^2$ 
(\ref{centraldeficit}) that is constant at early stages after the 
collision. Let 
\begin{equation} 
G_{ij} = e^{\kappa \varphi + Hct}\eta_{ij}~, \quad G_{00}=e^{\kappa '\varphi 
+ Hct}\eta_{00},
\label{metricinfl}
\end{equation}
where $t$ is the target time, $\varphi$ is the Liouville mode, 
$\eta_{\mu\nu}$ is the four-dimensional Minkowski metric, 
and $\kappa, \kappa '$ and $c$ are constants to be determined. 
As already discussed, the standard inflationary scenario in 
four-dimensional physics requires $Q = -3H$,
which partially stems from the identification
of the Liouville mode with time~\cite{emn}
\begin{equation}
\varphi = -t.
\label{liouvtime}
\end{equation}
The restriction (\ref{liouvtime}) is 
imposed dynamically~\cite{gravanis}
at the end of our computations. Initially, one should treat
$\varphi, t$ as independent target-space-time components. 

The Liouville dressing induces~\cite{ddk} $\sigma$-model terms of the form 
$\int_{\Sigma} R^{(2)} Q \varphi$, where $R^{(2)}$ is the world-sheet curvature.
Such terms provide non-trivial contributions to the dilaton background in 
the (D+1)-dimensional space-time $(\varphi,t,X^i)$:
\begin{equation}
\Phi (\varphi,t,X^i) = \frac{1}{2}Q \,\varphi + ({\rm const})' t + {\rm const}.
\label{seventy}
\end{equation}
If we choose 
\begin{equation}
({\rm const})'=\frac{1}{2}Q~, 
\label{const=Q}
\end{equation}
then (\ref{seventy}) implies a 
{\it constant} dilaton background during the inflationary era, in which the 
central charge deficit $Q^2$ is constant. We justify physically
the choices (\ref{seventy}) and (\ref{const=Q}) later in the article,
when we discuss a specific example of non-criticality induced by 
the collision of brane worlds. 

We now consider the Liouville-dressing equations~\cite{ddk}
(\ref{liouveq})  for the $\beta$ functions of the metric and antisymmetric
tensor fields (\ref{bfunctions}). For a constant dilaton field, the
dilaton equation yields no independent information, apart from expressing
the dilaton $\beta$ function in terms of the central charge deficit, as
usual. For the axion background (\ref{axion2}), only the metric yields a
non-trivial constraint (we work in units with $\alpha ' =1$ for
convenience):
\begin{equation} 
{\ddot G}_{ij} + Q{\dot G}_{ij} = -R_{ij} + \frac{1}{2}\beta^2 G_{ij},
\end{equation}
where the dot indicates differentiation with respect to the (world-sheet
zero mode of the) Liouville mode $\varphi$, and $R_{ij}$ is the
(non-vanishing) Ricci tensor of the (non-critical) $\sigma$ model with
coordinates $(t,{\vec x})$:  $R_{00}=0~, R_{ij}=\frac{c^2H^2}{2}e^{(\kappa
- \kappa ')\varphi}\eta_{ij}$. One should also take into account the
temporal ($t$) equation for the metric tensor (which is identically zero for 
antisymmetric backgrounds):
\begin{equation}
{\ddot G}_{00} + Q{\dot G}_{00} = -R_{00} = 0,
\label{tempgrav}
\end{equation}
where the vanishing of the Ricci tensor stems from the 
specific form of the background (\ref{metricinfl}).
We seek metric backgrounds of Robertson-Walker inflationary 
(de Sitter) form:
\begin{equation}
G_{00}=-1~, \quad G_{ij}=e^{2Ht}\eta_{ij}.
\label{desittermetric}
\end{equation}
Then, using (\ref{desittermetric}), (\ref{metricinfl}),
(\ref{constdil}) and (\ref{axion2}), and imposing
(\ref{liouvtime}) at the end, we observe that there is indeed a consistent
solution with:
\begin{equation}
Q = -3H = - \kappa ',~c=3,~\kappa = H,~\beta^2 = 5H^2,
\label{solution}
\end{equation}
corresponding to the conventional form of inflationary equations for
scalar fields.

\subsection{Current Stages of Cosmic Liouville Evolution:
Acceleration, Dark Energy and the String Coupling}

In the generic class of non-critical string models of interest in this
work, the $\sigma$ model always asymptotes, for long enough cosmic times,
to the linear dilaton conformal $\sigma$-model field theory
of~\cite{aben}. But it is important to stress that this is only an
asymptotic limit. In this respect, the current era of our Universe may be
viewed as being close to, but still not quite at, the relaxation
(equilibrium) point, in the sense that the dilaton is almost linear in the
$\sigma$-model-frame time, and hence varies logarithmically with the
Einstein-frame time (\ref{dil2}). It is expected that this slight
non-equilibrium will lead to a time-dependence of the unification gauge
coupling and other constants (e.g., the four-dimensional Planck length
(\ref{planckstring})), that characterize the low-energy effective field
theory, mainly through the time-dependence of the string coupling
(\ref{defstringcoupl})  as a result of the time-dependent linear dilaton
(\ref{lineardil}).

The asymptotic time regime of the Type-0
cosmological string model of~\cite{dgmpp} was obtained 
analytically, by solving the pertinent equations (\ref{liouvilleeq})
for the various fields. As already mentioned, at late times
the theory becomes four-dimensional, and the only non-trivial
information is contained in the scale factor and the dilaton, 
given that the topological flux field remains conformal in this approach,
and the moduli and initial tachyon fields decouple very fast in the 
initial stages after inflation in this model.
For times long after the initial fluctuations, 
such as the present epoch, when the linear approximation is valid, 
the solution for the dilaton in the $\sigma$-model frame
follows from the equations 
(\ref{liouvilleeq}) and takes the form:
\begin{equation} \label{dilaton} 
\Phi (t) =-{\rm ln}\left[\frac{\alpha A}{F_1}{\rm cosh}(F_1t)\right],
\end{equation}
with $F_1$ a positive constant, $\alpha$ a numerical constant of order one,
and 
\begin{equation}\label{defA2}
A = \frac{C_5 e^{s_{01}}}{\sqrt{2}V_6}~, 
\end{equation}
where $s_{01}$ is the equilibrium
value of the moduli field $\sigma_1$, associated with the large bulk dimension,
and $C_5$ the corresponding flux of the five-form flux field.
Notice the that $A$ is independent of this large bulk dimension.

For very large times $F_1 t \gg 1$ (in string units) 
one therefore approaches a 
linear solution for the dilaton 
$\Phi \sim {\rm const} -F_1 t$.
From (\ref{dilaton}), (\ref{defstringcoupl}) and (\ref{planckstring}),   
we thus observe that the asymptotic weakness of 
gravity in this Universe~\cite{dgmpp} is due to the smallness of 
the internal space $V_6$ as compared with the flux $C_5$ of the 
five-form field.
The constant $F_1$ is related to the central charge deficit 
of the underlying the non-conformal $\sigma$ model~\cite{dgmpp}:
\begin{equation}\label{ccd}
Q = Q_0 + \frac{Q_0}{F_1}(F_1 + \frac{d\Phi}{dt})~,
\end{equation}
where $Q_0$ is a constant, and 
the numerical solution of (\ref{liouvilleeq}), studied in \cite{dgmpp},
requires that 
\begin{equation}
Q_0/F_1 =(1 + \sqrt{17})/2 \simeq 2.56~,
\label{f1q0}
\end{equation}
which follows from the dilaton equation of motion. 
This connection of $F_1$ to $Q_0$ 
supports the above-described asymptotic 
conformal theory considerations of \cite{aben}, where the model relaxes to
for large times. In this spirit, we require that 
the value of $Q_0$ to which the central charge deficit (\ref{ccd}) asymptotes
must be, for consistency of the underlying string theory,
{\it one of the discrete values} obtained in \cite{aben}, 
for which factorization (unitarity) of the 
string scattering amplitudes 
occurs. Notice that this asymptotic 
string theory, 
with 
a constant (time-independent) central-charge deficit,  
$Q_0^2 \propto c^*-25 $ (or $c^*-9$ for superstring) 
is considered an {\it equilibrium} situation,
where an $S$-matrix can be defined for specific (discrete)
values of the central charge $c^*$~\cite{ddk,aben}.

Defining the Einstein frame time $t_E$ through (\ref{einsttime}),
we obtain in this case 
\begin{equation}\label{einstframe}
t_E=\frac{\alpha A}{F_1^2}sinh(F_1 t)~, \quad F_1t = {\rm ln}\left(\sqrt{1 
+ \gamma^2t_E^2} + \gamma t_E\right)~,
\end{equation}
where
\begin{equation}\label{defA} 
\gamma \equiv \frac{F_1^2}{\alpha A}~. 
\end{equation}
In terms of the Einstein-frame time,
one obtains a logarithmic time dependence~\cite{aben} for the 
dilaton (\ref{dilaton})~\footnote{Notice that in this subsection 
we work in $D=4$ space-time dimensions. For higher-dimensional models,
the normalisations given in the Introduction, see (\ref{dil2}),
should be used.}:
\begin{equation} 
\Phi _E = {\rm const} -{\rm ln}(\gamma t_E)~,
\label{einsteindil} 
\end{equation}
For large $t_E$, e.g., present or later cosmological time values, 
one has~\cite{dgmpp,emn04}
\begin{equation}\label{einstmetr} 
a_E(t_E) \simeq \frac{F_1}{\gamma}\sqrt{1 + \gamma^2 t_E^2}~.
\end{equation}
At very large (future) times $a(t_E)$ scales linearly
with the Einstein-frame cosmological time $t_E$~\cite{dgmpp},
and hence there is no cosmic horizon. From a field theory
viewpoint, this would 
allow for a proper definition of asymptotic
states and thus a scattering matrix. 
As we mentioned briefly above, however, 
from a stringy point of view,
there are restrictions in the asymptotic values of 
the central charge deficit $Q_0$, and only a  
discrete spectrum of values of $Q_0$ allow for a full stringy S-matrix
to be defined,
respecting modular invariance~\cite{aben}. 
The Universe relaxes asymptotically 
to its ground-state equilibrium situation,
and the non-criticality of the string caused by the initial fluctuation
disappears, yielding a critical (equilibrium) string Universe
with Minkowski metric and a linear-dilaton background. 
This is the generic feature of the models we consider here and 
in \cite{emn04}, allowing the conclusions 
to be extended beyond the Type-0 string theory to incorporate
also target-space supersymmetric strings/brane models, such as 
those in \cite{emw,brany}.

An important comment is in order at this point, regarding the 
form of the Einstein metric corresponding to (\ref{einstmetr}):
\begin{equation}\label{einstmetr2}
g_{00}^E=-1, ~~\quad g_{ii} = a_E^2(t_E) 
= \frac{F_1^2}{\gamma^2} + F_1^2t_E^2~.
\end{equation}
Although asymptotically, for $t_E \to \infty$, the above metric
asymptotes to the linearly-expanding Universe (\ref{metricaben}),
the presence of a constant $F_1^2/\gamma^2$  contribution
implies that the solution for large but finite $t_E$, such as the current 
era of the Universe, is different from that of \cite{aben}. 
Indeed, the corresponding 
$\sigma$-model-frame metric (\ref{smodeinst}) is not Minkowski flat,
and in fact the pertinent $\sigma$ model does not correspond 
to a conformal field theory. This should come as no surprise since,
for finite $t_E$ no matter how large, the $\sigma$-model 
theory requires
Liouville dressing. {\it It is only at the end-point of time/flow 
$t_E \to \infty$ that the underlying string theory becomes conformal,
and the system reaches equilibrium.}

The Hubble parameter of such a 
Universe for large $t_E$ is
\begin{equation}\label{hubble2} 
H(t_E) \simeq \frac{\gamma^2 t_E}{1 + \gamma^2 t_E^2} = 
\frac{F_1^2t_E}{a^2(t_E)}~.
\end{equation}
On the other hand, the Einstein-frame effective four-dimensional 
`vacuum energy density', defined through the running 
central-charge deficit $Q^2$,
upon compactification to four dimensions of the ten-dimensional
expression $2\int d^{10}x \sqrt{-g}e^{2\Phi}Q^2(t_E)$ in the 
Einstein frame, is~\cite{dgmpp}:
\begin{equation} 
\Lambda_E (t_E) = 2e^{2\Phi - \sigma_1 - 5\sigma_2}Q^2(t_E) 
\simeq \frac{2Q_0^2 \gamma^2}{F_1^2 ( 1 + \gamma^2 t_E^2)}
\sim \frac{13.11\gamma^2}{1 + \gamma^2 t_E^2}
\label{cosmoconst2} 
\end{equation}
in the normalization of (\ref{effaction}).
Here we used (\ref{ccd}) for $Q$ at large $t_E$, 
approaching its equilibrium value $Q_0$, and we have also used
(\ref{f1q0}). Thus, the dark energy 
density relaxes to zero for $t_E \to \infty$.
Notice an important feature of the relaxation form (\ref{cosmoconst2}),
namely that the proportionality constants in front 
are such that, for asymptotically large $t_E \to \infty$, 
$\Lambda (t_E \to \infty)$ is independent of 
the equilibrium conformal field theory central charge $Q_0$.

Finally, and most importantly for our purposes here, 
the deceleration parameter 
in the same regime 
of $t_E$ becomes:
\begin{equation} 
q(t_E) = -\frac{(d^2a_E/dt_E^2)~a_E}{({da_E/dt_E})^2}
\simeq -\frac{1}{\gamma^2 t_E^2}~.
\label{decel4}
\end{equation}
{\it As is clear from 
(\ref{einsteindil}), (\ref{defstringcoupl}), 
this expression can be identified, up
to irrelevant constant factors which by normalization 
are set to one, 
with the square of the string coupling 
(\ref{defstringcoupl})~\cite{emn04}}:
\begin{equation}
|q(t_E)| = g_s^2~.
\label{important}
\end{equation}
To guarantee the consistency of perturbation theory, one must have $g_s < 
1$,
which can be achieved in our approach 
if one {\rm defines} the  
{\it present era} by the time regime
\begin{equation}
\gamma^2 \sim \beta^2 t_E^{-2} 
\label{condition}
\end{equation} 
in the Einstein frame.
In view of its relation with the deceleration parameter at late epochs
(\ref{decel4}), $q=-1/\beta^2$, the numerical value of 
$\beta^2$ is determined by 
requiring agreement with the data~\cite{wmap}.
As we discuss below, phenomenologically $\beta^2 = {\cal O}(1)$. 

This is compatible with the time $t_E$ being large enough 
(in string units) 
for 
\begin{equation} 
|C_5|e^{-5s_{02}}/F_1^2 \sim |C_5|e^{-5s_{02}}/Q_0^2 \gg 1~, 
\label{largetwe}
\end{equation} 
as becomes clear from (\ref{defA2}),(\ref{defA}), (\ref{f1q0}). 
This condition can be guaranteed
{\it either} for small radii of five of the extra dimensions, 
{\it or} for a large value of the flux $|C_5|$ of the five-form 
of the Type-$0$ string, compared with $Q_0$. We discuss 
in the next subsection concrete examples of non-critical 
string cosmologies, in which the asymptotic value of the 
central charge $Q_0 \ll 1$ in string units.  
Recalling that the relatively large extra dimension
in the direction of the flux
$s_{01}$ decouples from this condition, we thus 
observe that there is the possibility of constructing 
effective five-dimensional models 
with a large uncompactified fifth dimension that are consistent with the 
condition
(\ref{condition}).  
Notice that, in the regime (\ref{condition}) of Einstein-frame times, 
the Hubble parameter
and the cosmological constant continue to be compatible
with the current observations, 
and in fact to depend on $\gamma \sim t_E^{-1}$ in the same way
as in their large-$\gamma t_E$ regime given above 
(\ref{hubble2}),(\ref{cosmoconst2}), but now the string coupling
(\ref{important}) is kept smaller than one and finite, of order
$1/2$, as also suggested  by grand unification phenomenology~\cite{gsw}.

We next turn to the equation of state of our Universe.
As discussed in~\cite{dgmpp}, it resembles a  
quintessence model 
with the dilaton playing the r\^ole of the quintessence
field. 
Hence the equation of state
for our Type-$0$ string Universe reads~\cite{carroll}:
\begin{equation}\label{eqnstate} 
w_\Phi = \frac{p_\Phi}{\rho_\Phi}=\frac{\frac{1}{2}({\dot \Phi})^2 - V(\Phi)}
{\frac{1}{2}({\dot \Phi})^2 + V(\Phi)}~,
\end{equation}
where $p_\Phi$ is the pressure and $\rho_\Phi$ is the energy density, 
and $V(\Phi)$ is the effective potential for the dilaton, which in our case
is provided by the central-charge deficit term. 
Here the dot denotes Einstein-frame differentiation. 
In the Einstein frame, in the normalization of (\ref{effaction}),  
the potential $V(\Phi)$ is given by 
\begin{equation}                       
V(\Phi) = \frac{\Lambda_E}{4} \sim \frac{6.56\gamma^2}{2(1 + \gamma^2t_E^{2})}~,
\label{potendilaton}
\end{equation} 
where $\Lambda_E$ is given in (\ref{cosmoconst2})
and we have used (\ref{f1q0}). 
Defining the present era 
by the condition (\ref{condition}), 
we obtain from (\ref{dilaton}),(\ref{einstframe}):
\begin{eqnarray}\label{dilpotkin}  
\frac{1}{2}\left(\frac{d\Phi}{dt_E}\right)^2 = \frac{1}{2}\beta^2 \cdot 
{\rm tanh}^2{\rm ln}\left( \sqrt{1 + \beta^2} + \beta \right) \cdot 
\frac{1}{t_E^2 (1 + \beta^2)}, \qquad V(\Phi) 
\sim \frac{6.56\beta^2}{2(1 + \beta^2)t_E^2}~. 
\end{eqnarray} 
This 
implies a \emph{constant} equation of state (\ref{eqnstate}) in the 
current era: 
\begin{equation} 
w_\Phi (t_E \gg 1) = \frac{{\rm tanh}^2{\rm ln}\left(\sqrt{1 + \beta^2}+ \beta \right)- 6.56}{{\rm tanh}^2{\rm ln}\left(\sqrt{1 + \beta^2}+ \beta \right)
+ 6.56}.
\label{eqnstatedilnum}
\end{equation}
We now remark that, if we use as the value of 
$q$ today the one inferred by best fits of FRW cosmology 
to the data on high-redshift supernovae and the CMB~\cite{snIa,wmap}:
\begin{equation}\label{qval}
q_{FRW,data}=-\frac{1}{\beta^2} \simeq -0.57~~({\rm today})~,
\end{equation} 
this corresponds by (\ref{eqnstatedilnum}) to an equation of state with
\begin{equation} 
w_\Phi = -0.82~, 
\label{eqnstatedilnum3}
\end{equation}
which is in the region allowed by the 
data~\cite{wmap,snIa,steinhardt}~\footnote{Although the data at present 
are not sufficient for an accurate measurement of $w(z)$, they
seem to indicate~\cite{steinhardt} negative
values smaller than $-0.6$ for $z \simeq 1$ and $w(z\to 0) \to -1$.}. 
On the other hand, an equation of state $w_\Phi = -0.78$, which is  
the upper bound   
given by the WMAP data~\cite{wmap,steinhardt}, yields   
a current-era 
deceleration (\ref{decel4}),(\ref{condition}) $q \sim -0.25$. 
All such values of $|q|< 1$ imply via (\ref{important})
perturbative values for the string coupling, 
close to the value used frequently in 
string-inspired particle phenomenology: $|q|= g_s^2 \sim 1/2$. 

The inclusion of 
matter modifies the situation, and allows for a more complete
expression of the equation of state in terms of the current
acceleration of the Universe for our model. As we discuss
in the next subsection, this is different 
from the analogous relation in conventional FRW cosmologies.
However, 
the function $q(z)$ is not yet measurable with sufficient
accuracy using supernovae alone: in order to infer a precise form of 
$q(z)$ from such measurements,
one has to make certain assumptions about the underlying 
dynamics.
In conventional FRW cosmologies, $q(z)$ 
is expressed in terms of the various 
energy-density components $\Omega_i(z)$ using the 
underlying Einstein cosmology. In a similar vein, in our 
model $q(z)$ can be expressed in terms
of the various energy-density components, in units of the critical
density, using the dynamics encoded in (\ref{eqsmotion}). However,
due to the off-shell Liouville modifications, the critical density 
for a spatially flat Universe and the relation of 
$q(z)$ to the various energy-density components are different from 
those in conventional
FRW models. Thus the above-used values of $\beta^2$ 
should not be taken for 
granted but only as indicative. For us, $\beta^2=1/|q|$ can only be determined
properly after a detailed direct fit of our model to the data.
We discuss these issues in the next Section, where we
show that the above considerations can be made compatible 
with the observations that suggest there was a past epoch of
deceleration at redshifts larger than $0.5$~\cite{deceldata}. 

These considerations concern the specific model of~\cite{dgmpp}.
One can be more generic when considering equations of state 
for dilatonic dark energy Liouville models, by simply
requiring that the present era is described by a linear 
dilaton solution (\ref{dil2}), asymptotic to a conformal
field theory with central charge deficit $Q^2$.
In this case, the dilaton potential and kinetic energy are given by 
\begin{eqnarray}\label{genpot}  
&& V(\Phi) = \frac{Q^2}{2}e^{2\Phi} = \frac{2}{t_E^2}e^{2\Phi_0}~, 
~~\Phi = \Phi_0 -{\rm ln}\frac{Qt_E}{2}~, \nonumber \\
&& \frac{d\Phi}{dt_E} = \frac{1}{t_E} ,
\end{eqnarray}
where $\Phi_0$ is a constant, denoting the initial value of 
the dilaton field in a generic situation. 
As we have seen previously (\ref{dilaton}),
this constant is 
determined in the model of~\cite{dgmpp} by the values of the flux field 
and the 
frozen moduli. In the general situation, where no microscopic model
is specified, this constant is free to be determined
by phenomenology, as we see below.  
In such a generic situation, the dilatonic dark-energy 
equation of state (in the same normalization of (\ref{effaction}), 
reads:
\begin{equation} 
w_\Phi (t_E \gg 1) \simeq \frac{1 - 4e^{2\Phi_0}}{1 + 
4e^{2\Phi_0}}~. 
\label{eqnstatedilnum2}
\end{equation}
One can easily obtain 
phenomenologically acceptable values of $w_\Phi$~\cite{wmap}
by adjusting the value of the constant $\Phi_0$.
For instance, for $e^{2\Phi_0} > \frac{7}{4}$ one obtains:
\begin{equation} 
w_\Phi \le -3/4 ,
\label{condition2}
\end{equation} 
in agreement with 
the WMAP value~\cite{wmap,steinhardt}. 

Such linear dilaton models can be made compatible with 
perturbative string couplings
$g_s < 1$, as required by string-inspired 
particle-physics phenomenology, provided one chooses the asymptotic central
charge $Q$ in the region 
\begin{equation}
Q t_E > \sqrt{7} .
\label{ccrestriction}
\end{equation}
Notice that the dark energy (\ref{genpot}) 
is independent of the value of the central charge, 
and its magnitude today in such models depends only on the age of the Universe.
The important question in such models is the precise form of the 
scale factor, which should be obtained as a specific solution of the 
appropriate dynamical equations (\ref{liouveq}), whose form depends
on the details of the underlying string theory. 
To be specific, in what follows  
we adopt~\cite{emn04} 
the class of string models that yield 
predictions similar to those in~\cite{dgmpp} 
as best 
describing the current era of our string Universe.

\subsection{Inclusion of Matter and Radiation}

So far, our model has not included ordinary matter or radiation, as only
fields from the gravitational string multiplet have been included. The
inclusion of ordinary matter is not expected to change the results
significantly, and we conjecture that the fundamental relation
(\ref{important}) will continue to hold, the only difference being that
probably the inclusion of ordinary matter will tend to reduce the string
acceleration, due to the fact that matter, being subjective to attractive
gravity, resists the acceleration of the Universe.

We now discuss in some detail the formalism that allows the inclusion of
matter in the Liouville framework.  The important thing to notice is that,
in the absence of matter, the Liouville-dressing approach of~\cite{emn},
together with the eventual (dynamical) identification of the Liouville
zero mode with the target time, as explained above, leads to the
generalized conformal invariance conditions (\ref{liouveq}) for the fields
of the gravitational multiplet of the string propagating in a
four-dimensional background~\footnote{Or five-dimensional, in the case of
compactified brane models in a single large bulk dimension.}.  These
\emph{are not} the ordinary equations of motion corresponding to a
four-dimensional gravitational effective action (\ref{effaction}), but
describe the dynamics of an \emph{off-shell relaxation process}.

Matter coupling to on-shell dilatonic gravity theory has been considered
in the past, see, e.g.,~\cite{damour,gasperini}, but in an \emph{on-shell}
formalism of critical strings, where the various target-space fields
satisfied classical equations of motion derived from a four-dimensional
action. As explained above, this is distinct from our Liouville cosmology
approach. Moreover, the analysis of~\cite{gasperini}, although dealing
with the possibility of a dilaton playing the r\^ole of a quintessence
field responsible for the current acceleration of the Universe,
nevertheless considers models in which the dilaton as well as its
potential increases to positive infinity, as the cosmic time elapses.  
This is exactly opposite to our situation here and in~\cite{dgmpp,emn04},
where the dilaton $\Phi \to -\infty$ asymptotically. In our situation, for
large cosmic times, the string coupling $e^\Phi \to 0$, and this is the
reason why in present and future eras of the Universe the
string-tree-level approximation is sufficient. This is to be contrasted
with the case of~\cite{gasperini}, where one should include asymptotically
(all) higher loop corrections to the string effective action, which are
not known at present.

We now discuss in some detail the proper inclusion of 
matter in our Liouville framework. 
The essential formalism 
is that of~\cite{aben}, in which all physically relevant quantities
should be reduced to the Einstein frame (\ref{smodeinst}) 
and Einstein cosmic time (\ref{einsttime}) 
framework. Examining the four-dimensional 
matter action (including radiation fields),
we observe that in string theory this action couples to the dilaton
field non-trivially, in a way that is specific to the various
matter species, as a result of purely stringy properties of the
effective action~\cite{gsw}. 
A generic $\sigma$-model-frame effective four-dimensional action
with dilaton potential $V(\Phi)$, which could even include higher-string-loop
corrections, has the form:
\begin{eqnarray}
&& S^{(4)} = \frac{1}{2\alpha '} \int d^4 x \sqrt{-G} [e^{-\Psi(\Phi)} R(G) 
+ Z(\Phi)(\nabla \Phi)^2 + 2\alpha ' V(\Phi)  \dots ] 
- \nonumber \\
&& \frac{1}{16\pi}\int d^4x \sqrt{G}\frac{1}{\alpha (\Phi)}F_{\mu\nu}^2 
- I_{\rm m}(\Phi, G, {\rm matter})~, 
\label{matteraction}
\end{eqnarray}
in the notation of \cite{gasperini}, with the various factors
$\Psi, Z, \alpha $ encoding information about higher 
string loop corrections. Also, $F_{\mu\nu}$ denotes 
the radiation field strength and $I_{\rm m}(\Phi, G, {\rm matter})$
represents  matter contributions, which  couple to the dilaton $\Phi$
in a manner dictated by string theory scaling laws~\cite{gsw} 
with shifts of the dilaton field $\Phi \to \Phi + {\rm const}$.
In our situation, where only the string tree level plays a r\^ole 
at late times, the various form factors simplify, e.g., $\Psi (\Phi) = 
2\Phi, Z(\Phi) = 4 e^{- 2 \Phi}$, \emph{etc.}. However, for purposes of 
generality,
in this section we keep the form (\ref{matteraction}).  
When higher loop corrections 
are important, these factors have a complicated form, for instance 
one has $e^{\Psi (\Phi)} = c_0 e^{-2\Phi} + c_1 + 
c_2e^{2\Phi} + \dots $, with $c_i$ constants, and the powers of the 
square of the string coupling $g_s^2=e^{2\Phi}$ count the numbers of 
closed string loops, as appropriate for the gravitational multiplet. 
For simplicity in this subsection we ignore the four-dimensional
antisymmetric tensor 
field, which, as discussed in~\cite{aben} and mentioned above, 
corresponds to an axion field.

According to our discussion in Section 2.1, the action $S^{(4)}$ 
coincides with the $I^{(D)}-{\cal I}_\phi$ 
part of the $D$-dimensional action (\ref{ddimaction}),
obtained from (\ref{lambdaaction}) when $D=4$,
upon the identification of the Liouville mode $\phi$ with the 
target time $X^0$:
\begin{equation}\label{4doffshell}
S^{(4)} = I^{(4)} - {\cal I}_\phi = 
\int d^4X e^{-\varphi} \{ C^{(4)}(X) - 25\}.
\end{equation} 
In contrast to the critical-string case considered 
in~\cite{damour,gasperini},  
the field variations of (\ref{4doffshell}) 
do not yield zero, 
but are such that they compensate the variations of the 
remaining (Liouville) part of (\ref{lambdaaction}), 
in order to yield the 
generalized conformal invariance conditions (\ref{liouvilleeq2}),
augmented by the inclusion of matter fields.
In particular, 
the set of couplings $\lambda^I$ in (\ref{lambdaaction}),
as well as the action $C^{(4)}(X)$ (c.f. (\ref{matteraction})),     
should now include matter fields in addition to the fields of the gravitational 
multiplet of the string. For simplicity, however, we may 
assume that, at least at the late epochs of the Universe
which are of interest to us here, 
the matter couplings are almost conformal, and the 
dominant reason for departure from criticality lies in the
fields of the gravitational multiplet. This is the case,
for instance, in the colliding-brane scenario discussed
in the next Section. This leaves the off-shell 
Liouville part of (\ref{lambdaaction}) in the form 
discussed in the previous Section. This will be understood 
in what follows. 

We have: 
\begin{equation} 
\frac{\delta S^{(4)}}{\delta g^i} = -\frac{\delta {\cal I}_\phi}{\delta g^i}~,
\label{vars}
\end{equation}
where $g^i=(\Phi, G, \dots) \equiv (\Phi, \lambda^I)$ and 
we took into account the fact that the action ${\cal I}^{(4)} \equiv
{\cal I}^{(4+1)}|_{\phi =X^0}$ is critical (the identification 
of the Liouville mode with the target time $X^0$ is done after the 
respective variation is taken). 
Near a fixed point 
one has  
$\frac{\delta S^{(4)}}{\delta \lambda^I} = {\ddot g}_I + Q{\dot g}_I$,
with the overdot denoting differentiation with respect 
to the Liouville zero mode. 
When passing to the Einstein frame (\ref{smodeinst}), and 
expressing the time in terms of the cosmic time $t_E$ (\ref{einsttime}),
the left-hand side of (\ref{liouvilleeq2}) in the supercritical
string case for the graviton fields $G_{\mu\nu}$ yields: 
\begin{eqnarray} 
&&{\dot G}_{\mu\nu} = e^{\Phi} \left(2\frac{d\Phi}{dt_E}g_{\mu\nu}^E
+ \frac{dg_{\mu\nu}^E}{dt_E} \right)~,
\nonumber \\
&& {\ddot G}_{\mu\nu} = 
2\left(\frac{d\Phi}{dt_E}\right)^2 g_{\mu\nu}^E + 
2\frac{d^2\Phi}{dt_E^2}g_{\mu\nu}^E + 3 \frac{d\Phi}{dt_E}
\frac{dg_{\mu\nu}^E}{dt_E} + \frac{d^2g_{\mu\nu}^E}{dt_E^2}~,
\nonumber \\
&& (0,0)-{\rm component}:~ {\tilde {\cal G}}_{00} \equiv 
{\ddot G}_{00} + Q{\dot G}_{00} = 
-2Q\frac{d}{dt_E}\left(e^\Phi\right) - 2(\frac{d\Phi}{dt_E})^2 - 
2\frac{d^2\Phi}{dt_E^2}~, ~~(g_{00}^E = -1)~, \nonumber \\
&& (i,i)-{\rm component}:~ {\tilde {\cal G}}_{ii} \equiv 
{\ddot G}_{ii} + Q{\dot G}_{ii} = \nonumber \\
&& 2a^2(t_E)\left((\frac{d\Phi}{dt_E})^2 + 3\frac{d\Phi}{dt_E}H + 
\frac{d^2\Phi}{dt_E^2} + (1-q)H^2\right) 
+ 2Qa^2(t_E)e^\Phi\left(\frac{d\Phi}{dt_E} + H\right)
~, \nonumber \\
&& H \equiv a^{-1}(t_E)\frac{da(t_E)}{dt_E}~, ~~ q \equiv 
-\frac{a(t_E)\frac{d^2}{dt_E^2}a(t_E)}{(\frac{da(t_E)}{dt_E})^2}~,
\label{liouvmatter}
\end{eqnarray}
with $q$ the deceleration parameter (\ref{decel4}). 
The dilaton variation of the function ${\cal I}_\phi$ of (\ref{ddimaction}), 
on the other hand, reads: 
\begin{eqnarray} 
&&\frac{\delta {\cal I}_\phi}{\delta \Phi} \equiv {\cal I}'_\phi = 
6\int d^{D-1}X~e^{-\varphi}\left({\dot \varphi}^2 - {\ddot \varphi}
+ \frac{1}{8}({\dot \lambda}^I)^2\right) = \nonumber \\
&& 6V^{(3)}\{ \left(2\frac{d\Phi}{dt_E} + H \right)^2
- \frac{d\Phi}{dt_E}\left(2\frac{d\Phi}{dt_E} + H \right) + 
\frac{dH}{dt_E} + 2\frac{d^2\Phi}{dt_E^2} + 2\left(\frac{d\Phi}{dt_E}\right)^2
+ 3H\frac{d\Phi}{dt_E} + \frac{3}{2}H^2\}~, \nonumber \\
\label{iprimephi}
\end{eqnarray}
in the notation of (\ref{lambdaaction}), where $V^{(3)}$ denotes the 
three-dimensional spatial volume.
A complete analysis of matter effects requires solving the equations
emerging from considering the variations (\ref{vars}),(\ref{liouvmatter})
and (\ref{iprimephi}) with respect to the metric field in the Einstein and
cosmic time frames~\cite{aben}. This depends on the specific form of
matter action considered.

At this stage we remind the reader 
of a few crucial technical details on the equivalence of the generalized
conformal invariance conditions (\ref{liouveq}) to 
target-space dynamical equations. The Zamolodchikov metric
in theory space,
${\cal G}_{ij} = z^2{\overline z}^2 \langle V_i(z)V_j(0)\rangle $,
where $\langle \dots \rangle $ denotes a $\sigma$-model 
average including Liouville contributions,  
acts as a link between the 
$\sigma$-model $\beta$ functions and field variations of the
target-space effective actions $S[g]$: 
\begin{equation}\label{effactions}
{\cal G}_{ij}\beta^j = \frac{\delta S[g]}{\delta g^i} ,
\end{equation}
where, in what follows, the $g^i$ 
denote various background target-space fields other than the dilaton,
which is treated separately. 
To order $\alpha '$, standard analysis~\cite{miramontes,emn} 
shows that one can find a renormalization scheme on the 
world sheet in which the Zamolodchikov metric,
in the case of the graviton 
and dilaton backgrounds we are restricting ourselves here,  
becomes near a fixed point 
\begin{equation}
{\cal G}_{ij} = e^{-2\Phi}\left(\delta_{ij} + {\cal O}(g^2)\right)~, \qquad
\delta_{ij} \to \frac{1}{2}\left(-G^{\mu\nu}G^{\alpha\beta} + G^{\mu\alpha}G^{\nu\beta} 
+ G^{\mu\beta}G^{\nu\alpha} \right) ,
\label{miramontes}
\end{equation}
where $G_{\mu\nu}$ is a $\sigma$-model-frame target-space metric
(which, in our case, is four-dimensional after appropriate compactification or
restriction on a three-brane). The
exponential dilaton term arises from world-sheet zero-mode
contributions to the $\sigma$-model average at tree level, and
includes (linear) Liouville-zero-mode- ($\phi_0$-) dependent
terms in the non-critical string case. 
This Liouville 
dependence is crucial~\cite{emn} in ensuring the 
following property for the Zamolodchikov metric:
$d{\cal G}_{ij}/d\phi_0 = Q {\cal G}_{ij}$, with $Q^2$ 
the central-charge deficit, 
which guarantees the derivation 
of the Liouville terms on the left-hand side of 
(\ref{liouveq}) from a target-space action, as seen above, and therefore 
the 
canonical quantization of the Liouville-dressed couplings/fields $ g^i$ 
in string-theory space (upon summation over world-sheet topologies). 

The form (\ref{miramontes}) implies that a contraction with a Ricci (or
any other symmetric)  tensor in target space, which is contained in the
graviton $\beta$ function to ${\cal O}(\alpha ')$, results in an Einstein
tensor on the right-hand side of (\ref{effactions}), as appropriate for
proper target-space dynamics. In a similar manner, upon contraction with
(\ref{miramontes}), one obtains appropriate Einstein-like tensor
structures for the Liouville modifications apppearing in the left-hand
side of (\ref{liouveq}) for the graviton case.
On the other hand, considering the variation with respect to the 
four-dimensional graviton 
$g^i  \equiv G_{\mu\nu}$, and passing to the Einstein-frame 
(\ref{smodeinst}), we obtain: 
\begin{equation}\label{einsteinframeaction}
\frac{\delta S[g]}{\delta G_{\mu\nu}} = 
e^{-2\Phi}\frac{\delta S[g]}{\delta g^E_{\mu\nu}} ,
\end{equation}
where the precise form of the exponential factor is exclusive to the four 
target space-time dimensions we consider here. As a result 
of (\ref{miramontes}), (\ref{einsteinframeaction}), the 
exponential factors 
cancel out in (\ref{effactions}). The above 
results and 
properties will be understood in what follows.

Defining the Einstein-like tensor 
\begin{eqnarray}\label{jcaldef}
{\cal J}_{\mu\nu} \equiv {\tilde {\cal G}}_{\mu\nu} 
- \frac{1}{2}g_{\mu\nu}^E 
\left (g^{\nu\lambda}_E {\tilde {\cal G}}_{\nu\lambda} \right)~,
\end{eqnarray} 
and assuming a normal fluid form for matter or radiation, with stress tensor
$T_\mu^{\nu E} = {\rm diag}\left(-\rho, p\delta_i^j\right)$
in the Einstein frame (\ref{smodeinst}),(\ref{einsttime}),
we then obtain the following gravitational and dilaton equations of motion 
(in units $M_P^2 = 1/8\pi G_N = 2$, where $M_P$ is the four-dimensional 
Planck constant): 
\begin{eqnarray} 
&& 6H^2 = \rho + \rho_\Phi + {\cal J}_{00}~, \nonumber \\
&& 4\frac{d}{dt_E}H + 6H^2 = -p - p_\Phi - a^{-2}(t_E){\cal J}_{ii}~,\qquad i = 1,2,3, 
\nonumber \\
&& \frac{d^2\Phi}{dt_E^2} + 3H\frac{d\Phi}{dt_E} 
+  V'(\phi) + \frac{1}{2}[\Psi'(\Phi)(\rho - 3p) + \sigma + \sigma_\phi ]=0~,\nonumber \\
&& \sigma \equiv  -2\frac{1}{V^{(3)}\sqrt{-g_E}}
\frac{\delta (I_{\rm m} + \int (16\pi\alpha (\Phi))^{-1}F^2)}{\delta \Phi}~, 
\nonumber \\
&& \sigma_\phi \equiv  -2\frac{1}{V^{(3)}\sqrt{-g_E}}{\cal I}'(\Phi)
= -12 \{ 4\left(\frac{d\Phi}{dt_E}\right)^2  + 6H\frac{d\Phi}{dt_E}
+ 2 \frac{d^2\Phi}{dt_E^2} + \frac{dH}{dt_E}+ \frac{5}{2}H^2\}~, 
\nonumber \\
&&\rho_\Phi \equiv \frac{1}{2}\left(\frac{d\Phi}{dt_E}\right)^2 + V(\Phi)~,
\quad p_\Phi \equiv \frac{1}{2}\left(\frac{d\Phi}{dt_E}\right)^2 - V(\Phi)~,
\label{eqsmotion}
\end{eqnarray}
where the prime denotes differentiation
with respect to $\Phi$, 
${\cal I}_\phi'(\Phi)$ is defined in (\ref{iprimephi}), and we 
use canonically-normalized dilaton fields.
Notice that (\ref{eqsmotion}) differ from the 
corresponding on-shell equations in~\cite{gasperini} by the 
Liouville \emph{out-of--equilibrium} 
contributions ${\cal J}$ and $\sigma_\phi$,
which are exclusive to our treatment~\cite{emn,emninfl,dgmpp}. 

The equations (\ref{eqsmotion}) lead, after standard manipulations, 
to the coupled (non-) conservation equations of 
matter and dilaton energy density, 
in the presence of the non-equilibrium contributions:
\begin{eqnarray}\label{eqsmotion2}  
&&\frac{d\rho_\Phi}{dt_E} + 3H(\rho_\Phi + p_\Phi) 
+ \frac{1}{2}
\frac{d\Phi}{dt_E}[\Psi'(\Phi)(\rho - 3p) + \sigma + \sigma_\phi ] =0~, 
\nonumber \\ 
&& \frac{d\rho}{dt_E} + 3H(\rho + p) 
- \frac{1}{2}\frac{d\Phi}{dt_E}[\Psi'(\Phi)(\rho - 3p) + \sigma + \sigma_\phi ] 
+ \left(\frac{d}{dt_E} + 3H\right){\cal J}_{00} + 3Ha^{-2}(t_E){\cal J}_{ii} 
= 0~, \nonumber \\
\end{eqnarray}
with the values of ${\cal J}_{00}$ and ${\cal J}_{ii}$ (common for all
$i=1,2,3$ in our case) given by (\ref{jcaldef}),(\ref{liouvmatter}). 
As we see from (\ref{eqsmotion2}) the covariant 
conservation of the matter stress tensor (the first three terms in the 
second of eqs. (\ref{eqsmotion2})) breaks down, due not only to the presence
of a dilaton field, but also to the off-shell
Liouville contributions given by the ${\cal J}$-dependent terms, 
which express the non-equilibrium nature of the
Liouville cosmology. 

To solve (\ref{eqsmotion2}) in the various epochs of the 
Universe, it is convenient first to split the energy density of matter
as well as the function $\sigma$ 
into radiation $\rho_r$, baryonic $\rho_b$ and dark-matter $\rho_d$ 
components, and to use the simple equation of state (\ref{eqnstate})
for the dilaton fluid: 
\begin{eqnarray}\label{step1} 
&& \rho = \rho_r + \rho_b + \rho_d \equiv \rho_r  + \rho_m~, \nonumber \\
&&\sigma = \sigma_r + \sigma_b + \sigma_d \equiv \sigma_r + \sigma_m, 
\nonumber \\ 
&& p_b = p_d =0, \quad p_r = \frac{1}{3}\rho_r,  \quad 
p_\Phi = w_\Phi \rho_\Phi~.
\end{eqnarray} 
Using (\ref{step1}), one can split the matter evolution equation (second 
of eqs. (\ref{eqsmotion2})) into various components, which for an expanding 
Universe can be cast in the form:
\begin{eqnarray} 
&& \frac{d\rho_r}{d\chi} + 4\rho_r 
- \frac{1}{2}\frac{d\Phi}{d\chi}[\sigma_r + \sigma_\phi]+
\frac{d}{d\chi}{\cal J}_{00} + 3({\cal J}_{00} + a^{-2}(t_E){\cal J}_{ii}) =0~, 
\nonumber \\
&& \frac{d\rho_A}{d\chi} + 3\rho_A 
- \frac{1}{2}\frac{d\Phi}{d\chi}[\Psi '(\phi)\rho_A + \sigma_A + \sigma_\phi]+
\frac{d}{d\chi}{\cal J}_{00} + 3({\cal J}_{00} + a^{-2}(t_E){\cal J}_{ii}) =0~, 
\quad A=b,d~, \nonumber \\
&& \frac{d\rho_\Phi}{d\chi} + 3(1 + w_\Phi)\rho_\Phi 
+ \frac{1}{2}\frac{d\Phi}{d\chi}[\Psi '(\phi)\rho_m + \sigma + \sigma_\phi]=0~,
\label{components}
\end{eqnarray} 
with 
\begin{equation}
\chi ={\rm ln}(a/a_{\rm init})=-{\rm ln}(1 + z) + {\rm ln}(a(0)/a_{\rm init}), 
\end{equation}
where $z$ is the redshift, 
$a_{\rm init}$ is an initial scale, and $a(0)$ is the present 
value of the scale factor, evaluated at redshift zero. 

Solving the above equations rigorously is a complicated task, and depends
on the details of the matter theory.  In general, one may
relate~\cite{gasperini} the various $\sigma_i$, $i=r,b,d$ with the
corresponding energy densities $\rho_i$, through proportionality factors
that depend on the dilaton field $\Phi$.  However, for our purposes we may
assume that at the current era of the Universe's evolution the dark matter
component dominates over ordinary matter and that the dilaton is
approximated by its logarithmic evolution (\ref{dil2}) in cosmic
Einstein-frame time.  We also assume that the current scale factor is also
approximately given by the expression (\ref{einstmetr}). These assumptions
guarantee that the relation (\ref{important}) between the (square of) the
string coupling and the acceleration of the Universe is valid today.

Using the familiar (model-independent) relation of the scale factor 
with the redshift $z$:   
$a(z)= a(0)(1 + z)^{-1}$, we may then determine the region of 
$z$ for which the approximation (\ref{einstmetr}) is consistent.
Recalling that in our model the current era of the Universe  
is defined by the relations (\ref{condition}),(\ref{decel4}), 
with $\beta^2 = -1/q(z=0)$ where $q(z=0)$ is the acceleration of the 
Universe today at $z=0$, it is straightforward to arrive at:
\begin{equation}
a(0) = (F_1/\gamma)(1 + \beta^2)^{1/2}~. 
\label{a0def}
\end{equation}
Making the (wrong) hypothesis 
that the formula for (\ref{einstmetr}) is valid all the way down to $t_E = 0$,
we would then find that $z$ should lie in the region:
\begin{equation}
0 < z < z_{\rm init}~; \qquad z_{\rm init} = \sqrt{1 + \beta^2} - 1~,  
\label{zregion}
\end{equation}
in order that the form (\ref{einstmetr}) of the space-time metric 
be valid. For $q(z=0)=\beta^{-2} = -0.57$ (c.f., (\ref{qval})), we 
would then have $z_{\rm init} \simeq 0.66$. 
 
We assume $\sigma_d \simeq  \eta  \rho_d$ for the dark matter, 
with $\eta $ an approximately constant 
proportionality factor $\eta  = {\cal O}(1)$ 
for the present and future eras. 
These assumptions lead to simplifications of some of the equations.
With $\Psi \simeq 2\Phi$ (the string tree-level 
approximation), one obtains after some elementary 
algebraic manipulations:
\begin{eqnarray} 
&& {\cal J}_{00}  \simeq  -\frac{7H}{t_E} + H^2(1-q) \simeq 
-\frac{6F_1^2}{a^2(t_E)}~,
~~
{\cal J}_{ii}  \simeq  a^2(t_E)\left(\frac{H}{t_E} -H^2(1-q)\right) 
\simeq 0~, ~i=1,2,3 , \nonumber \\
&& \frac{d}{d\chi}{\cal J}_{00} + 3\left({\cal J}_{00} + a^{-2}(t_E){\cal J}_{ii}\right) \simeq 
 -\frac{6F_1^2}{a^2(t_E)} = {\cal J}_{00}~ \rightarrow ~|{\cal J}_{00}| \simeq
\left(\frac{a_{\rm init}}{a(t_E)}\right)^2 = 
\left(\frac{a_{\rm init}}{a(0)}\right)^2(1 + z)^2~, \nonumber \\
\label{jcalvalues}
\end{eqnarray}
for the configuration (\ref{dil2}), (\ref{einstmetr}), (\ref{hubble2}) and
(\ref{decel4}), assumed to characterize the current era of the Universe. 

Equation (\ref{components}) yields, then, for the dark matter 
energy density, (which is assumed  to dominate the matter sector: $\rho_m 
\simeq \rho_d$, $\sigma \simeq \sigma_d$):
\begin{equation}\label{darkmatter} 
\frac{d\rho_d}{d\chi} + 3\rho_d + 
\frac{d\rho_\Phi}{d\chi} + 3(1 + w_\Phi)\rho_\Phi + {\cal J}_{00} = 0~.
\end{equation} 
This is a rather complicated equation to solve in 
general~\footnote{Notice that most of the
complications arise from the presence of the 
off-shell Liouville modifications ${\cal J}_{00}$.
In their absence, i.e., in `conventional' dilaton cosmologies, 
one can solve this equation straightforwardly and 
obtain the standard scaling for the various energy-density 
components $\rho_d \sim \rho_d^0 a^{-3}$, 
$\rho_\Phi \sim \rho_\Phi^0 a^{-3(1 + w_\Phi)}$, 
with $\rho_d^0 + \rho_\Phi^0 \simeq 1$ 
in the case of dominant dark matter.
This is no longer true when ${\cal J}_{00} \ne 0$, and, as we shall
see below, one obtains in that case 
a mixed scaling for the matter energy density.}.
Its solution, in conjunction with the rest of eqs.~(\ref{eqsmotion}),
will provide a 
scaling for the energy densities of matter
and dark energy with the scale factor, which is modified in general. 

A simplification can occur, however, if one 
concentrates on the epoch of large cosmic time (present era),
and uses the asymptotic behaviour of the dilaton dark energy density 
$\rho_\Phi$, dictated by (\ref{potendilaton}), 
(\ref{dilpotkin})~\footnote{Notice that, despite the $a^{-2}$ scaling of 
$\rho_\Phi$ and the off-shell Liouville term ${\cal J}_{00}$,
none of these contributions is equivalent to a (negative) 
curvature contribution. This is due to the fact that 
the dilaton dark energy and the off-shell Liouville modifications
enter the relevant dynamical equations (\ref{eqsmotion}) in a different
manner than the curvature term. This 
is consistent with the fact that our brane/string Universe is spatially 
flat by construction.}, i.e., $\rho_\Phi\sim \rho^0_{\Phi} a^{-2}(t_E)$. 
Assuming then mixed scaling behaviours
\begin{equation}
\rho_d \sim \rho_{\rm dust}^0a^{-3} + \rho_{\rm exotic}^0 a^{-2}~, 
\label{mixedscaling}
\end{equation}
where the first term is compatible with dust
properties, and the second expresses the 
entanglement with the off-shell Liouville environment, 
and the value (\ref{eqnstatedilnum3}) for the 
equation of state,  
in agreement  
with recent WMAP data~\cite{wmap,steinhardt}, 
we observe that (\ref{darkmatter}) 
is  satisfied, provided  
\begin{equation}
 \rho_{\rm exotic}^0 = 6F_1^2 + 1.46\rho^0_{\Phi}~.
\label{rhoexotic}
\end{equation}
Note that, for the model of \cite{dgmpp}), 
$\rho^0_\Phi$ is in most cases also of 
order $F_1^2$ (c.f., (\ref{dilpotkin}),
which in turn is of the order of the square of the 
asymptotic central charge (\ref{f1q0}). The latter can 
be very small, e.g., of order
$10^ {-60}$ in string units, in 
models~\cite{gravanis} involving compactification on magnetized
tori (c.f. (\ref{sigmamodelq0}), (\ref{horder}) below), which 
guarantees compatibility of the order of the 
magnetically-induced target-space 
supersymmetry breaking with realistic 
phenomenological considerations. In this way, $\rho_{\rm exotic}^0$ 
can be very small, 
and hence both terms in $\rho_d$ may be of comparable magnitude today.
Specifically, from (\ref{dilpotkin}) it follows that 
$\rho_\Phi \simeq 3.78 \frac{F_1^2}{a^2}$, which implies that 
$\rho_\Phi^0 \simeq 3.78F_1^2$. Thus, (\ref{rhoexotic})
would yield in that case: 
\begin{equation}
\rho_{\rm exotic}^0 \simeq 3.25\rho_\Phi^0~.
\label{rhoexotic2}
\end{equation}
Thus, we may write for the (dark) matter energy density today
\begin{equation}
\rho_d^0 \simeq \rho_{\rm dust}^0  + \lambda \rho_\Phi^0~,~~ 0 < 
\lambda ={\cal O}(1-10) .
\label{lambdadef}
\end{equation}
We stress once
more that this mixed scaling in the matter energy density 
is due not only to the entanglement with the dilaton, 
but also to the non-trivial r\^ole of the off-shell Liouville
${\cal J}$-dependent contributions. 
This is an exclusive
feature of our non-equilibrium Liouville-string 
approach to cosmology~\cite{emn,emn04,dgmpp}, which does not apply in  
conventional on-shell treatments~\cite{damour,gasperini}. 
It has its roots in viewing target time as a 
world-sheet renormalization-group dynamical scale in non-critical
string theory, which is a cornerstone of our approach. 

As for the dilaton, a form linear in the $\sigma$-model frame is assumed
throughout, which in turn determines the time dependences of the
quantities $\sigma, \sigma_\phi$ via the respective dilaton equation
(\ref{eqsmotion}).  Since the form of matter action is not in general
fully known in our generic low-energy considerations, and depends on the
details of the underlying microscopic string/brane model, we do not
analyse this equation further here.  The existence of self-consistent
solutions to the graviton equations (\ref{eqsmotion})  including matter
and radiation, inferred by the above analysis, justifies \emph{ a
posteriori} our assumption that the important relation (\ref{important}),
which is based on the solution~\cite{emn04,dgmpp} (\ref{einstmetr2}),
(\ref{einsteindil})  for the space-time metric and the dilaton fields,
survives the inclusion of matter.
 
We now present the formalism for fitting our 
Liouville Cosmology to cosmological data in a rather
model-independent way. Consider the first of the Einstein
equations (\ref{eqsmotion}) for our Liouville cosmology. In the present
era, we may assume the following asymptotic behaviour for the Liouville
part ${\cal J}$ (c.f. (\ref{jcalvalues})): ${\cal J}_{00} \simeq
-6H^2(1-q)$. From the first of eqs.~(\ref{eqsmotion}), we may then
conclude that, as a result of the Liouville out-of-equilibrium
contributions, the \emph{critical} total mass (energy)  density of the
fluid, $\rho_c$ required to have a spatially flat Universe, is no longer
$6H^2$, as in the conventional on-shell Einstein cosmologies, but
\begin{equation}
\rho_c = 6H^2(2 - q).
\label{criticaldens}
\end{equation}
One may then define modified $\Omega'_i$ fractions:
\begin{equation}
\Omega' \equiv \frac{\rho_i}{\rho_c} = \frac{\rho_i}{6H^2(2 - q)}~,~~ 
i={\rm matter},~{\rm dilaton}~\Phi~{\rm dark~energy} ~\emph{etc.}.
\label{newomegas}
\end{equation}
With this definition the first of equations (\ref{eqsmotion}) 
would imply the standard relation for a spatially flat Universe today:
\begin{equation}\label{sumomegas}
\Omega'_{\rm Matter} + \Omega'_\Phi = 1~.
\end{equation}
Notice that the critical density (\ref{criticaldens}) scales with 
$a = a(0)(1 + z)^{-1}$ as:
\begin{equation}
\rho_c(z) = \frac{6F_1^2}{a^2}\left( 2 - \xi ( 1 + z)^2\right)~, 
~~\xi \equiv \frac{F_1^2}{\gamma ^2 a^2(0)} = \frac{1}{1+ \beta^2}~, 
\beta^2 = -1/q(z=0)~, 
\label{rcscale}
\end{equation}
for the (rather generic) string model of~\cite{dgmpp} used here,
where we took into account (\ref{a0def}). 
The reader should also recall that $0 < z < 0.66$  for the validity of the
approximations leading to the above analysis. 
Thus, we have the following scaling with the redshift:
\begin{eqnarray}
\Omega'_\Phi (z) &=& \frac{\rho_\Phi^0}{6F_1^2 (2 - \xi (1 + z)^2)}
 \equiv {\Omega'_\Phi}^{0}\frac{2-\xi}{2-\xi(1 +z)^2}~,
\nonumber \\
\Omega'_{\rm Matter} & = & \frac{1}{6F_1^2(2 - \xi (1 + z)^2)}\left(
\frac{\rho_{\rm dust}^0}{1+z} + \lambda \rho_\Phi^0 \right)~,~~
\label{omegaprimescaling}
\end{eqnarray}
where 
$\lambda = {\cal O}(1-10)$ depending on model details, and  
${\Omega'_\Phi}^{0}$denotes the corresponding quantity today, i.e. at $z =0$. 

Alternatively, one may use the first of eqs. (\ref{eqsmotion}) 
to express the critical density in terms 
of the various energy-density components instead of the parameter $\xi$.
This allows a more convenient 
model-independent formalism to be used for comparison 
with data. In this way,  
using the mixed scaling  (\ref{mixedscaling}) for the matter sector 
(including the dominant dark matter) of 
our spatially flat Universe,  
$\rho_{\rm Matter} = \rho_{\rm dust}^0 (1 + z)^3 + 
\rho_{\rm exotic}^0(1 + z)^2$ with $\rho_{\rm dust}^0 + \rho_{\rm 
exotic}^0 
+ \rho_\Phi^0 = \rho^0_c$, one obtains 
the following scaling of $\Omega_i'$ with the redshift $z$:
\begin{eqnarray}\label{scalings}
 \Omega'_\Phi (z) = \frac{1}{1 + \rho_{\rm Matter}/\rho_\Phi} &=& 
\frac{1}{1 + \frac{{\Omega '}_{\rm dust}^0}{{\Omega '}_\Phi^0}(1 + z) 
+ \frac{{\Omega '}_{\rm exotic}^0}{{\Omega '}_\Phi^0}}~, 
\nonumber \\
\Omega'_{\rm Matter} (z) = 1 -  \Omega'_\Phi (z) &=& 
\frac{1}{1 + \frac{1}{\frac{{\Omega '}_{\rm dust}^0}{{\Omega '}_\Phi^0}
(1 + z) + \frac{{\Omega '}_{\rm exotic}^0}{{\Omega '}_\Phi^0}}}~. 
\end{eqnarray} 
These expressions can be used to fit the astrophysical data and derive
values for the cosmological parameters of our Q-cosmology model. 

Since the dilaton, matter and radiation energy densities scale
differently, there is a (past) era of this Liouville Universe,
corresponding to redshifts larger than a critical value, $z > z^*$, in
which matter effects dominate over the dilaton dark energy, leading to a
decelerating phase of the Universe.  In fact, such a past early era when
there was deceleration of the Universe was present also in the model
of~\cite{dgmpp}, even in the purely gravitational and moduli sector. Such
a feature is simply pronounced by the inclusion of matter, since the
latter feels the attractive feature of gravity. The
critical $z^*$ is shifted from an early era in the purely gravitational
case of~\cite{dgmpp}) towards the current epoch: $z^* \to \sim {\cal
O}(1)$, as a result of the inclusion of matter in the model. The past
deceleration in our Universe is a feature confirmed by astrophysical
data~\cite{wmap,deceldata}, which indicate a value $z^* \sim 1/2$.  
However, our model is too generic, at this stage, to claim a specific
prediction for $z^*$.

\section{Concrete Non-critical String Examples: Colliding Branes}

The above considerations are rather generic for models which relax
asymptotically to the linear-dilaton conformal field theory solutions
of~\cite{aben}, and from this point of view are physically interesting. We
have not yet specified the microscopic theory underlying the deviation
from criticality.  For this purpose, one needs specific examples of such
deviations from the conformal invariant points in string theory space. One
such example with physically interesting consequences is provided by a
colliding-brane-world scenario, in which the Liouville string $\sigma$
model describes stringy excitations on the brane worlds for relatively
long times after the collision, so that string perturbation theory is
valid. This Section is devoted to a detailed discussion of such a
scenario~\cite{gravanis,emw}.

\subsection{Example I: Colliding Type-IIB Five-Branes}

\begin{figure}[htb]
\begin{center}
\epsfxsize=3in
\bigskip
\centerline{\epsffile{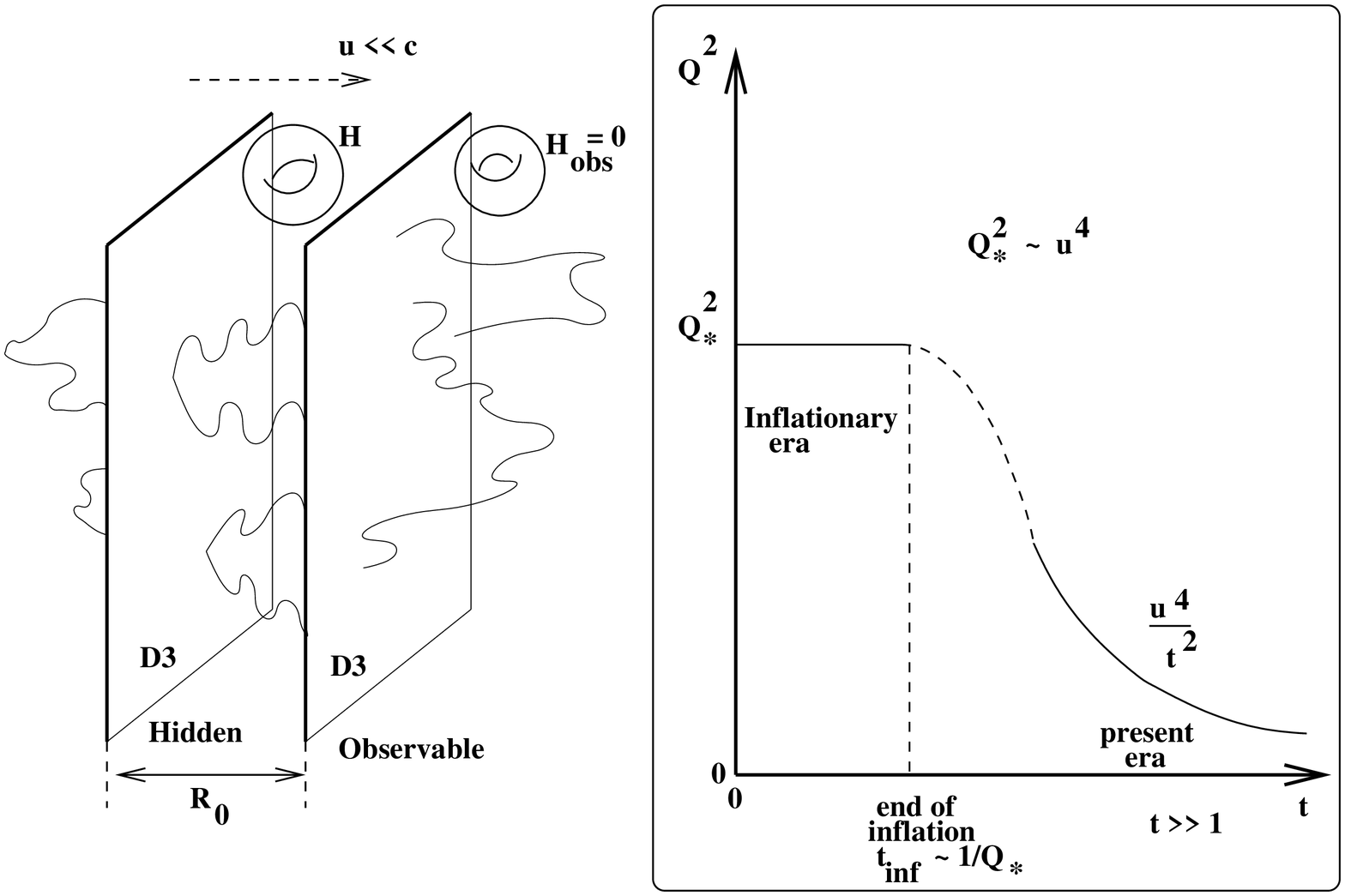}}
\caption{{\it A scenario 
in which the collision of two Type-II five-branes provides 
inflation and a relaxation model for cosmological vacuum energy.
\label{infla}}}
\end{center} 
\end{figure}

We now concentrate on particular examples of the previous general
scenario~\cite{emninfl}, in which the non-criticality is induced by the
collision of two branes, as seen in Fig.~\ref{infla}. We first discuss the
basic features of this scenario. For our purposes below we assume that the
string scale is of the same order as the four-dimensional Planck scale.
However, this is an assumption which can be relaxed in view of recent
developments in strings with large compactification directions, as was
mentioned in the Introduction.

Following~\cite{gravanis}, we consider two five-branes of Type-II string
theory, in which the extra two dimensions have been compactified on tori.
On one of the branes (assumed to be the hidden world), the torus is
magnetized with a field intensity ${\cal H}$.  Initially our world is
compactified on a normal torus, without a magnetic field, and the two
branes are assumed to be on a collision course with a small relative
velocity $v \ll 1$ in the bulk, as illustrated in Fig.~\ref{infla}. The
collision produces a non-equilibrium situation, which results in electric
current transfer from the hidden brane to the visible one.  This causes
the (adiabatic) emergence of a magnetic field in our world.

The instabilities associated with such magnetized-tori compactifications
are not a problem in the context of the cosmological scenario discussed
here. In fact, as discussed in~\cite{gravanis}, the collision may also
produce decompactification of the extra toroidal dimensions at a rate much
slower than any other rate in the problem. As discussed
in~\cite{gravanis}, this guarantees asymptotic equilibrium and a proper
definition of an $S$-matrix for the stringy excitations on the observable
world. We come back at this issue at the end of this Section.

The collision of the two branes implies, for a short period afterwards
while the branes are at most a few string scales apart, the exchange of
open-string excitations stretching between the branes, where their ends
are attached. As argued in~\cite{gravanis}, the exchanges of such pairs of
open strings in Type-II string theory result in an excitation energy in
the visible world. The latter may be estimated by computing the
corresponding scattering amplitude of the two branes, using string-theory
world-sheet methods~\cite{bachas}: the time integral for the relevant
potential yields the scattering amplitude. Such estimates involve the
computation of appropriate world-sheet annulus diagrams, due to the
existence of open string pairs in Type-II string theory. This implies the
presence of `spin factors' as proportionality constants in the scattering
amplitudes, which are expressed in terms of Jacobi $\Theta$ functions. For
the small brane velocities $v \ll 1$ we are considering here, the
appropriate spin structures start at {\it quartic order } in $v$, for the
case of identical branes, as a result of the mathematical properties of
the Jacobi functions~\cite{bachas}. This in turn
implies~\cite{gravanis,emw} that the resulting excitation energy on the
brane world is of order $V = {\cal O}(v^4)$, which may be thought of as an
initial (approximately constant) value of a {\it supercritical}
central-charge deficit for the non-critical $\sigma$ model that describes
stringy excitations in the observable world after the collision:
\begin{equation}
Q^2 = \left(\sqrt{\beta} v^2  + {\cal H}^2\right)^2 > 0,
\label{initialdeficit}
\end{equation} 
where, in the model of~\cite{emw,brany}, the proportionality factor $\beta$, 
computed using string amplitude computations, is of order
\begin{equation} 
\beta \sim 2\sqrt{3} \cdot 10^{-8} \cdot g_s~, 
\label{betaorder}
\end{equation} 
with $g_s $ the string coupling, which is of order $g_s^2 \sim 0.5$ for
interesting phenomenological models~\cite{gsw,ibanez}.  The
supercriticality, i.e., the positive definiteness of the central charge
deficit (\ref{initialdeficit}) of the model, is essential~\cite{aben} for 
a
time-like signature of the Liouville mode and hence its interpretation as
target time.

At times long after the collision, the branes slow down and the central 
charge deficit is no
longer constant but relaxes with time $t$.  In the approach
of~\cite{gravanis}, this relaxation has been computed by using world-sheet
logarithmic conformal field theory methods~\cite{kogan}, taking into
account recoil (in the bulk) of the observable-world brane and the
identification of target time with the (zero mode of the) Liouville field.
In that work it was assumed that the final equilibrium 
value of the central-charge deficit was zero, i.e., the theory approached
a critical string. This late-time varying deficit $Q^2(t)$ 
scales with the target time (Liouville mode) as follows (in units 
of the string scale $M_s$):
\begin{equation} 
Q^2 (t) \sim \frac{({\cal H}^2 + v^2)^2}{t^2} .
\label{cosmoconst}
\end{equation} 
Some explanations are necessary at this point.
In arriving at (\ref{cosmoconst}), one identifies 
the world-sheet renormalization group scale ${\cal T} ={\rm ln}(L/a)^2$,
where $(L/a)^2$ is the world-sheet area, which appears in the 
Zamolodchikov $c$-theorem
used to determine the rate of change of $Q$ with ${\cal T}$,  
with the zero mode of a normalized 
Liouville field $\phi_0$, such that $\phi_0 = Q{\cal T}$. This normalization
guarantees a canonical kinetic term for the Liouville field in the 
world-sheet action~\cite{ddk}. Thus, $\phi_0$ is identified
with $-t$, where $t$ is the target time. This will always 
be understood in what follows.

On the other hand, in other models~\cite{dgmpp}
that we discuss below, 
the asymptotic value of the central-charge 
deficit may not be zero, in the sense that the asymptotic theory 
is that of a 
dilaton field that is linear in time, with a Minkowski metric in the 
$\sigma$-model frame~\cite{aben}. This theory is still a conformal
model, but the central charge is a constant $Q_0$, and in fact the 
dilaton is of the form $\Phi = Q_0 t + {\rm const}$, 
where $t$ is the target time in the $\sigma$-model frame.
Conformal invariance, as already mentioned previously, 
suggests~\cite{aben}
that $Q_0$ takes on 
one of a {\it discrete} set of values, in the way explained
in~\cite{aben}.
In such a case, following the same method as in the $Q_0=0$ case 
of \cite{gravanis},
one arrives at the asymptotic form
\begin{equation} 
Q^2 (t) \sim Q_0^2 + {\cal O}\left(\frac{{\cal H}^2 + v^2)}{t}Q_0\right) 
\label{cosmoconst3}
\end{equation} 
for large times $t$. 

\paragraph{}
\begin{figure}[tb]
\begin{center}
\includegraphics[width=4cm]{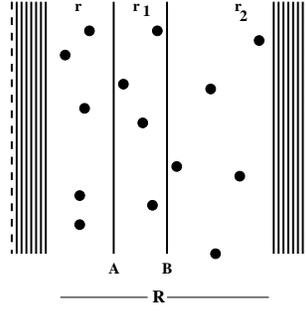}
\end{center}
\caption{\it A model for supersymmetric D-particle foam
consisting of two stacks each of sixteen parallel coincident D8-branes, 
with orientifold planes (thick dashed lines) attached to them~\cite{emw}.
The space does not extend beyond the orientifold planes.
The bulk region of ten-dimensional space in which the D8-branes 
are embedded is punctured by D0-branes (D-particles, dark blobs). 
The two parallel stacks are sufficiently far from each other
that any Casimir contribution to the vacuum energy is negligible.
If the branes are stationary,
there is zero vacuum energy, and the configuration 
is a consistent supersymmetric string vacuum. To obtain excitations
corresponding to interesting cosmologies, one should move one (or more) of 
the 
branes from each 
stack, let them collide (Big Bang), bounce back (inflation),
and then relax to their original position, where they collide again
with the remaining branes in each stack (exit from inflation, reheating).}
\label{fig:nonchiral}
\end{figure}

\subsection{Example II: Orientifold/Eight-Brane/D-particles 
Colliding-Brane Model}

The colliding-brane model of~\cite{gravanis} 
can be extended to incorporate 
proper supersymmetric vacuum configurations 
of string theory ~\cite{emw}. As illustrated in Fig.~\ref{fig:nonchiral},
this model consists of two stacks of D8-branes with the same tension, separated
by a distance $R$. The transverse bulk space is restricted to lie between
two orientifold planes, and is populated by D-particles. It was shown
in~\cite{emw} that, in the limit of static branes and D-particles, this
configuration constitutes a zero vacuum-energy supersymmetric ground state
of this brane theory.

The bulk low-energy effective theory in such configurations 
is known to be the ten-dimensional  Type-IIA supergravity,
whose bosonic part is given in the string frame by~\cite{polchinski}:
\begin{equation}
S = S_{NS} + S_R + S_{CS},
\end{equation}
where
\begin{eqnarray}\label{actioniia}
S_{NS} &=& \frac{1}{2\kappa^2_{10}}\int d^{10} x \sqrt{- G }
e^{-2\Phi}\left( R + 4 |\nabla_\mu \Phi|^2 -\frac{1}{2}|
H_3|^2
\right),\\
S_R &=& -\frac{1}{4\kappa^2_{10}}\int d^{10} x \sqrt{-
  G}\left(|F_2|^2+|\tilde F_4|^2  \right),\\  
S_{CS} &=& -\frac{1}{4\kappa^2_{10}}\int B_2 \wedge dC_3 \wedge dC_3 ,
\end{eqnarray}
in standard notation. 
However, in order to incorporate eight-dimensional 
branes, one needs actually a modified version of the Type-IIA
supergravity, which we now proceed to describe briefly, along with its
possible compactifications, since we are eventually interested 
in four space-time dimensional theories of phenomenological interest.

\subsection{Dual Formulation of Type-IIA Supergravity}

Type-IIA string theory contains all even-dimension D-branes from zero to 
eight dimensions $p$~\cite{polchinski}. D$p$-branes couple to R-R 
$p+1$-forms, but the action
for Type-IIA supergravity only contains 1-form (D0-brane) and 3-form
(D2-brane) gauge potentials - how are the other D-branes incorporated
into the action? In~\cite{Bergshoeff:2001pv}, a \emph{dual}
formulation of Type-IIA supergravity was constructed, which
contains higher-dimensional R-R potentials and hence allows objects
like the D8-brane to be incorporated. The dual Type-IIA supergravity allows 
for the construction of Type-IA supergravity, and has the action:
\begin{align}\label{dualaction}
 S_{\text{bulk}} =
  & - \frac{1}{2\kappa_{10}^2}\int d^{10} x \sqrt{-g}
    \Big\{
    e^{-2\Phi} \big[
    R\big(\omega(e)\big) + 4\big( \partial{\Phi} \big)^{2}
    +\tfrac{1}{2} H \cdot H
    -2\partial^{{\mu}}{\Phi} \CHI{1}_{{\mu}}
        + H \cdot \CHI{3}+ \notag \\
  & +2 \bar{{\psi}}_{{\mu}}{\Gamma}^{{\mu}{\nu}{\rho}}
    {\nabla}_{{\nu}}{\psi}_{{\rho}}
    -2 \bar{{\lambda}}{\Gamma}^{{\mu}}
    {\nabla}_{{\mu}}{\lambda}
    +4 \bar{{\lambda}} {\Gamma}^{{\mu}{\nu}}
    {\nabla}_{{\mu}}{\psi}_{{\nu}}
    \big]
    +   \sum_{n=0,1,2}
     \tfrac{1}{2} \G{2n} \cdot \G{2n}
    + \G{2n} \cdot \PSI{2n}  +  \notag \\
  & - \star \, \big[
      \tfrac{1}{2} \, \G{4} \G{4} B
    - \tfrac{1}{2} \, \G{2} \G{4} B^2
    + \tfrac{1}{6} \, \G{2}{}^2 B^3
    + \tfrac{1}{6} \, \G{0} \G{4} B^3 -\tfrac{1}{8} \, \G{0} \G{2} B^4+ \notag \\
  & +\tfrac{1}{40} \, \G{0}{}^2 B^5
    + {\bf e}^{- B} {\bf G}
    d (\A{5} - \A{7} + \A{9}) \big] \Big\}  +\mbox{ quartic fermionic 
terms}\,,
\end{align}
in conventional notation~\cite{Bergshoeff:2001pv}. However, this dual 
formulation only
describes branes of dimension 4, 6 and 8 because of the problem of
consistently introducing all of the available R-R forms. A
\emph{democratic} formulation was also constructed, which contained all
potentials, but this version has no proper action~\cite{Bergshoeff:2001pv}. 
Since the D-foam model of~\cite{emw} contains D8-branes, O8-planes and
D0-branes, along with fundamental strings when the D0-branes are
between an odd number of D8-branes, neither the dual nor the democratic
action is appropriate. One needs a combined action, which has been
constructed in~\cite{Kallosh:2001zc}: 
\begin{align}\label{kalloshact} 
S =  &   - \frac{1}{2\kappa_{10}^2}\int d^{10} x \sqrt{-G}
    \Big\{  e^{-2\Phi} \big[ R  + 4\big(
  \partial{\Phi}   \big)^{2} +\tfrac{1}{12} (H^{(3)})^2  \big] +
  \tfrac{1}{2} (\G{0})^2  +  \tfrac{1}{2} (\G{2})^2 - \star  \big[
    \G{0}  d \A{9}    \big] \Big\}        \non\\ 
  & -T_8 (n-8) \int d^{10} x
  \left(e^{-\Phi} \sqrt{|G_{(9)}|}+\alpha
  \frac{1}{9!}\epsilon^{(9)} \A{9} \right) \left[\delta(x^9) -
    \delta(x^9-\pi R) \right]  \non\\ 
   -T_2 &\int d^{10} x 
  \left[\left(e^{-\Phi}
    \sqrt{-G_{tt}}-bA_t\right)+\frac{e^{-\phi}}{\sqrt{G_{zz}}}\left(\sqrt{-G_{tt} 
    G_{zz}} - 
    \frac{a}{2!}\epsilon^{\nu\lambda} B_{\nu\lambda}\right)\right] \left[\sum_k N_k
    \delta^8 (\vec x-\vec x_k)  \right]  ,
\end{align}
where $\vec x$ denotes an eight-dimensional vector, 
$\G{n}, A^{(n)}$ are appropriate gauge flux fields, $G$ denotes the
ten-dimensional $\sigma$-model-frame target-space metric, 
$G_{00}, G_{zz}$ are 
the temporal and bulk components of this metric respectively,
and $G_{(9)}$ is a nine-dimensional metric.
The second line describes the D8-branes and orientifold planes
and the third the combined action for the D0-brane and fundamental
strings. These brane-bulk actions describe all of the dynamics relevant to 
the branes of interest.

\subsection{Towards Realistic Compactifications}

To compactify  such an action, e.g., on $T^4/\mathbb{Z}_2$,
the  fields which survive the orbifold projection must be
determined. An example of this for D6-O6 branes is given 
in~\cite{Villadoro:2005cu}. Once the remaining field content is found,
calculation of the dimensionally-reduced  Bianchi identities then
leads to the lower-dimensional effective potential and  corresponding
superpotential. Instead of a normal Kaluza-Klein dimensional
reduction, Scherk-Schwarz fluxes~\cite{Scherk:1979zr} can be 
added~\cite{Dall'Agata:2005ff}, which have the advantage of allowing a
greater range of vacua: see~\cite{ DeWolfe:2005uu, 
Villadoro:2005cu,Camara:2005dc} and references 
therein~\footnote{As we saw in the generic
analysis of~\cite{dgmpp}, reviewed above, 
such flux fields play an important r\^ole 
in ensuring the stabilisation of large bulk dimensions.
A similar scenario is envisaged for the compactified version
of the eight-brane model of~\cite{emw}.}.     

The overall structure of Type~IA string theory/supergravity is
$\mathcal{M}^{\hspace{1pt} 9} \times S^1/\mathbb{Z}_2 I_9\Omega$,
corresponding to an orbifold of the Type-IIA theory with an orientifold
projection in the ninth dimension. A realistic compactification would
result in either a Randall-Sundrum type scenario~\cite{randal}, i.e., a
3-brane embedded in five dimensions, or a conventional intersecting
D-brane model~\cite{Blumenhagen:2002wn}, with the unusual feature of using
D8-branes instead of D6-branes~\cite{Honecker:2002hp}.
For Randall-Sundrum-II (RS-II) scenarios, it has been 
suggested~\cite{Bergshoeff:2001pv} that a metric product of $AdS_5$ and  some
Euclidean 5-manifold would give 3-branes in 5-dimensional Minkowski
space, with the bulk solution being uplifted 
from~\cite{Freedman:1999gp}~\footnote{We recall that 
the strong-coupling limit of Type-IA string theory is equivalent to the 
solution of Horava and Witten (HW)~\cite{Horava:1996ma, Lukas:1998yy, 
Stelle:1998yv}.}.

The range of compactification choices can be summarized as follows:

\begin{itemize}
\item{Compactify D8-O8 on $AdS_5 \times \mathcal{M}^5$ to get an 
RS-II scenario~\cite{Bergshoeff:2001pv},}
\item{Compactify on appropriate torii to get an intersecting brane
model~\cite{Honecker:2002hp},}
\item{Compactify on K3 $\times S^d$, giving another intersecting brane
model based upon Calabi-Yau manifolds~\cite{Blumenhagen:2002wn}.}
\end{itemize}
The important requirement is to obtain $D=4$, $\mathcal{N}=1$
supersymmetry on the brane, which in the (compactified) model
of~\cite{emw} would correspond to the static D-brane/D-particle
configuration.  We recall that Type-IIA supergravity has 32
supersymmetries~\cite{polchinski}. In the the model of~\cite{emw}, the
bulk space has $\mathcal{N}=2$, and on the brane there
is~\cite{Bergshoeff:2001pv} $\mathcal{N}=1$. It should be noted that, for
simplicity in this case, we are assuming that all of the branes are
located on the orientifolds, and not in the bulk. Toroidal
compactification of supergravity does not break any supersymmetries, so
compactification on $T^5$ would give $D=5$, $\mathcal{N}=4$ supersymmetry
in the bulk, with $\mathcal{N}=2$ on the brane. Changing this to
$T^5/\mathbb{Z}_2$ breaks half of the supersymmetries resulting in
$\mathcal{N}=2$ in the bulk and $\mathcal{N}=1$ on the brane. More
complicated compactifications would change the precise way in which the
supersymmetries are broken, as in the example suggested
by~\cite{Bergshoeff:2001pv}, where the metric is the product of $AdS_5$
and a Euclidean 5-manifold. This would also result in $D=4$,
$\mathcal{N}=1$ on the brane. Inspired by the analysis on the colliding
five-brane model of~\cite{gravanis}, in which the five-branes were
compactified on magnitized tori to yield three-brane worlds with broken
supersymmetry, as a result of internal magnetic fields, it would be
desirable to discuss similar magnetized compactifications for the
eight-branes of~\cite{emw}. This may be subtle due to the
presence of the orientifolds, but some progress has already been made in 
this
direction~\cite{dudas}.

In this work we deal no further with the important issue of
compactification, but postpone a detailed analysis to a future
publication. 

\subsection{Supersymmetry Breaking and Vacuum Energy in the \\
Post-Inflationary Era}

We now discuss briefly issues related to the supersymmetry breaking that
would result from brane motion in the model.

\subsubsection{Supersymmetry Breaking via Internal Magnetic Fields}

In the colliding-brane scenario of~\cite{brany}, which 
uses the orientifold configuration of~\cite{emw} shown in
Fig.~\ref{fig:nonchiral}), one may imagine 
that the exit from inflation and the reheating phase
corresponds to a second collision, when the moving brane world
returns to its initial position
and hits the original stack of branes again. In such a case the recoil 
velocity 
of the brane world vanishes, but one may still have a magnetic 
field ${\cal H}$ on the brane world, corresponding to a contribution 
to the four-dimensional 
energy density on the brane of order ${\cal H}^2$, for compactification
radii of the extra dimensions of order $M_s$. One may identify
therefore 
\begin{equation} 
Q_0^2 \sim {\cal H}^2 
\label{sigmamodelq0} 
\end{equation}
in the $\sigma$-model 
frame, leading to a dilaton of the form $\Phi \sim {\cal H} t + {\rm 
const}$.
For consistency with the results of~\cite{aben}, one would then discretize
${\cal H}$ with one of the values dictated by conformal invariance
in this asymptotic $\sigma$ model. 

An important issue that we have already mentioned, but would like to
stress again, is that in the Einstein frame the constants in the
expression for the dilaton are such that the dark-energy density (c.f.,
(\ref{cosmoconst2})  below)  relaxes to zero with the cosmic time $t_E$ in
the Einstein frame (c.f. (\ref{einstframe} below)  as $1/t_E^2$, in a
manner independent of the magnitude of $Q_0$.  In this way, the magnitude
of the supersymmetry breaking in target space induced by the presence of
${\cal H}$~\cite{bachas,gravanis} may be large enough to be of
phenomenological interest, whilst the observed value of the vacuum energy
may be acceptably small, as we now explain.

The reason why the magnetic field ${\cal H}$ in the extra 
dimensions~\cite{gravanis}
breaks target-space supersymmetry~\cite{bachas} is that bosons and 
fermions on the brane worlds couple differently
to ${\cal H}$.  This is nothing other than a Zeeman-type 
energy-splitting effect. 
In our problem, where the magnetic field is turned on
adiabatically, the resulting mass difference between bosonic and fermionic
string excitations is found to be~\cite{gravanis}:
\begin{equation} 
\Delta m^2_{\rm string} \sim 2q_e|{\cal H}|{\rm cosh}\left(\epsilon \varphi + 
\epsilon t\right)\Sigma_{45},
\label{masssplit}
\end{equation} 
where $q_e$ is the electric charge, 
$\Sigma_{45}$ is a standard spin operator in the plane of the torus,
and $\epsilon \to 0^+$ is the regulating parameter 
of the Heaviside operator $\Theta_\epsilon (t) = -i\int_{-\infty}^\infty 
\frac{d\omega}{\omega -i\epsilon}e^{i\omega t}$
appearing in the D-brane recoil formalism~\cite{kogan}.  
The dependence in (\ref{masssplit}) implies that the formalism selects 
dynamically a 
Liouville mode which flows opposite to the target time $\varphi = -t$,
as mentioned earlier, as a result of minimization of the effective 
field-theoretic potential of the various stringy excitations.

In the scenario of~\cite{gravanis}, where the 
dilaton remains constant asymptotically in time, 
the mass splitting (\ref{masssplit}) with $\varphi = -t$
is the only contribution to supersymmetry breaking as far
as excitations are concerned. Since in that scenario the
dark energy in target space relaxes to zero asymptotically 
(\ref{cosmoconst}) while the mass splittings remain 
finite, provided ${\cal H}$ remains constant one 
has a supersymmetry {\it obstruction}~\cite{witten},
rather than breaking, on the brane world, 
since the cosmological constant of the vacuum state is 
still zero, as required by a supersymmetric theory, but the excitation
spectrum is not supersymmetric. 
By choosing appropriately 
\begin{equation}
q_e|{\cal H}| \sim 10^{-30} ~\quad {\rm (in~string-scale~units)}, 
\label{horder}
\end{equation}
we may arrange for the 
supersymmetry-breaking/obstruction 
scale to be of the order of a few TeV. Such 
contributions would therefore be significantly 
subdominant, compared with the
velocity contribution, in the expressions (\ref{initialdeficit}), 
(\ref{cosmoconst}) during the inflationary era. (We recall 
that phenomenological analyses
such as those in~\cite{brany,emw}, 
yield recoil velocities as large as   
$v \sim {\cal O}(10^{-1})$ towards the end of inflation.)
However, we note here that if the string scale is itself of the 
order of a few TeV, then $q_e|{\cal H}|$ in 
(\ref{masssplit}) may be chosen of order one in string units
in order to reproduce supersymmetry breaking at TeV scales.

There is an issue with the scenario of~\cite{gravanis} concerning the
recoil velocity of the brane worlds after inflation. In~\cite{gravanis} it
was assumed that the brane worlds eventually stop moving in the bulk, as a
result of gravitational radiation, i.e., emission of closed strings from
the brane towards the bulk. This would imply that there were no
velocity-dependent contributions to the supersymmetry breaking in the bulk
asymptotically.  We stress that, if the branes are moving relative to each
other, the asymptotic vacuum energies on the brane world are non-zero, but
depend on some power of the recoil velocity (in the case of identical
recoiling branes this is $v^4$~\cite{emw}),
breaking the (bulk) supersymmetry.

On the other hand, in scenarios with an asymptotically linear dilaton
one has~\cite{aben}, as a result of the presence of the background charge 
$Q$, {\it tachyonic shifts} $-Q^2$ in the masses of {\it bosons}, while the
fermion masses remain {\it unaffected}.
Such shifts induce {\it additional} contributions to 
supersymmetry-breaking
mass splittings (\ref{masssplit}) asymptotically~\footnote{Note that, 
since during inflation the dilaton remains constant,
there are no extra shifts in the boson masses due to the central charge 
deficit in that era.}:
\begin{equation} 
\Delta m^2_{\rm susy-br} \sim q_e|{\cal H}| + O\left({\cal H}^2\right),
\label{masssplit2}
\end{equation} 
and in this type of breaking one has~\cite{bachas}
${\rm Str}m^2 =0$, where ${\rm Str}$ denotes the supertrace.

In the scenario of~\cite{brany}, one uses the supersymmetric vacuum
configuration of Fig.~\ref{fig:nonchiral}, where one or more of the branes
of one stack collide with branes of the other stack before returning to
their original position, where they collide for a second time, and
eventually stop. The end of the inflationary era in this framework
corresponds to this {\it second collision}.  We assume for simplicity that
there is only one collision between the Big Bang and the exit from
inflation, where our brane world collides with its original stack of
branes and stops.
This second collision results in a phase transition
and {\it reheats} the Universe, as a result of entropy production due to the 
collision. This second collision is much milder than the initial one, 
because the recoiling brane world 
may lose energy not only via its collisions with D-particles
in the foam, but also due to gravitational radiation, i.e., closed string 
emission in the bulk. The precise mechanism for reheating is still open:
one possible contribution is the gravitational collapse
of the bulk D-particles in the model of~\cite{emw} into black holes,
due to distortions of their populations following the second collision.
Evaporation of such bulky black holes on the brane worlds would
result in Hawking radiation, represented by open string excitations
attached to the brane worlds, thereby contributing to reheating.

In such scenarios, the asymptotic value of $Q^2 \sim {\cal H}^2$, since
there is no recoil velocity of the brane after the second collision. For
the order of magnitude of the magnetic field chosen above (\ref{horder}),
such contributions are negligible compared with the Zeeman mass splittings
(\ref{masssplit}). Moreover, for such values of the magnetic field, the
equilibrium central charge (\ref{sigmamodelq0}) is of order $Q_0^2 \sim
10^{-60}$ (in string units), and the value of $\gamma $ in such a Universe
is compatible with the current-era condition (\ref{condition}), provided
(c.f., (\ref{defA}),(\ref{defA2}), (\ref{largetwe})): $|C_5|e^{-5s_{02}}
\sim 1 $ (in string units), which is a natural value for the flux field in
the model of~\cite{dgmpp}.  This guarantees a present-era vacuum energy
(\ref{cosmoconst}) of the observed order, compatible with a
phenomenologically-viable scenario for supersymmetry-breaking mass
splittings (\ref{masssplit2}). On the other hand, if the string scale is
of the order of a few TeV, then the age of the Universe today is $t_E \sim
10^{44}$ in string units, and since $Q_0^2 \sim {\cal H}^2 \sim 1$ in
these units, one needs very large five-dimensional fluxes
$|C_5|e^{-5s_{02}} \sim 10^{44}$ to ensure the condition
(\ref{condition}).

The basic features of the low-energy limit of the non-supersymmetric
Type-0 string theory that we used in~\cite{brany} and in~\cite{dgmpp},
can be extended appropriately to the supersymmetric brane/orientifold
compactification model of~\cite{emw}, without affecting the basic
characteristics of the model, such as the existence of one large extra
dimension, the presence of flux bulk fields, tachyons and extra moduli
fields which freeze out quickly, and play no r\^ole in the phenomenology
of the Universe in the present era. As far as tachyons are concerned,
these fields existed in Type-0 string theory as a result of the explicit
breaking of supersymmetry due to projecting the partners out of the string
spectrum. In the supersymmetric models of colliding-brane
worlds~\cite{emw,brany}, the motion and collision of the brane worlds
breaks supersymmetry explicitly, both on the branes and in the bulk, as a
result of the non-zero relative velocities. This also results in tachyonic
excitations in the string spectrum, reflecting the instability of the
configuration. This instability is essential in cosmological situations,
such as the one we encounter here. The same analysis as in~\cite{dgmpp}
can then be performed for the bosonic sector of the low-energy field
theory in this case, to demonstrate the existence of solutions of
cosmological relevance, in which the tachyon fields decouple quickly,
leading to a similar late-stage analysis and results like those in
\cite{dgmpp,emn04}, as reviewed in the previous Section.

We would like to call the reader's attention to one final point. As
mentioned above, magnetized toroidal compactifications are known to have
Nielsen-Olesen instabilities~\cite{bachas}. It may well
happen~\cite{gravanis}, therefore, that as a result of the collision(s) a
{\it decompactification} process takes place at a rate slower than any
other time scale in our physical Universe, which implies, however, that
the compactification radius $R \to \infty$ asymptotically in cosmic time,
whilst the magnetic field energy ${\cal H}^2R^{p}$, for $p$ compact
dimensions on the brane worlds, remains {\it finite}. This would imply
vanishing magnetic fields asymptotically, and hence restoration of
supersymmetry.

However, the compactification on magnetized internal manifolds is not the
only way for supersymmetry to be broken in such cosmologies. As already
mentioned, in the models of~\cite{emw} the motion of the brane world
constitutes another source of breaking of supersymmetry. Moroever, as we
also discuss below, the thermalization of the bulk and brane worlds soon
after the collision could in principle result in yet another (independent)
contribution to supersymmetry breaking. However, the finite recoil
velocity of the colliding brane world and the temperature will be related,
and hence there will be only one independent type of supersymmetry
breaking in the scenario of~\cite{emw}~\footnote{Notice that the model
of~\cite{emw} does not involve compactification, and hence the
considerations on phases with broken supersymmetry pertain to
eight-dimensional brane worlds, moving in the ninth bulk dimension. Upon
subsequent compactification it is possible to have additional sources of
supersymmetry breaking, including the ones associated with possible
internal magnetic fields, as discussed elsewhere.}.

\subsubsection{Moving Branes and Supersymmetry Breaking}

The colliding-brane scenario can be realized~\cite{brany} in this
framework by allowing (at least one of) the D-branes to move, keeping the
orientifold planes static. One may envisage a situation in which the two
branes collide, at a certain moment in time corresponding to the Big Bang
- a catastrophic cosmological event setting the beginning of observable
time - and then bounce back. The width of the bulk region is assumed to be
long enough that, after a sufficiently long time following the collision,
the excitation energy on the observable brane world - which corresponds to
the conformal charge deficit in a $\sigma$-model
framework~\cite{gravanis,emw} - relaxes to tiny values.

It is expected that a ground state-configuration will be achieved when the
branes reach the orientifold planes again (within stringy length
uncertainties of order $\ell_s=1/M_s$, the string scale). In this picture,
since observable time starts ticking after the collision, the question how
the brane worlds started to move is merely philosophical or metaphysical.
The collision results in a kind of phase transition, during which the
system passes through a non-equilibrium phase, in which one loses the
conformal symmetry of the stringy $\sigma$ model that describes
perturbatively string excitations on the branes. At long times after the
collision, the central charge deficit relaxes to zero~\cite{gravanis},
indicating that the system approaches equilibrium again. The dark energy
observed today may be the result of the fact that our world has not yet
relaxed to this equilibrium value. Since the asymptotic ground state
configuration has static D-branes and D-particles, and hence has zero
vacuum energy as guaranteed by the exact conformal field theory
construction of~\cite{emw,brany}, it avoids the fine-tuning problems in
the model of~\cite{gravanis}.

Thus, the bulk motion of either the D-branes or the
D-particles~\footnote{The latter could arise from recoil effects following
scattering with closed-string states propagating in the bulk.} results in
non-zero `vacuum' (or, rather, `excitation') energy~\cite{emw}, and hence
the \emph{breaking of target-space supersymmetry}, proportional to some
power of the average (recoil) velocity squared, which depends on the
precise string model used to described the (open) stringy matter
excitations on the branes.  Sub-asymptotically, there are several
contributions to the excitation energy of our brane world in this picture.
One comes from the interaction of the brane world with nearby D-particles,
i.e., those within distances at most of order ${\cal O}(\ell_s)$, as a
result of open strings stretched between them.  The other contribution
comes from the collision of the identical D-branes.

A detailed analysis, using world-sheet methods for the computation
of the various potentials felt by the D-branes/D-particles in the 
colliding-brane model of~\cite{emw}, yields two types of effective
potentials. One is a potential in the bulk space,
felt by closed-string excitations from the gravitational multiplet
that are allowed to propagate in the bulk. The bulk potential 
is given by: 
\begin{equation}
{\cal V}_{\rm sym} \simeq V_8\frac{(30R-64r)v^4}{2^{13}\pi^9{\alpha '}^5}-
N\left(\frac{v}{\alpha '}\right)^{1/2},
\label{symmpot1}
\end{equation}
where the distances $R, r$ are defined in Fig.~\ref{fig:nonchiral},
$v$ is the recoil velocity of our brane world, and 
$N$ is the number of D-particles near the moving brane 
world, which are the only type of D-particles that contribute
significantly to the potential~\cite{emw}. A symmetric configuration
of branes has been considered in Fig.~\ref{fig:nonchiral} 
for concreteness and simplicity. For a sufficiently dilute
gas of nearby D-particles, one may assume that this latter contribution is
the dominant one. In this case, one may ignore the D-particle/D-brane
contributions to the vacuum energy, and hence apply the
previous considerations on inflation, based on the ${\cal O}(v^4)$ central
charge deficit, with $v$ the velocity of the brane world in the bulk.

The other type of potential, 
generated in the moving-brane scenario of~\cite{emw}, 
is an effective potential felt by the brane world itself, as 
a result of its interactions with the other branes
and D-particles. This second type of potential
is felt by the open-string excitations whose ends are attached to the 
brane, which constitute the Standard Model matter and radiation, 
living on the brane world. The brane potential is~\cite{emw}:
\begin{equation}
V_{\rm brane} =  -V_8\frac{31(R-2r)v^4}{2^{13}\pi^9{\alpha '}^5}~,
\label{branepotopen1}
\end{equation}
where $r_1 = R - 2r$ denotes the relative separation of the branes
in the symmetric situation of Fig.~\ref{fig:nonchiral} with $r_2~=~r$.
Notice that the potential is negative, 
which expresses the fact that 
the brane world feels an attractive force towards its original stack,
and the configuration is stabilized when $v \to 0$.
In Section 4 we return to a physical interpretation of 
the above potentials, which determine the various phases of 
our early Q-cosmology. 

From the point of view of the low-energy bulk action (\ref{actioniia}) and
(\ref{kalloshact}), the bulk potential (\ref{symmpot1})  would correspond
to a non-zero contribution to the scalar potential of the Type-IIA
supergravity theory, proportional to a central charge deficit:
$e^{-2\Phi}Q^2$, expressing the non-criticality of the associated $\sigma$
model describing bulk string excitations. The fact that the potential
(\ref{symmpot1}) changes sign, depending on the value of $r$, will lead to
a rich phase structure, as we discuss in Section 4. However, due to the
fact that we consider here a brane excitation, the system does not sit at
a global minimum of the potential, but rather in a local (metastable)
extremum.  We return to this important point in Section 4, when we discuss
the various phases of the bulk theory. As we show there, the compactified
Type-IIA theory may not be characterized by such a global minimum as a
result of purely stringy properties (lack of certain T-duality
symmetries~\cite{kounnas}).

For the effective low-energy of the open-string excitations on the brane,
a similar excitation `vacuum energy' is provided by the potential
(\ref{branepotopen1}), but with subtleties because its value is always
negative. As we discuss in Section 4, this may be interpreted as
thermalization of the brane world, throughout the inflationary period and
its exit phase (this mechanism is an alternative to the usual description
of reheating). Moreover, the presence of matter on the brane world causes
back-reaction onto the space-time, along the lines discussed earlier.
Matter is assumed to satisfy classical equations in an effective
four-dimensional supergravity field theory on the brane world. There are
of course subtleties associated with specific compactification scenarios,
which we do not discuss here.

A final comment concerns the r\^ole of the D-particles in the above
models. The presence of these space-time defects, which inevitably cross
the D-branes as the latter move in the bulk, even if the D-particle
defects are static initially, distorts slightly~\cite{papavass} the
inflationary metric on the observable brane world at early times after the
collision, during an era of approximately constant central charge deficit.
However, this effect does not lead to significant qualitative changes.
Moreover, the existence of D-particles on the branes affects the
propagation of string matter on the branes, in the sense of modifying
their dispersion relations by inducing local curvature in space-time, as a
result of recoil following collisions with string matter. However, it was
argued in~\cite{emnequiv} that only photons are susceptible to such
effects in this scenario, due to the specific gauge properties of the
membrane theory at hand.  The dispersion relations for chiral matter
particles, or in general fields on the D-branes that transform
non-trivially under the Standard Model gauge group, are protected by
special gauge symmetries in string theory, and as such are not modified.

\section{Finite Temperature in the Liouville Framework} 

We now discuss thermalized strings in the context of our Liouville
formalism, and describe the thermal phase diagram of our early Universe.

\subsection{Brane Collisions and Hot Universes}

In our colliding-brane scenario, each brane collision
thermalizes the string excitation spectrum
on the brane worlds and in the bulk, 
as a result of the conversion of the kinetic energy
of the moving branes into thermal energy. In the 
scenario with two moving colliding branes of~\cite{emw} (c.f., 
Fig.~\ref{fig:nonchiral}), the string excitations may be thermalized
immediately after the collision. 
Indeed, as the detailed computations of~\cite{emw} have shown,
the effective potential of the configuration when the 
two branes lie a distance $r_1 = R-2r$ apart (with the symbols as in 
Fig.~\ref{fig:nonchiral}, restricting ourselves to the 
symmetric case $r_2=r$ for simplicity) is given by (\ref{symmpot1}).

We observe from (\ref{symmpot1}) that for a {\it sufficiently dilute gas}
of D-particles, the potential is positive for $r \ll R/2$.
This implies that, for relatively 
long times after the collision when the distance of our brane world
from its original stack satisfies the above constraint $r \ll R/2$,  
closed-string excitations in the bulk
feel this positive vacuum energy, which means that 
the corresponding $\sigma$ model is {\it supercritical}.
It must therefore be dressed
by a time-like Liouville field, which is eventually identified 
with the target time. In fact, in the analysis of~\cite{gravanis},
for reasons associated with the convergence of the 
world-sheet path integral, we considered the initial time coordinate
$X^0$ (before Liouville dressing) as {\it space-like} ({\it Euclidean time}).
This was important, because dressing with a time-like Liouville
field implied a $(D+1)$-dimensional 
target-space metric (in our normalization here)~\cite{gravanis}:
\begin{equation}
ds^2_{D+1} = -2(d\varphi)^2 + (dX^0)^2 + d{\vec x}^2 .
\end{equation}
Upon the identification (\ref{liouvtime}) $\varphi = -t$, 
where now $t=X^0$ is a Euclidean
target time, one obtains a {\it Minkowski-signature} 
$D$-dimensional space-time in a dynamical way. 
Although in~\cite{gravanis} we viewed the use of Euclidean time merely
as a mathematical peculiarity of the world-sheet path integral,
it may be given a physical meaning in the context of 
the colliding-brane scenarios, as follows. 

Assuming that the adiabatic analysis of~\cite{emw} is valid soon after the
initial collision, and ignoring again the contributions from D-particles,
assuming them sufficiently dilute, we observe that in that early epoch of
the Universe the potential (\ref{symmpot}) is negative, since in that era
$30R/64 < r < R/2$ (c.f., Fig.~\ref{fig:nonchiral}). The closed-string
excitations find themselves described by a {\it subcritical} $\sigma$
model, which can become critical upon Liouville dressing by a {\it
space-like} Liouville mode. This correspond to {\it thermalization} as a
result of the collision, during which the initial kinetic energy of the
D-branes is transformed into thermal energy. In fact, if we assume that
the initial relative velocity of the D-branes is of the same order as the
recoil velocity, which in~\cite{emw,brany} was estimated to be of order
$10^{-3} < v < 10^{-1}$ in units of $c=1$, then we observe that the
induced temperature $\frac{1}{2}{\cal M} v^2 \sim k_B T$, where ${\cal M}$
is the D-brane mass, could be close to the Hagedorn temperature of the
corresponding string theory, $T \simeq T_H \sim \frac{1}{2\pi
\sqrt{2\alpha '}}$ in order of magnitude~\footnote{Slight differences in
the proportionality factors occur between the various string theories.}.  
Thus, during the collision phase, the branes and the bulk (closed) string
excitations find themselves at a finite (high) temperature.

It is interesting to describe the stringy excitations under such
conditions.  In what follows we review the r\^ole of the Liouville
formalism in describing generic strings at finite temperature. We commence
our analysis with the heterotic string case, which is the simplest and
among the most interesting cases for phenomenology.

\subsection{Liouville Approach to Finite-Temperature Strings: the Case of
Heterotic Strings}

Historically~\cite{kounnas}, there has been interest in obtaining a
description of a hot, stable phase of strings at temperatures beyond the
Hagedorn phase transition at $T_H \sim 1/2\pi \sqrt{\alpha '}$. In our
case, we are interested in the description of strings much below such high
temperatures. However, it is instructive for our purposes to review first
the Hagedorn phase, as studied for heterotic strings in~\cite{kounnas}. We
then return to the brane model of~\cite{emw}, characterized by a bulk
low-energy Type-IIA effective supergravity theory in the next Section.

The easiest approach to discussing strings at finite temperature $T$ is to
compactify the time direction on a circle of radius $R =1/\pi T$ and
discuss the mass spectrum of the winding modes of the string. Using
appropriate T-dualities the authors of~\cite{kounnas} have discussed the
instabilities arising from the fact that some of these T-winding modes of
the string become tachyonic above the Hagedorn temperature of a gas of
strings.  This defines the high-temperature phase of strings, and in our
case we could identify it with the epoch soon after the initial collision,
where the separation between the branes is small.

The presence of a non-zero temperature leads in general to additional
contributions to supersymmetry breaking beyond the ones discussed
so far. An important result~\cite{rostant} in the context of strings
is that $D$-dimensional superstrings at finite temperature look like 
$(D-1)$-dimensional superstrings with {\it spontaneously} broken 
supersymmetry.  Restricting our attention to the (bulk) closed string 
winding sector,
this observation implies~\cite{kounnas} that 
the corresponding low-energy effective 
supergravity field theory is characterized by a 
a non-zero (negative) value of the 
(global) minimum of its corresponding  scalar potential, 
proportional to the 
square of the gravitino mass:
\begin{equation}
V_{\rm min} = -2m_{3/2}^2\kappa^{-2} =-1/2S\kappa^{-2}~,
\label{minscalar}
\end{equation} 
where $S=e^{-\Phi}$ denotes the dilaton field in the supergravity multiplet,
and $\kappa $ is the (ten-dimensional) gravitational constant.  
From a stringy $\sigma$-model viewpoint, such a minimum
corresponds to the propagation of strings in a space-time
with a tree-level non-constant cosmological term, providing a
runaway-dilaton potential.
A detailed analysis of heterotic superstrings in
high-temperature phases has been performed in~\cite{kounnas}, where
it was shown that there exists a conformal field theory description
of this high-temperature phase, corresponding to a $\sigma$ model 
where the central charge has been lowered by four units.

Indeed, the conformal field theory is nothing other than the
strongly-coupled Liouville theory~\cite{ddk}, which is not yet very well
understood. This Liouville conformal theory corresponds in target space to
a {\it subcritical} superstring whose $\sigma$-model-frame metric is the
flat Minkowski one: $G^\sigma_{\mu\nu}=\eta_{\mu\nu}$, with a dilaton
linear in a {\it space-like} coordinate, playing the r\^ole of the
Euclidean compactified time $\Phi = QX^0$, where $Q$ is a background
charge. The space-like nature of the Liouville mode (temperature) is due
to the fact that there is a central-charge {\it deficit} and not a {\it
surplus} as in the time-like Liouville case of~\cite{aben,emn} discussed
in previous Sections. The physical metric, corresponding to a canonically
normalized Einstein term in the effective action, is again given by
(\ref{smodeinst}), and the corresponding target-space effective action by
(\ref{effaction}).  From the cosmological term of this action, and its
identification with the the minimum of the supergravity scalar potential
(\ref{minscalar}), one can compute the central-charge deficit $\delta c
\propto Q^2 <0$ of the conformal theory (paying particular attention to
the appropriate normalization factors~\cite{kounnas}), essentially by
identifying it with the numerical coefficient of $1/S$ in the expression
(\ref{minscalar}):
\begin{equation} 
Q^2 = \frac{\delta c}{8\alpha '} = -\frac{1}{2\alpha '} < 0,
\end{equation}
implying a central charge deficit: $\delta c = -4$ for the superstring.
Taking into account the fact that for flat $\sigma$-model-frame metric
backgrounds $\delta c = D-10$, where $D$ is the space-time dimensionality
of the free superstring, we therefore observe that in the high-temperature
phase the thermalized closed-string system corresponds to a non-critical
superstring living in 5+1 dimensions. In such a phase it was remarked
in~\cite{kounnas} that five-branes condense. Such features may turn out to
be quite important for cosmological model building. For instance, this
would imply that in the original model of colliding five-branes
of~\cite{gravanis}, immediately after the collision one would have
condensation of the five-branes, which does not happen in the model
of~\cite{emw}.

An important feature of such finite-temperature
superstrings is the existence of a space-like supersymmetry
at a perturbative level which characterizes the hot phase~\cite{kounnas}.
Indeed, before Liouville dressing, the finite temperature 
contributes to supersymmetry breaking mass shifts 
between the bosonic ($M_B$) and fermionic ($M_F$) 
excitations of the corresponding
supergravity theory:
\begin{equation} 
(M_B)^2_{i\bar j} = (M_F)^2_{i\bar j} - m_{3/2}^2\delta_{i\bar j}
\label{masshift}
\end{equation}
in the mass-matrix notation of~\cite{kounnas}.
After the Liouville dressing by the space-like linear dilaton, the 
fermion masses remain unaffected, but the 
boson masses undergo mass shifts~\cite{aben}. However, this time,
due to the subcriticality of the string, which is to be contrasted
with the case of~\cite{aben}, the mass shifts are not tachyonic,
but real:
\begin{equation}\label{qgravino}
\delta (M_B)^2 = Q^2 = m_{3/2}^2~, \quad \delta (M_F)^2 = 0 
\end{equation}
in our normalization. 
These mass shifts are additive to (\ref{masshift}), which 
implies that {\it at a perturbative level} supersymmetry
is restored in this hot phase of strings. 
In our case, therefore, this means that, at a perturbative level, 
supersymmetry breaking is still be given by the 
magnetic terms as described above. 

The supersymmetry is however broken, or rather {\it
obstructed}~\cite{witten} at a {\it non-perturbative} level, due to the
fact that masses in three space dimensions produce conical singularities,
and as such they break supersymmetry at the level of the excitation
spectrum, although the vacuum may still be supersymmetric.  Such
non-perturbative breaking has been discussed explicitly in~\cite{kounnas},
and we do not discuss it further here. We mention, though, that this
corresponds to an instability of the high-temperature phase, because this
breaking of supersymmetry produces tachyonic states.

The string must leave this unstable phase and re-enter a phase where such
instabilities eventually disappear, and the string system relaxes to an
equilibrium situation (with unbroken supersymmetry, modulo the effects of
the magnetic field, if compactification on magnetized manifolds is
considered).  We now discuss how this may be understood from a world-sheet
view point, in the context of our cosmological model of colliding branes
presented in~\cite{emw}.  Since the effective low-energy theory in the
bulk in this model is Type-IIA supergravity, we first review a
finite-temperature analysis of this special theory, with the aim of
repeating the analysis of~\cite{kounnas} for this case. There are
important physical differences, however, associated with the lack of
global minima in Type-IIA theories, which we outline in due course.  We
commence our analysis with a finite-temperature study of the effective
Type-IIA supergravity theory, which characterizes the low-energy bulk
dynamics in the model of~\cite{emw}.

\subsection{Type-IIA Supergravity at Finite Temperature}

As already mentioned, in~\cite{emw} we have a system of D8-branes and
orientifolds, in the configuration known as Type-IA string theory. Between
the two stacks of D8-branes, the bulk space corresponds to Type-IIA
supergravity. When some of the D8-branes move into the bulk, the overall
bulk potential induced by this motion can become negative~\cite{emw},
which can be interpreted as the system moving into a finite-temperature
phase. Finite-temperature field theory is realized by the Euclidean
compactification of the time dimension, and its effects can be calculated
using the Scherk-Schwarz mechanism~\cite{Scherk:1979zr}.

When one of the D8-branes from each stack moves into the bulk, there are
two potentials which must be taken into account. First there is the bulk
potential (\ref{symmpot}), describing the overall energy of the system,
where the potential is positive as long as the distance between the brane
and its originating stack, $r$, is less than $15R/32$. Secondly there is
the potential on the moving brane itself (as we are dealing with a
symmetric case, we consider the left-hand brane). This case is more
complex and will be discussed later on.

As stressed above, the important result when considering-finite
temperature supergravity is that $D$-dimensional superstrings at finite
temperature look like ($D-1$)-dimensional superstrings with spontaneously
broken supersymmetry~\cite{rostant,kounnas}. Thus, spontaneous
supersymmetry breaking via the Scherk-Schwarz mechanism is equivalent to
considering the system at finite temperature.  The Scherk-Schwarz
mechanism~\cite{Scherk:1979zr} works by generalizing the standard
dimensional reduction procedure~\cite{Antoniadis:1997wu}, in which all of
the fields are taken to be independent of the compact coordinates.
Instead, the fields are given a specific dependence on the internal
coordinates of the compact manifold, namely twisting the boundary
conditions of the compact dimensions by a global symmetry of the action.
This twist induces a shift in the mass terms of the lower-dimensional
fields.

In finite-temperature QFT~\cite{Atick:1988si}, bosons are
periodic and fermions anti-periodic in the compact Euclidean time
dimension~\cite{kounnas}:
\begin{equation}
\Phi(t+ 2L \pi R) = (-)^{La} \Phi(t),
\end{equation}
where for a $2\pi$ rotation $L=1$, and  $a=0,1$ for bosons and
fermions respectively. The modular 
invariance of Type~II string theory requires further constraints to be
placed on the periodicity conditions~\cite{rostant, kounnas,Atick:1988si}:
\begin{equation}
(-)^{aL+bn}
\end{equation}
where $m,n$ are winding numbers and $a$ and $b$ are the fermionic spin
structures along the world-sheet torus. The resulting shift in the
lattice momenta along the compact coordinate is~\footnote{The following 
discussion is taken from~\cite{kounnas}.}:
\begin{equation}\label{momenta shift}
p_{L,R} = \frac{m+a/2}{R}\pm \frac{nR}{2},
\end{equation}
with an additional sign factor of $(-)^{ab}$ which reverses the GSO
projection in the odd winding-number sectors. By redefining $m$, $a$
can be identified with the $D$-dimensional helicity operator ${\mathcal Q} =
\mathbb{Z}+a/2$, so that $(D-1)$-dimensional thermal states are mapped
to a supersymmetric theory on $S^1$ without the momentum shift
(\ref{momenta shift}). The helicity vector $\vec{{\mathcal Q}} = 
({\mathcal Q}_L , {\mathcal Q}_R)$ is
defined in terms of the left- and right-moving string helicities, and
the vector $\vec{e}=(1,1)$ for the Type II string. The inner product
is Lorentzian, $\vec{A} \cdot \vec{B} = A_L B_L - A_R B_R$. 

Thus there is a mapping between the $(D-1)$-dimensional supersymmetric
theory with quantum numbers $(n,m,{\mathcal Q})$ and the $(D-1)$-dimensional
thermal theory, which results in the quantum numbers of the
supersymmetric theory being shifted:
\begin{equation}
\left( \begin{array}{c}
 n \\ m \\ {\mathcal Q} 
\end{array}\right)
\longrightarrow 
\left( \begin{array}{c}
 n \\ m + {\mathcal Q} \cdot e +\frac{1}{2}ne\cdot e \\ {\mathcal Q} -ne
\end{array}\right).
\end{equation}
Clearly, all of the previously massless fermionic states  with $n = m = 0$
have their masses shifted to non-zero values, which means  that
supersymmetry is broken, with a supersymmetry-breaking  mass 
\begin{equation}
m_{3/2} = \frac{{\mathcal Q} \cdot e}{R}.
\end{equation}
In the case of Type-IIA string theory, the
vector product ${\mathcal Q} \cdot e = 1/2$~\cite{kounnas}, giving
a supersymmetry-breaking mass of  $m_{3/2} = 1/(2R)$. As already
noted, a $D$-dimensional theory at finite temperature is equivalent to
a $(D-1)$-dimensional theory with broken supersymmetry, so the radius
$R$ can be identified with the temperature of the system, $2\pi R=
T^{-1}$, giving   
\begin{equation}
\label{gravitino}
m_{3/2} = \pi T,
\end{equation}
from which it is clear that supersymmetry is restored when
$R\rightarrow \infty$, i.e., at zero temperature.

\subsection{Effective Potentials in Type-IIA Supergravity} 

The spontaneously broken $(D-1)$-dimensional theory can be used in certain
cases to determine the value of the gravitino mass, via minimization of
the scalar potential. As discussed in~\cite{kounnas}, for the case they
considered of $D=5$ heterotic theory at temperature, a global minimum was
found which could be used to calculate $m_{3/2}$ in terms of the dilaton
field. For the case of Type-IIA strings, a similar analysis was performed,
but the form of the scalar potential was such that there was no global
minimum.

This is understandable in view of the duality symmetries which occur
at finite temperature. At finite temperature the heterotic string
possesses a duality which relates the original Hagedorn temperature to
an upper Hagedorn temperature, above which the tachyon 
disappears~\cite{Atick:1988si, O'Brien:1987pn}:
\be
R \rightarrow \alpha^\prime/R, \hspace{20pt} T \rightarrow (4\pi^2
  \alpha^\prime T)^{-1}.
\ee 
This temperature duality of the heterotic string is directly related
to a duality in the scalar potential found by~\cite{kounnas}, the
existence of which appears to determine the existence of the global
minimum. For the five-dimensional Type-IIA string theory considered
in \cite{kounnas}, there is no such duality, thus no
global minimum. These considerations, however, concern the 
compactified theories~\footnote{The spontaneously broken supersymmetric 
$9$-dimensional effective theory,
representing the ten-dimensional supergravity at finite temperature, 
has a scalar superpotential proportional, as usual, to the 
square of the gravitino mass.}.
In the compactified Type-IIA case, the scalar potential assumes the form:
\begin{equation}
V = -\frac{1}{S} \cdot a(Z, \Omega, \dots) \propto -m_{3/2}^2,
\label{scpot}
\end{equation} 
where $S = e^{-\Phi}$ is the dilaton, and the (positive) function 
$a (Z, \Omega, \dots )$, with $Z$, $\Omega, \dots $ appropriate
moduli fields in the supergravity multiplet, 
is given in~\cite{kounnas} for the D=5 case. 
As discussed in~\cite{kounnas}, minimization with respect to the 
$\Omega$ field leads to a runaway potential in the $Z$ direction,
thereby leading to the absence of a global minimum, in accordance with the 
above-mentioned duality argument. 

The absence of a global minimum is not an unwelcome situation for the
cosmological model of~\cite{emw}, where the collision of branes causes an
excitation of the brane world, which no longer sits at its stable minimum
and becomes metastable. The excitation energy is determined in this case
by the bulk potential (\ref{symmpot}), which in turn is identified with
the central charge deficit of an appropriate non-critical $\sigma$ model,
describing (perturbative) string (bulk) excitations.

\subsection{Colliding-Brane Scenario, Non-Critical Strings and Effective 
Potentials in Type-IIA Theories} 

We now examine the previous case in some detail, with the aim of
understanding from a world-sheet framework the various hot and cold phases
of the theory.

\subsubsection{Thermal Type-IIA Phase following the Collision}

We return to the colliding brane scenario described in~\cite{emw},
in particular in the phase shortly after the first collision in the 
configuration of Fig.~\ref{fig:nonchiral}, 
when the relative separation $r_1$ of the colliding branes is 
\begin{equation}\label{negativepotregion}
r_1 \le \frac{R}{16} .
\end{equation}
In this region the bulk effective potential (\ref{symmpot}) is negative.

In the colliding-brane scenario~\cite{emw}, we are dealing essentially
with a \emph{non-equilibrium} situation.  The bulk potential
(\ref{symmpot}), therefore, should \emph{not} be viewed as indicating a
minimum value of a superpotential of the low-energy supergravity theory in
the bulk. Indeed, as we discussed in the previous Section, the effective
potential of Type-IIA supergravity does not have a global minimum.
Instead, we view the potential (\ref{symmpot}) as a \emph{non-equilibrium}
excitation energy of the vacuum due to the collision of the brane worlds.
From the point of view of the low-energy effective theory this is a
\emph{metastable vacuum} (local minimum), which is potentially interesting
in the cosmological context considered here.

Following the analysis in~\cite{emw}, we may associate 
the negative potential (\ref{symmpot}) with a central charge
deficit $Q^2 = C - c^* < 0$ of a \emph{subcritical} 
$\sigma$ model describing (perturbative) string 
bulk excitations in this hot phase. 
The analysis of~\cite{emw} assumed configurations of the bulk D-particles 
that were sufficiently 
dilute that the dominant contribution to the 
central-charge deficit, identified as the ten-dimensional
energy density corresponding to the potential (\ref{symmpot}),
is: 
\begin{equation}\label{ccd2}
 |Q^2| \simeq 1.2 \cdot 10^{-8} v^4 g_s^2,
\end{equation}
where $g_s$ is the string coupling, and 
$v$ is the brane-world recoil velocity, which
is constrained by WMAP~\cite{wmap}
data to be at most of order~\cite{emw,brany}: $v \le 0.8$
for the symmetric model of colliding branes of~\cite{emw}, as
depicted in Fig.~\ref{fig:nonchiral}.
This last relation is obtained upon compactifying (formally) 
the model into one large dimension along the ninth (bulk) direction,
and five small directions of order $\sqrt{\alpha '}$)~\footnote{The 
compactification issue is a non-trivial one in our case, and the resulting 
four-dimensional
supergravity may present complications. For our purposes here we only
present generic qualitative arguments, postponing a detailed compactification
analysis for a future publication.}.

Since, according to~\cite{rostant}, $D$-dimensional strings at finite
temperature are equivalent to $(D-1)$-dimensional strings with
spontaneously broken supersymmetry, we may view the effective target-space
supergravity theory corresponding to the hot phase of the colliding branes
of~\cite{emw} as living in 9 target dimensions, and corresponding to the
effective action of a non-critical string with an anti-de-Sitter
(negative) cosmological constant whose magnitude is given by (\ref{ccd}).
The pertinent nine-target-dimensional 
$\sigma$-model theory needs Liouville dressing~\cite{ddk}
to restore conformal symmetry, but with a 
\emph{space-like} Liouville mode. The pertinent
dressed $\sigma$ model is characterized 
by a flat Minkowski target-space metric $G_{\mu\nu}=\eta_{\mu\nu}$ 
and a background dilaton linear in 
the Liouville coordinate~\cite{aben}, which is viewed as a 
Euclidean time $X^0_E$: 
\begin{equation}\label{lindil}
\Phi = -\frac{1}{2}QX^0_E~.
\end{equation} 
The corresponding target space of the dressed theory is again 
ten-dimensional: $(9,X_E^0)$, and the corresponding effective
action in the $\sigma$-model frame is given by 
\begin{equation} 
S_{\rm \sigma-{\rm frame}} = \int_0^\beta dX^0_ E~d^{9}X 
\sqrt{G} e^{-2\Phi}\left( R - Q^2
+  4(\nabla _\mu \Phi)^2 + \dots \right) ,
\end{equation}
where $\beta = 1/2\pi T$ is the inverse temperature, which should
be compared with the appropriate parts of (\ref{actioniia}).
We see that the difference from (\ref{actioniia}) is the 
presence of a dark energy term proportional to $e^{-2\Phi}Q^2$,
which plays the r\^ole of a non-zero contribution to the
appropriate scalar potential, and is responsible for supersymmetry
breaking. Additional contributions/modifications 
will result from compactification, but for the purposes
of this Section we restrict ourselves to the 
uncompactified thermal case. 

The equations of motion obtainable from this action are equivalent to the 
conformal invariance conditions of the Liouville-dressed 
ten-target-dimensional stringy $\sigma$ model.
The dilaton equation (equivalently the vanishing of the 
ten-euclidean-dimensional dilaton $\beta$ function) reads~\cite{gsw}:
\begin{equation} 
R + 4 (\partial _\mu \Phi)(\partial ^\mu \Phi) - 4\Box \Phi = Q^2 ,
\end{equation}
from which we see that the linear-dilaton background~\cite{aben}
(\ref{lindil}) in a flat $\sigma$-model-frame target metric 
satisfies this equation, as expected from the fact that 
the Liouville dressing restores the conformal symmetry. 
This implies that this background is at least a \emph{local minimum}
of the action. Upon compactification of the type IIA theory, we know from 
the work of~\cite{kounnas} that there may be no global minimum, thereby 
making the above-mentioned extremum of the action a metastable vacuum.
This is a welcome fact, because this will lead to the cosmological
evolution of our brane world, and its eventual exit from this hot phase.

One may go one step further, and derive a relation between the temperature
of the hot phase and the recoil velocity by requiring a perturbative
space-like supersymmetry between bosonic and fermionic degrees of freedom,
as in the heterotic string case (\ref{masshift}),(\ref{qgravino}). Indeed,
since we \emph{have postulated} that the string theory describing the
excitations in the bulk of this situation is a \emph{subcritical
Liouville} string~\cite{emn,aben}, we know from the generic analysis
of~\cite{aben} that in such a non-critical string the bosonic masses will
acquire a shift by $Q^2$, as compared with the $Q=0$ case, while the
fermion masses remain unshifted (c.f., (\ref{qgravino})). We now
\emph{require} that there should be no supersymmetry breaking at the
perturbative level in the bulk theory, exactly as happens in the heterotic
string case. We base this postulate on duality symmetries between the
heterotic and Type-IIA theories. It means that the finite-temperature mass
shift of the gravitino (\ref{masshift})  should compensate the Liouville
shift (\ref{qgravino}). This would result to a restoration of a bulk
space-like supersymmetry, at the perturbative level. From the point of
view of the original model of~\cite{emw}, this supersymmetry restoration
would be compatible with the anti-de-Sitter nature of the bulk geometry in
the regime where the effective potential (\ref{symmpot}) is negative.

From (\ref{gravitino}) and (\ref{qgravino}), then, we may determine a 
relationship between the central charge deficit $Q^2$ 
of the Liouville $\sigma$ model, describing bulk string excitations,
and the temperature $T$. Furthermore, as we mentioned above, 
the analysis of~\cite{emw} relates the central charge deficit to the brane 
recoil velocity $v$ (\ref{ccd2}). The result of such an analysis is 
therefore:
\begin{equation} 
m_{3/2}^2 = \pi^2 T^2 = Q^2 \simeq 1.2 \cdot 10^{-8} v^4 g_s^2.
\label{qt}
\end{equation}
From the point of view of the 
spontaneously-broken nine-dimensional target-space theory,
this gravitino mass is proportional to the scalar potential at a 
local minimum. Upon compactification of the theory, this minimum 
is not a global one, as can be seen by an analysis 
similar to that of~\cite{kounnas}, 
mentioned previously. The metastable vacuum state 
of the thermal vacuum of the compactified Type~IIA theory 
can then be found by solving the appropriate dilaton 
equation. The cleanest method is to use the equation of motion 
in the Einstein frame (\ref{smodeinst}), 
where the gravitational curvature term
in the effective action has a canonical normalization.
The pertinent dilaton equation in a conformally flat target-space
background reads:
\begin{equation}\label{eqbox}
4\Box \Phi = -Q^2e^{2\Phi}.
\end{equation} 
The solution of such equations (of the compactified theory), together with
the Einstein equations, determines the metastable 
thermal Type-IIA vacuum corresponding 
to our case, with an excitation energy proportional to $Q^2 < 0$, 
given in (in magnitude) by (\ref{ccd2}). Notice that a dilaton 
of the form (\ref{dil2}) in the Euclidean-time Einstein frame 
satisfies the above equation for $D = 4$ uncompactified dimensions. 
The non-trivial issue in the higher-dimensional case of~\cite{emw}
is to find, upon compactification, the dependence of a 
dilaton satisfying (\ref{eqbox})
on the radii of the compact dimensions/moduli~\cite{kounnas}. 
We do not consider this issue further here.

We now remark that the equality 
(\ref{qt}) allows us to determine a recoil-velocity 
dependence of the temperature of the early phase of the brane Universe 
after the collision, in the adiabatic situation considered here:
\begin{equation}\label{tv2}
T \sim 10^{-4} 2\sqrt{2} g_s v^2/(2\pi \sqrt{2\alpha '})~\le~ 
1.28 \cdot 10^{-4}~T_H,
\end{equation}
where $T_H = 1/2\pi\sqrt{2\alpha '}$ is the respective Hagedorn temperature,
and we assumed standard weakly-coupled strings with $g_s^2 \sim 1/2$.
The fact that the temperature turns out to be proportional to $v^2$ 
is in agreement with the arguments given above on the 
transformation of most of the (non-relativistic) 
kinetic energy of the colliding branes
into thermal energy, in the adiabatic approximation we use here.

We see from (\ref{tv2}) that this early phase of the brane Universe, soon
after the collision, could be characterized by quite a high temperature,
up to $10^{13}$ GeV, if we accept that a typical string scale corresponds
to an energy of $10^{18}$~GeV~\cite{gsw}. Of course, the above estimate
has been obtained by saturating the upper bounds for the recoil velocity
that fit the WMAP data~\cite{emw,brany}, and in practice one may have
somewhat lower temperatures. In fact, lower temperatures may be required
in order to avoid massive gravitino overproduction. Such constraints would
restrict further the upper bound on the recoil velocities in the
(compactified version of the) model of~\cite{emw}.

However, despite the perturbative supersymmetry restoration,
one would have non-perturb- ative thermal instabilities, 
for the same reason as in the heterotic case examined 
above~\cite{kounnas}, associated with supersymmetry obstruction.
Such non-perturbative instabilities would result in 
the presence of tachyonic states in the string spectrum,
which could provide the initial cosmological instability.
As discussed in~\cite{dgmpp}, it seems to be a generic feature
of such tachyonic states to decouple quickly in the cosmological
Liouville evolution. In addition to these non-perturbative
instabilities, compactification of Type-IIA theories
leads to extra instabilities, due to the above-mentioned
lack of a global minimum in the low-energy effective scalar
potentials arising from thermal supersymmetry breaking.
The metastable nature of the hot phase of the Type-IIA vacuum 
leads to an exit from this phase, which is succeeded by a 
cold inflationary phase that we now proceed to discuss. 

\subsubsection{Inflationary Phase}
 
Some time after the initial collision,
the recoiling D-brane world's bulk potential (\ref{symmpot})
becomes {\it positive}. 
From a conformal field theory point of view, and in 
the adiabatic approximation we assumed in~\cite{emw},
this phase might be described by {\it an analytic continuation}
of the above linear-space-like dilaton solution:
\begin{equation}
Q \to iQ~, \quad X_E^0 \to i~t .
\label{analcont}
\end{equation} 
The corresponding $\sigma$-model-frame metric, which in the 
hot phase was a flat Minkowski metric, becomes now
conformally flat:
\begin{equation}
G_{\mu\nu} = e^{-2\Phi}\eta_{\mu\nu}~,~~\Phi = \frac{1}{2}Q~t  .
\label{cflat}
\end{equation}
In the model of~\cite{emw} this could be the full ten-dimensional metric,
although appropriate compactifications can restrict the indices to 
the four dimensions relevant for
a three-brane, as seen in Fig.~\ref{infla}, which represents
the inflationary model~\cite{brany}
reviewed in Section 2.3 above. The normalizations in (\ref{cflat}) 
pertain to the four-dimensional case of the three-branes, and would
change for higher-dimensional branes~\cite{emw}.

It is a curiosity that, setting $Q=-3H$, the physical (Einstein-frame)  
dilaton and metric fields (\ref{smodeinst}) of the hot phase, which remain
{\it real} under (\ref{analcont}), become equivalent to the corresponding
Liouville-undressed fields (\ref{cflat}) (c.f.,
(\ref{metricinfl}),(\ref{seventy})) when one sets the Liouville field
$\varphi = 0$. However, this is only a coincidence, since in the
inflationary phase it is the $\sigma$-model-frame metric that acquires the
conformally-flat form. As a $\sigma$-model-frame metric, (\ref{cflat}) is
\emph{not} conformal invariant, since its one-loop $\beta$ function is
non-vanishing: $\beta_{\mu\nu}^G = R_{\mu\nu} = Q^2 G_{\mu\nu} \ne 0$. The
central charge deficit $Q^2$ in this case is given by the potential
(\ref{symmpot})  in the regime in which is positive, which is treated as a
constant in the adiabatic approximation.

Because of the positive central charge deficit, the string system requires
Liouville dressing by a {\it time-like} field, $\varphi$, which is an
extra time-like coordinate, in addition to $t$.  The eventual
identification (\ref{liouvtime}), which in this scenario is dictated by
dynamical reasons~\cite{gravanis}, ensures that there is only one time
variable in the formalism, and leads to an eventual constant dilaton
during the inflationary phase. This phase in which the Universe cools down
is nothing other than the inflationary phase, described in Section 2.3
above. The analytic continuation procedure (\ref{analcont})  describes
simply a phase transition of the bulk superstrings from the hot phase to a
cold inflationary one, within the colliding-brane system of
Fig.~\ref{fig:nonchiral}.

We recall that, as a result of the non-perturbative breaking of
target-space supersymmetry in the hot phase, there are tachyonic states in
the spectrum, which trigger the initial cosmological instability. However,
as discussed in~\cite{dgmpp}, such states decouple relatively quickly in
the cosmic evolution.

\subsubsection{Exit from the Inflationary Phase: Reheating and Possible
Subsequent Collision(s)}

In a similar vein, one may discuss the phase transition associated with
the second collision, and the subsequent reheating of the Universe.  
However, the physics of reheating is not understood at a satisfactory
level in this framework. In the context of the model of~\cite{emw}, one
has to understand technical details associated with internal magnetic
fields in the respective orientifold compactification~\cite{dudas}, as
well as issues with the potential felt by open-string excitations on a
brane world. These issues still raise many open questions, but, for
completeness, we now present some relevant speculations.

The potential $V_{\rm brane}$ felt by our brane world in the 
model of~\cite{emw},
in the configuration of Fig.~\ref{fig:nonchiral}, is by itself negative
even during the inflationary phase~\cite{emn}, 
\begin{equation}
V_{\rm brane} =  -V_8\frac{31(R-2r)v^4}{2^{13}\pi^9{\alpha '}^5}~,
\label{branepotopen}
\end{equation}
where $r_1 = R - 2r$ denotes the relative separation of the branes
in the symmetric situation of Fig.~\ref{fig:nonchiral} with $r_2~=~r$.
This negative value can be understood by recalling that 
the brane world feels an attractive force towards its original stack,
and the configuration is stabilized when $v \to 0$.

An issue arises at this point, concerning the boundary conformal field
theory of open-string excitations, with their ends attached to the brane.
In view of this negative potential, one may think of dressing the
open-string $\sigma$ model with a {\it space-like} Liouville mode to
restore conformal symmetry, already during the inflationary phase.  In
view of the corresponding situation in the closed string
sector~\cite{kounnas}, discussed above, one is tempted to take the view
that such a negative brane potential represents some sort of
thermalization of open string excitations on the brane world during the
inflationary era.

Indeed, for $r < R/2$, i.e., very soon after the initial brane collision
{\it both} the brane and bulk Universes are {\it hot} and in thermal
equilibrium. As the two bouncing brane worlds of Fig.~\ref{fig:nonchiral}
move further apart, the closed-string excitations in the bulk cool down,
since the available space becomes larger and their collisions rarer,
whilst the brane Universe remains initially hot, because the inflationary
expansion is only a `mirage' due to the brane motion.

The order of magnitude of the brane potential (\ref{branepotopen}) is the
same as the bulk one (\ref{symmpot}), which implies that, after the exit
from the inflationary phase, our brane Universe remains thermalized with a
temperature at intermediate energy scales $\sim 10^{13}$ GeV, according to
the calculation above~\footnote{Such temperatures may characterize
no-scale supergravity models~\cite{noscale}.}. This may provide an
alternative to conventional reheating scenarios in the following sense:
although the bulk cools down significantly, and the Universe exits from
inflation, undergoing an appropriate phase transition, expressed by the
change in sign of the bulk potential (\ref{symmpot}), the brane world
remains thermalized after the exit from inflation at temperatures of order
$10^{13}$ GeV. This is simply a result of the initial collision, without
the need for other reheating mechanisms.

The usual constraints on gravitino overproduction in spontaneously-broken
supergravity models, such as those pertaining to the brane Q-cosmologies
of~\cite{emw}, restrict the allowed temperature to values much smaller
than $10^{13}$~GeV. This in turn implies an upper bound on the brane
recoil velocities, according to the discussion following (\ref{tv2}).  
However, one should bear in mind that in brane models the produced
gravitino will escape in the bulk, since it is an excitation of the closed
superstring multiplet, and therefore these constraints may not be so
strict as in conventional supergravity cosmologies.

This approach may provide an explicit realization of the ideas in the
ekpyrotic scenario for inflation and reheating of the
Universe~\cite{ekpyrotic}. Immediately after inflation there is a
difference in temperature between the bulk and the brane worlds.  As time
passes, this difference in temperature will cause significant closed
string (gravitational)  emission from the brane to the bulk, in order to
equilibrate the situation with a common (low) temperature in both brane
and bulk worlds.

Due to energy conservation, this causes non-adiabatic motion, with the
brane world decelerating towards an eventually zero velocity. This would
correspond to the exit phase from the inflationary epoch, given that the
central charge deficit of the pertinent stringy $\sigma$ model would
vanish asymptotically and, according to the discussion in Section 2.3, the
space-time metric would tend to that of a static flat Minkowski
space-time. In some models, however, e.g., those with internal magnetic
flux contributions, such as the Type-0 models discussed previously, the
asymptotic state may be that of a linear dilaton, leading to a linearly
expanding Einstein-frame Universe. In addition, as a result of brane
recoil effects also discussed above, one would have contributions to the
dark energy, relaxing asymptotically either to zero or to a constant
contribution (set, for instance, by the internal magnetic field
contributions in magnetized compactifications~\cite{bachas}), as in
(\ref{cosmoconst}) or (\ref{cosmoconst3}), (\ref{sigmamodelq0}).  In all
such models, current-era cosmology can be made compatible with
observations by fixing the various stringy
parameters~\cite{dgmpp,brany,emn04}.

\subsubsection{Open Issues: a Second Collision? Nucleosynthesis?
...} 

There are several open issues regarding the fate of the inflationary
Universe.  It is an open, and certainly model-dependent question whether
the gravitational radiation from the brane to the bulk makes the brane
world stop before the second collision takes place (in which case the
latter will never occur).  Indeed, if the brane gravitational radiation
causes a significant reduction of the brane velocity $v$ before the second
collision with the stack of D-branes in Fig.~\ref{fig:nonchiral}, it is
possible that the velocity contribution to the bulk potential
(\ref{symmpot}) diminishes significantly, in such a way that the
D-particle term overcomes the positive $v^4$ term. In such a case the bulk
potential becomes unstable (negative) and the bulk string system may
thermalize in the way described earlier. The thermalization may also have
a conformal field theory description, with the bulk background central
charge deficit being given by $-Nv^{1/2}$.  If the second collision takes
place before the brane stops due to radiation, there may be a local
disturbance in the population of the D-particles near the brane worlds
during the (second)  collision, such that $N$ is significantly higher than
before, when the brane was moving in the bulk. Such local disturbances may
even cause the massive D-particles to collapse forming black holes, whose
Hawking evaporation leads to additional thermal contributions to the brane
world after the second collision.

It is an open issue whether the potential energy of these re-thermalized
bulk closed strings becomes of similar order as the brane potential
(\ref{branepotopen}).  If such were the case, one would reach thermal
equilibrium between brane and bulk worlds, the gravitational radiation
towards the bulk could counterbalance the breaking up of bulk closed
strings on the brane, and the brane world would stop decelerating before
the second collision take place. The brane world could then either move
again adiabatically in the bulk with a very small velocity, in which case
there could still be a long time before it starts accelerating again due
to the influence of the other branes, either until the second collision
takes place, or until it stops.  The low (equilibrium) temperature in
either case could be identified with the CMB temperature of the present
era of the Universe.

\begin{figure}[tb]
\begin{center}
\epsfxsize=3.5in
\bigskip
\centerline{\epsffile{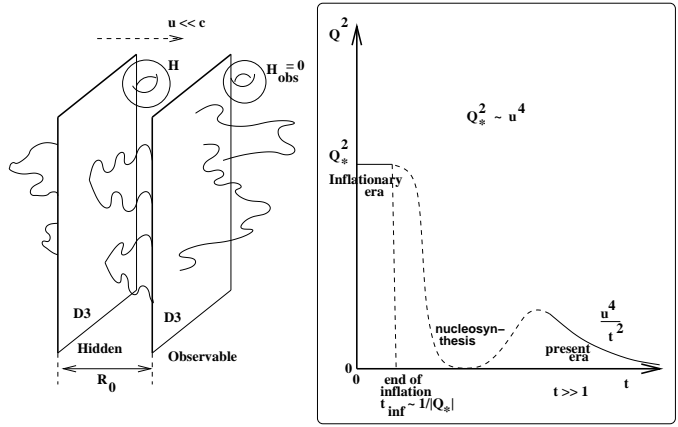}}
\caption{{\it A nucleosynthesis-friendly Liouville cosmology scenario,  
according to which the second collision of a  moving brane world 
with the 
(left) stack of branes in Fig.~\ref{fig:nonchiral}
occurs after the phase where the brane almost stops due to radiation.
This scenario provides 
a relaxation model for the cosmological vacuum energy (central-charge 
deficit of the Liouville $\sigma$ model), 
which passes first through a metastable phase where it almost vanishes 
(up to magnetic field contributions) during the nucleosynthesis era,
and then raises again,  as a result of the second collision,  
but at a much lesser height than in the initial collision.
\label{fig:nucleosynthesis}}}
\end{center} 
\end{figure}

Another open issue concerns the mechanism for \emph{nucleosynthesis} in
such a scenario.  Nucleosynthesis requires a delicate balance between the
expansion of the Universe and the rate of nuclear reactions for the
formation of the light elements, which appears to work very well in
scenarios with a negligible cosmological constant.  It may therefore be
desirable that the reduction in brane velocity due to radiation occurs
around the nucleosynthesis era, so that in such a case the brane Universe
has only a very small vacuum energy. For instance, in the class of models
with compactified branes in magnetized internal manifolds~\cite{gravanis}
it could be that only the magnetic field supersymmetry-breaking
contributions to the vacuum energy are present during the nucleosynthesis
era. At the end of nucleosynthesis a second collision of the brane world
with the stack of branes in Fig.~\ref{fig:nonchiral} takes place,
resulting in an increase of the central-charge deficit (vacuum energy),
but at a much lesser height than in the initial collision (due to the much
smaller velocities involved), as seen in Fig.~\ref{fig:nucleosynthesis}).  
Eventually the central charge relaxes again to zero asymptotically,
providing the vacuum energy in the present epoch.

The question then arises as to what precisely causes the current energy
density of the Universe in such a case.  So far we have argued that a
linear (in the string frame) dilaton may act as a quintessence field, in
accordance with current cosmological phenomenology.  However, it is our
opinion that, in order to answer this question completely, one should also
incorporate in the above discussion the {\it recoil} fluctuations on the
brane world, which echo the initial brane collision.  As mentioned above,
such effects would provide {\it positive} contributions to the present-era
central charge deficit of the corresponding stringy $\sigma$ model, for
asymptotically long times after the initial collision. The recoil
contributions depend on the recoil velocity of the branes during the
(adiabatic)  bouncing inflationary phase, but they diminish with the
cosmic time, relaxing towards either zero or some other small positive
(equilibrium) value, determined for instance by the internal magnetic
field (c.f., (\ref{cosmoconst}) or (\ref{cosmoconst3}),
(\ref{sigmamodelq0}) respectively). These recoil contributions may be
responsible for parts of the dark energy density of the observable
Universe, which exceed those due to the linear-dilaton quintessence.  
They could also be in accordance with current astrophysical
observations~\cite{snIa,wmap}. Such recoil contributions may overcome any
negative thermal contributions in the bulk, so that the bulk energy never
becomes negative, in contrast to the negative brane energy for non-zero
velocities.

These and other related issues are currently under investigation, and we
hope to be able to report some more complete results soon.

\section{Conclusions and Outlook} 

We have examined in this work various cosmological models based on
non-critical Liouville strings - Q-Cosmologies - with various asymptotic
configurations of the dilaton, and have speculated on the inflationary
phase, on the possibility of exit from it and reheating, as well as the
large-times eras of the Universe (current and future). A particularly
interesting case from a physical point of view is that of a linear dilaton
that is asymptotically linear in cosmic time, which is known to correspond
to a true conformal field theory~\cite{aben}.  In such a model we have
observed that the string coupling is identified (up to irrelevant
constants of order one)~\cite{emn04} with the deceleration parameter of
the Universe, through equation (\ref{important}). We have argued that the
present-era phenomenology of the model, including matter, is compatible
with the astrophysical data in a quite natural way, for suitable values of
the adjustable parameters in the model.

We stress once more the importance of being
non-critical in order to arrive at (\ref{important}). In critical strings,
which usually assume the absence of a four-dimensional dilaton, such a
relation cannot be obtained, and the string coupling is not directly
measurable with cosmological data. The logarithmic variation with the 
cosmic time of the dilaton
field at late times implies a slow variation of the string coupling
(\ref{important}), ${\dot g_s}/g_s = 1/t_E \sim 10^{-60}$ in the present
era, and hence a corresponding variation of the gauge coupling constants.
However, this variation is too small to be seen currently.

The use of Liouville strings to describe the evolution of our Universe is
natural, since non-critical strings are associated with non-equilibrium
situations which undoubtedly occurred in the early Universe.  We have
discussed in this framework the phase diagram of a Liouville cosmological
string model of two colliding-brane worlds. We have seen that, immediately
after the collision, the bulk string Universe passes through a hot,
metastable phase, before entering an inflationary cold phase. On the other
hand, the brane Universe (our world) remains thermalized throughout the
two phases, at a relatively high temperature, causing gravitational
radiation from the brane to the bulk, which tends to equilibrate the
temperature, which eventually decelerates the motion of the brane world in
the bulk. From the point of view of an observer on the brane, however, the
brane Universe may at present seem to be accelerating, with the
acceleration provided by the dilaton field of the string multiplet, as
mentioned above.

Exit from the inflationary phase is still an unresolved issue, although
scenarios have been conjectured, involving for instance a second collision
of the brane world of the model of~\cite{emw} with the stack of D-branes
in Fig.~\ref{fig:nonchiral}. This could provide extra contributions to the
reheating of the brane world, as a result of the gravitational collapse of
D-particle populations to form bulk black holes, which subsequently emit
Hawking radiation.

There are many phenomenological tests of this class of cosmologies that
can be performed, which the generic analysis presented here is not
sufficient to encapsulate. Tensor perturbations in the cosmic microwave
background radiation is one of them. The emission of gravitational degrees
of freedom from the hot brane to the cold bulk, during the inflationary
and post-inflationary phases is something to be investigated in detail. A
detailed knowledge of the dependence of the equation of state on the
redshift is something that needs to be looked at in the context of
specific models. The constant equation of state obtained here is only an
asymptotic feature of an era where the gravitational sector dominates.  
Moreover, issues regarding the delicate balance of the expansion of the
Universe and nucleosynthesis, which requires a very low vacuum energy,
must be resolved in specific, phenomenologically semi-realistic models,
after proper compactification to three spatial dimensions, in order that
the conjectured cosmological evolution has a chance of success.

Finally, the compactification issue \emph{per se} is a most important part
of a realistic stringy cosmology. In our discussion above, we have
presented a rather simplified compactification on magnetized internal
manifolds, in Type-II five-brane models, which also provided
phenomenologically realistic ways of breaking target-space supersymmetry
in cold Universes, compatible with the very small value of the vacuum
energy that has been reported in the Universe today. However, in the
context of the model of~\cite{emw}, involving eight-branes and
orientifolds, such compactifications may present subtleties that require
extra attention~\cite{dudas}.

We hope to be able to report on these and other related issues in future
work. We are far from claiming a detailed understanding in this framework
of several important facts of modern cosmology, such as the Universe's
current acceleration, dark energy, the various phase transitions in the
past history of the cosmos, {\it etc.}. Nevertheless, we believe that 
Liouville
strings are probably the only viable way, in the context of string theory,
to discuss rigorously cosmological string backgrounds, especially those
involving accelerated Universes and, in general, dark-energy contributions
to the Universe's energy budget.

In this last respect, we stress once more that the non-equilibrium
Liouville approach to cosmology advocated in this article is based
exclusively on the treatment of target time as an irreversible dynamical
renormalization-group scale on the world sheet of the Liouville string
(the zero mode of the Liouville field itself). This irreversibility is
associated with fundamental properties of the world-sheet renormalization
group, which lead in turn to the loss of information carried by
two-dimensional degrees of freedom with world-sheet momenta beyond the
ultraviolet cutoff~\cite{zam} of the world-sheet theory. This fundamental
microscopic time irreversibility may have other important consequences,
associated with fundamental violations of CPT
invariance~\cite{emninfl,emn,mavromatosdecoh} in both the early Universe
and the laboratory, providing other tests of these ideas.

\section*{Acknowledgments}

We thank V. A. Mitsou for useful discussions about the cosmological data. 
N.E.M. thanks the University of Valencia and IFIC for hospitality during
the final stages of this work. The work of D.V.N. is supported by D.O.E.
grant DE-FG03-95-ER-40917. That of M.W. is supported by an EPSRC (U.K.)  
Research Studentship.

\end{document}